\documentclass[11pt,a4paper]{article}
\usepackage{authblk}
\usepackage{amsmath,amssymb,amsfonts,bm,amsthm}

\DeclareFontFamily{U}{mathx}{}
\DeclareFontShape{U}{mathx}{m}{n}{ <-> mathx10 }{}
\DeclareSymbolFont{mathx}{U}{mathx}{m}{n}
\DeclareFontSubstitution{U}{mathx}{m}{n}

\usepackage{xpatch}
\xpatchcmd{\proof}{\itshape}{\bfseries}{}{}

\DeclareMathAccent{\widecheck}{0}{mathx}{"71}
\usepackage{cite}
\usepackage{mathrsfs}
\usepackage{mathalfa}
\usepackage{graphicx}
\usepackage{color}
\usepackage{lscape}
\usepackage{extarrows}
\usepackage{enumitem}
\usepackage{empheq}
\usepackage{listings}
\usepackage{graphicx}
\usepackage[dvipsnames]{xcolor}
\usepackage{chngcntr}
\usepackage{hyperref}
\usepackage{appendix}
\usepackage{tikz}
\usepackage{comment}
\usepackage{braket}
\usepackage{bbm}
\usepackage{pifont}
\usepackage{physics}
\usepackage[english]{babel}
\usepackage[skins,breakable]{tcolorbox}

\usepackage{shuffle}
\usepackage{tocloft}
\usepackage[inline,marginal,final]{showlabels}

\hypersetup{
	colorlinks=true, 
	linkcolor=violet,  
	urlcolor=blue,    
	anchorcolor=green, 
	citecolor=violet   
}

\newcommand{\as}{\alpha}
\newcommand{\ri}{\RN{1}}
\newcommand{\rii}{\RN{2}}
\newcommand{\ii}[3]{\int_{#1}^{#2}\mathrm{d}#3\,}

\newcommand{\id}{\mathbbm{1}}

\newcommand{\fin}{\mathrm{fin}}

\newcommand{\eq}[1]{Eq.\,(\ref{#1})}

\newcommand{\Fig}[1]{Fig.\,\ref{#1}}

\theoremstyle{plain}
\newtheorem{theorem}{Theorem}
\newtheorem{proposition}{Proposition}
\newtheorem{corollary}{Corollary}
\newtheorem{lemma}{Lemma}

\theoremstyle{definition}
\newtheorem{definition}{Definition}

\theoremstyle{remark}
\newtheorem{remark}{Remark}

\newcommand{\RN}[1]{%
	\textup{\uppercase\expandafter{\romannumeral#1}}%
}

\title{Clustering Theorem for Bose-Hubbard class Gibbs states}
\author{Xin-Hai Tong\thanks{\href{mailto:xinhai@iis.u-tokyo.ac.jp}{xinhai@iis.u-tokyo.ac.jp}}}
\affil{Department of Physics, The University of Tokyo, 5-1-5 Kashiwanoha, Kashiwa-shi, Chiba 277-8574, Japan}
\author{Tomotaka Kuwahara\thanks{\href{mailto:tomotaka.kuwahara@riken.jp}{tomotaka.kuwahara@riken.jp}}}
\affil{Analytical Quantum Complexity RIKEN Hakubi Research Team,
	RIKEN Center for Quantum Computing (RQC), 2-1 Hirosawa, Wako, Saitama 351-0198, Japan}
\affil{PRESTO, Japan Science and Technology (JST), Kawaguchi, Saitama 332-0012, Japan}
\author{Zongping Gong\thanks{\href{mailto:gong@ap.t.u-tokyo.ac.jp}{gong@ap.t.u-tokyo.ac.jp}}}
\affil{Department of Applied Physics, The University of Tokyo, 7-3-1 Hongo, Bunkyo-ku, Tokyo 113-8656, Japan}

\begin{document} 
\maketitle

\begin{abstract}
We establish the exponential clustering of correlation functions for the high-temperature Gibbs states of Bose-Hubbard type models. To overcome the technical difficulties arising from the unboundedness of bosonic operators, we develop the interaction-picture cluster-expansion technique. This method also allows us to systematically bound the moments of the local particle number. This result provides an analytical justification for the low-boson-density condition frequently assumed in the study of bosonic many-body systems. As direct mathematical consequences of the clustering property, we derive a uniform upper bound on the specific heat density and establish a bosonic thermal area law with improved temperature dependence.
\end{abstract}

\tableofcontents

\section{Introduction}\label{sec_intro}
Away from criticality, the exponential clustering of spatial correlation functions is a fundamental property of locally interacting lattice models with uniformly bounded interactions. For quantum spin and fermionic models characterized by finite-dimensional Hilbert spaces, this clustering property has been well-established in various settings, starting from the seminal work of Araki \cite{araki1969gibbs} in one dimension, and later extended to higher dimensions at high temperatures \cite{kliesch2014locality,frohlich2015some} or for gapped ground states \cite{hastings2006spectral,nachtergaele2006lieb}. 

However, analogous results for bosonic systems remain a long-standing open problem. The main difficulty arises from the infinite-dimensional local Hilbert spaces and the unboundedness of the operators of interest (e.g., creation and annihilation operators). These features make powerful techniques, such as the standard cluster expansion, ineffective. In this article, we primarily focus on the high-temperature Gibbs state for a class of bosonic models defined on a finite simple graph $(V,E)$, which is constituted by inhomogeneous nearest-neighbor (according to graph distance) hopping and squeezing, along with a quadratic on-site repulsive potential. This class includes the standard Bose-Hubbard model\label{reply_intro_page}, i.e.,
\begin{equation}\label{standard_BHM}
	\begin{aligned}[b]
		H_{\mathrm{BHM}}=-J\sum_{x\sim y}(a_{x}^{\ast}a_{y}+\mathrm{h.c.}) 
		+\sum_{x\in V}\qty[\frac{U}{2}n_{x}(n_{x}-1)-\mu n_{x}],\quad  n_{x}\coloneq a_{x}^{\ast}a_{x}
	\end{aligned}
\end{equation}
with real hopping strength $J$ and chemical potential $\mu$, and a repulsive strength $U>0$. Here, $a_{x}^{\ast}$ and $a_{x}$ are the usual creation and annihilation operators of a particle at site $x\in V$, and $x\sim y$ denotes nearest-neighbor pairs. Extensive studies have revealed the importance of the Bose-Hubbard model in various branches of physics, see \cite{cazalilla2011one,jaksch1998cold,bloch2008many} and references therein. 

To deal with these bosonic models, we develop an interaction-picture cluster expansion technique. We establish the following results beyond a temperature threshold independent of the system size (see Section \ref{section_mainresults_precise} for precise statements): 
\begin{itemize}
	\item[(i)] The magnitude of the thermal correlation of two operators decays exponentially with the distance between their supports.
	
	\item[(ii)] The thermal expectation value of the $s$-th order moment of the local particle number grows at most factorially with $s$. 
\end{itemize}
Here, the second claim, usually called the low-boson-density inequality, has been frequently introduced as a preliminary assumption in many rigorous studies on bosons \cite{kuwahara2024effective,yin2022finite,kuwahara2021lieb,kuwahara2024enhanced}, yet a rigorous justification has been lacking unless the state is trivial (e.g., a product state). As corollaries, at high temperatures, these results directly imply a bosonic version of the thermal area law \cite{wolf2008area} with an improved temperature scaling compared to previous work \cite{lemm2023thermal}, as well as a uniform upper bound for the specific heat density.

Historically, the clustering of correlations in bosonic settings has attracted much attention over the past several decades. For continuous Bose gases whose Hamiltonians consist of a Laplacian plus a two-body interaction, Ginibre \cite{Ginibre1971} utilized the Feynman-Kac formula to represent the correlation functions in terms of Wiener integrals. This probabilistic framework allows one to apply standard measure-theoretic tools to establish the clustering of correlations at high temperatures, and it was also adapted to unbounded lattice spin systems by Park and Yoo \cite{Park1995}. However, due to the lack of a continuous Laplacian to serve as a smoothing kernel, such methods cannot be directly applied to lattice bosonic models. For bosonic lattice models such as the standard Bose-Hubbard model \eqref{standard_BHM}, by investigating the analyticity of the free energy density and introducing an effective boson norm, Ueltschi \cite{Ueltschi1999} managed to obtain partial results, achieving exponential clustering for two observables that grow at most linearly with the local particle number. By adapting the Feynman-Kac method to represent the partition function via random walks, similar results have been established for the hard-core Bose-Hubbard model \cite{Fernandez2006}. Later on, coherent-state functional integrals were introduced in a series of papers by Balaban, Feldman, Knörrer, and Trubowitz, see \cite{Balaban2008,Balaban2008a,Balaban2010,Balaban2010a,Balaban2010b}. With the help of this framework, a conditional proof of the clustering property for bosonic lattice models was achieved by Lohmann \cite{Lohmann2015} for sufficiently small chemical potentials, under a technical ``small-field hypothesis". More recently, Salmhofer \cite{Salmhofer2021} developed an elegant regularized coherent-state functional integral for lattice bosons, explicitly pointing out its relation to the aforementioned probabilistic framework. These results may also be useful for studying the properties of correlation functions.
\label{reply_literature_compare}

It is worth emphasizing that all the aforementioned results regarding bosonic models focus on Hamiltonians where the total particle number is conserved, a restriction that is not required for our method and results. Furthermore, our approach not only ensures the optimal temperature scaling of the low-boson-density inequality but also yields an improved thermal area law. Finally, we derive an explicit temperature scaling for the pre-exponential factor in the correlation bounds, which precisely captures the intrinsic thermal fluctuations of unbounded bosonic operators at high temperature.

The rest of the paper is organized as follows. In Section \ref{sec-setup}, we introduce the setup, notations, and precise statements of the main results. In Section \ref{sec_main_technique}, we provide a detailed explanation of our newly developed techniques. In Sections \ref{sec_proof_main_theorem} and \ref{sec_proof_techinical_lemma}, we present detailed proofs for the main theorems and the supporting lemmas, respectively. Applications of these results are detailed in Section \ref{sec_application}. In Section \ref{sec_discussion}, we present some further discussions. We summarize this paper in Section \ref{sec_conclusion} with some future prospects. Throughout this article, for the statements of supporting propositions and lemmas, we use capital letters $C, C_1, C_2 \dots$ to denote generic constants that may change from statement to statement. For the proofs of all results other than the main theorems and corollaries, we use lowercase letters $c, c_{1}, c_{2} \dots$ for generic constants that may change from line to line.

\section{Mathematical Setup and Preliminaries}\label{sec-setup}

\subsection{Graphs}
Let $\mathfrak{d} > 1$ be a fixed integer. Consider a finite, simple, connected graph $(V, E)$ with maximum degree bounded by $\mathfrak{d}$, i.e., $\deg(x) \le \mathfrak{d}$ for all $x \in V$. Here, the finite vertex set $V$ represents the lattice sites, and the edge set $E$ consists of unordered pairs of distinct vertices. We denote the cardinality of subsets $\Lambda \subseteq V$ and $G \subseteq E$ by $|\Lambda|$ and $|G|$, respectively. Throughout this paper, letters $x, y, z$ denote sites, and $\lambda$ denotes an edge. \label{reply_degree}

Edges can be naturally viewed as two-element vertex subsets. For edges $\lambda, \lambda' \in E$, we write $\lambda \cap \lambda' \neq \emptyset$ if they share a common vertex, and $\lambda \subseteq \Lambda$ if both endpoints of $\lambda$ lie in $\Lambda \subseteq V$. For any edge subset $G \subseteq E$, we define its complement $G^c \coloneq E \setminus G$, its extension $\overline{G} \coloneq \{\lambda \in E \colon \exists \lambda' \in G \text{ s.t. } \lambda \cap \lambda' \neq \emptyset\}$, and its boundary $\partial G \coloneq \overline{G} \setminus G$. Let $V_G \coloneq \{x \in V \colon \exists \lambda \in G \text{ s.t. } x \in \lambda\}$ be the set of vertices incident to the edges in $G$. As will be clear later, we refer to $|V|$ as the system size.

The distance $\operatorname{dist}(x, y)$ between sites $x, y \in V$ is defined as the standard graph distance, namely, the minimum number of edges in a path (a sequence of linked edges) connecting them. We say $x$ and $y$ are nearest neighbors, denoted by $x \sim y$, if $\operatorname{dist}(x, y) = 1$. For subsets (also referred to as regions or subsystems) $X, Y \subseteq V$, the distance between them is defined as $\operatorname{dist}(X, Y) \coloneq \min_{x \in X, y \in Y} \operatorname{dist}(x, y)$.

A vertex subset $\Lambda \subseteq V$ is connected if any $x, y \in \Lambda$ can be joined by a path with all vertices entirely in $\Lambda$. Two vertex subsets $X, Y \subseteq V$ are connected by an edge subset $G \subseteq E$ if there exists a path entirely within $G$ that connects a vertex in $X$ to a vertex in $Y$. An edge subset is connected if any pair of its edges can be connected by a sequence of edges entirely within the subset.\label{reply_def_growth_constant}

The graph has a growth constant $\sigma$ independent from the system size, defined as the smallest positive number such that for any $m \in \mathbb{N}_{>0}$ and any edge $\lambda \in E$, the number of connected edge subsets of size $m$ containing $\lambda$ is bounded by $\sigma^m$. Precisely, \begin{equation}\label{def_growth_constant}
	\begin{aligned}[b]
		\sup_{\lambda\in E}|\{G\subseteq E \text{ is connected }\colon G\ni \lambda,|G|=m\}|\leq \sigma^{m}.
	\end{aligned}
\end{equation}
As shown by Klarner \cite{Klarner1967} and discussed in Penrose \cite{Penrose1994}, $\sigma$ satisfies the bound $\sigma \leq e\mathfrak{d}$, where $e$ is Euler's number.

A prototypical example is a graph generated by a finite subset of a $d$-dimensional hypercubic lattice, with the vertex set denoted by $V_{\text{fc},d} \subset\subset \mathbb{Z}^d$. The edge set $E_{\text{fc},d}$ consists of all nearest-neighbor bonds, and the graph distance corresponds to the Manhattan distance. For this lattice, the maximum degree $\mathfrak{d} = 2d$ is sufficient. In this case, the growth constant is bounded by $2de$ \cite{MejiaMiranda2011}.

\subsection{Bose-Hubbard Class of Hamiltonians}\label{sec_bhclass_hamiltonian}
We work extensively with the bosonic Fock space. For a finite lattice $V$, we take $\ell^2(V)$ as the single-particle Hilbert space. The corresponding bosonic Fock space is defined as \label{reply_sec_bhclass_hamiltonian}
\begin{equation}\label{}
	\begin{aligned}[b]
		\mathcal{F}[\ell^2(V)] \coloneq \bigoplus_{n=0}^{\infty} P_{+}[\ell^{2}(V)^{\otimes n}],
	\end{aligned}
\end{equation}
where $P_{+}[\ell^{2}(V)^{\otimes n}]$ is the $n$-fold symmetric tensor product of $\ell^2(V)$ and we set it to be $\mathbb{C}$ for $n=0$. This total Fock space is separable by construction. The subspace of states with a finite total number of particles, denoted by
\begin{equation}\label{}
	\begin{aligned}[b]
		\mathcal{D}_{\mathrm{fin}} \coloneq \{\psi \in \mathcal{F} \colon \exists\, n_0 \in \mathbb{N}_{0} \text{ s.t. } \psi_n = 0 \text{ for all } n \geq n_0\},
	\end{aligned}
\end{equation}
forms a dense subset of $\mathcal{F}[\ell^2(V)]$.

For any single-particle state $\xi \in \ell^2(V)$, the creation and annihilation operators, $a^*(\xi)$ and $a(\xi)$, satisfy the canonical commutation relations (CCR):
\begin{equation}\label{}
	\begin{aligned}[b]
		[a(\xi), a(\xi')] = [a^*(\xi), a^*(\xi')] = 0  ,\quad [a(\xi), a^*(\xi')] = ( \xi, \xi' ) \id,
	\end{aligned}
\end{equation}
where $( \cdot, \cdot )$ denotes the inner product on $\ell^2(V)$. These operators are defined on the dense subspace $\mathcal{D}_{\mathrm{fin}}$. See more detailed discussion on bosonic Fock space in \cite{Bratteli2012}. Let $\delta_x \in \ell^2(V)$ denote the localized state at site $x \in V$ such that $\delta_{x}(y)=\delta_{x,y}$. We naturally identify the creation and annihilation operators of a particle at $x$ as $a^*_x \coloneq a^*(\delta_x)$ and $a_x \coloneq a(\delta_x)$, respectively.

Using the well-known exponential property of Fock spaces, $\mathcal{F}(\mathfrak{h}_1 \oplus \mathfrak{h}_2) \cong \mathcal{F}(\mathfrak{h}_1) \otimes \mathcal{F}(\mathfrak{h}_2)$, and the direct sum decomposition $\ell^2(V) = \bigoplus_{x \in V} \mathbb{C}$, we can identify the total Fock space with the tensor product of local Hilbert spaces: $\mathcal{H} \coloneq \mathcal{F}[\ell^2(V)] \cong \bigotimes_{x \in V} \mathcal{H}_x$. Here, each local space $\mathcal{H}_x \cong \ell^2(\mathbb{N}_{0})$ is associated with a specific site $x \in V$. Under this isomorphism, the local particle number operator $n_x \coloneq a^*_x a_x$ is diagonal, and its eigenvectors $\{\ket{k}_x\}_{k \in \mathbb{N}_{0}}$ form the canonical orthonormal basis of $\mathcal{H}_x$. The total particle number operator is given by $N \coloneq \sum_{x \in V} n_x$. Since $N$ admits a direct sum decomposition of finite-dimensional blocks over the finite-particle sectors, it is essentially self-adjoint on $\mathcal{D}_{\mathrm{fin}}$. \label{reply_sec_bhclass_hamiltonian_2}

For any subset $X \subseteq V$, we define the local Hilbert space as $\mathcal{H}_X \coloneq \bigotimes_{x \in X} \mathcal{H}_x$. A product basis vector in this space is characterized by a configuration $\bm{k} = (k_x)_{x \in X} \in \mathbb{N}_{0}^X$, denoted as $\ket{\bm{k}} \coloneq \bigotimes_{x \in X} \ket{k_x}_x$. Since $V$ is finite, any such basis vector possesses a finite total particle number. The domain of finite-particle vectors over $X$ is defined as the algebraic span 
\begin{equation}\label{Dfin_span_X}
	\mathcal{D}_{\fin}(X) = \operatorname{span} \{ \ket{\bm{k}} : \bm{k} \in \mathbb{N}_{0}^{X} \}.
\end{equation}
Consequently, the dense domain in the full space is naturally recovered as $\mathcal{D}_{\mathrm{fin}} = \mathcal{D}_{\mathrm{fin}}(V)$.

Throughout this paper, we denote the identity operator on $\mathcal{H}_x$ by $\id_x$, and the identity on $\mathcal{H}_X$ by $\id_X \coloneq \bigotimes_{x \in X} \id_x$. Unless otherwise specified, all unbounded operators considered in this work are implicitly assumed to share $\mathcal{D}_{\mathrm{fin}}$ as a common core, which is sufficient to include any polynomial functions of $\{a^*_x, a_x\}_{x \in V}$. Finally, we follow the convention of using $\mathcal{B}(\mathcal{H})$ to denote the algebra of all bounded linear operators on $\mathcal{H}$.

Next, given uniformly bounded real parameters \begin{equation}\label{uniformbound}
	\begin{aligned}[b]
		|J_{xy}|\leq J, \quad|\widetilde{J}_{xy}|\leq J, \quad |\mu_{x}|\leq \mu,\quad 0<U_{\min}\leq U_{x}\leq U_{\max}, \quad\forall x,y \in V,
	\end{aligned}
\end{equation}
the inhomogeneous Bose-Hubbard class of Hamiltonians on the graph $(V, E)$ is defined by
\begin{equation}\label{bh}
	\begin{aligned}[b]
		H&=-\sum_{x\sim y}(J_{xy}a_{x}^{\ast}a_{y}+\widetilde{J}_{xy}a_{x}^{\ast}a^{\ast}_{y}+\mathrm{h.c.}) 
		+\sum_{x\in V}\qty[\frac{U_{x}}{2}n_{x}(n_{x}-1)-\mu_{x}n_{x}] 
		\\&\eqcolon -\sum_{\lambda\in E}h_{\lambda}+\sum_{x\in V}W_{x}\eqcolon -I+W.
	\end{aligned}
\end{equation}
For convenience, for each edge $\lambda = \{x, y\}$, we define $h_{\lambda} \coloneq J_{xy} a_x^* a_y + \widetilde{J}_{xy} a_x^* a_y^* + \mathrm{h.c.}$ to collect the corresponding hopping and squeezing terms. \label{reply_bh_def} We emphasize that the squeezing terms are also well-motivated in physics (due to parametric driving \cite{McDonald2018}) and have been experimentally realized \cite{Slim2024}. 

Note that the on-site potential term $W$ is essentially self-adjoint on $\mathcal{D}_{\mathrm{fin}}$. By the following proposition, the operator $H$ is also mathematically well-defined.

\label{reply_hamiltonian_ESA}
\begin{proposition}\label{pro_self_adjoint_bh}
	The Hamiltonian $H$ \eqref{bh} over a finite graph $(V, E)$ is essentially self-adjoint on the finite-particle subspace $\mathcal{D}_{\mathrm{fin}}$.
\end{proposition}
\begin{proof}[Sketch of Proof]
	The proof relies on a standard application of the Kato-Rellich theorem, noting that the on-site potential grows quadratically with $n_x$, thus dominating the hopping and squeezing terms. Detailed derivations are provided in Appendix \ref{app_well_define}.
\end{proof}

We denote the domain of an operator $A$ by $\mathcal{D}(A)$ and its closure by $\overline{A}$. For symmetric operators $A, B$ defined on a common dense domain $\mathcal{D}_{\mathrm{fin}}$, we write $A \le B$ if $B - A$ is a positive operator, i.e., $(\psi, (B - A)\psi) \ge 0$ for all $\psi \in \mathcal{D}_{\mathrm{fin}}$. If both $A$ and $B$ are essentially self-adjoint on $\mathcal{D}_{\mathrm{fin}}$ and lower bounded (i.e., $A, B \ge cI$ for some $c \in \mathbb{R}$), the inequality $(\psi, A\psi) \le (\psi, B\psi)$ extends to all $\psi \in \mathcal{D}(\overline{A}) \cap \mathcal{D}(\overline{B})$, since $\mathcal{D}_{\mathrm{fin}}$ is a core for both operators. The following proposition ensures the stability (boundedness from below) of both the Hamiltonian $H$ and the on-site potential $W$. \label{reply_proposition_2}

\begin{proposition}\label{pro_stability}
	The Hamiltonian $H$ over a finite graph $(V, E)$ and the on-site potential $W$ satisfy the lower bounds $H \ge U_{\min} \sum_{x \in V} n_x^2 / 4 - C_{\mathrm{sta}, V}$ and $W \ge U_{\min} \sum_{x \in V} n_x^2 / 4 - C_{\mathrm{sta}, V}$ for some constant $C_{\mathrm{sta}, V}$ depending only on $|V|$, $U_{\min}$, and $\mu$.
\end{proposition}
\begin{proof}
	This follows immediately from the standard operator inequalities $\pm(a_x^* a_y + \mathrm{h.c.}) \le n_x + n_y$ and $\pm(a_x^* a_y^* + \mathrm{h.c.}) \le n_x + n_y + \id$, combined with elementary algebraic calculations. This result is physically expected, as the on-site quadratic interaction term $W$ naturally dominates the linear hopping and squeezing terms.
\end{proof}

In the remainder of this work, by a slight abuse of notation, we will use $O$ to denote both an essentially self-adjoint operator on $\mathcal{D}_{\mathrm{fin}}$ and its unique self-adjoint closure $\overline{O}$, provided no confusion arises. For any subsystem $\Lambda \subseteq V$, the local Hamiltonian is naturally defined as $H_{\Lambda} \coloneq -I_{\Lambda} + W_{\Lambda}$, where $I_{\Lambda} \coloneq \sum_{\lambda \subseteq \Lambda} h_{\lambda}$ and $W_{\Lambda} \coloneq \sum_{x \in \Lambda} W_x$. The subsystem Hamiltonian $H_{\Lambda}$ shares the same self-adjointness and stability properties as the total Hamiltonian $H$. We also denote the subsystem particle number operator by $N_{\Lambda} \coloneq \sum_{x \in \Lambda} n_x$.

\subsection{Correlation Function and Norms}\label{sec_corrfunction_norm}
The main focus of this work is the clustering of correlations in the bosonic Gibbs state at high temperatures. \label{reply_page_boltweight_traceclass} In light of Proposition \ref{pro_self_adjoint_bh}, we can define the Boltzmann weight at inverse temperature $\beta > 0$ via the spectral theorem:  \begin{equation}\label{boltzmann_factor_spectrum}
	\begin{aligned}[b]
		e^{-\beta H}=\int_{\sigma(H)}\mathrm{d}E(\lambda)\,e^{-\beta \lambda},
	\end{aligned}
\end{equation}
where $\sigma(H)$ is the spectrum of $H$ and $E(\cdot)$ is the associated projection-valued measure. Throughout this paper, we restrict our attention to the positive temperature regime. The following proposition ensures that $e^{-\beta H}$ is also trace-class.

\begin{proposition}\label{pro_trace_class}
	For the Bose-Hubbard Hamiltonian defined in \eqref{bh} over a finite graph $(V, E)$, the operator $e^{-\beta H}$ is trace-class for any $\beta > 0$.
\end{proposition}
\begin{proof}[Sketch of Proof]
	The proof relies on bounding the Hamiltonian $H$ from below by a quadratic function of the local particle numbers $\{n_x\}_{x \in V}$. Detailed derivations are provided in Appendix \ref{app_well_define}.
\end{proof}

By Proposition \ref{pro_trace_class}, the Gibbs state is well-defined and we denote it by $\rho_{\beta} \coloneq e^{-\beta H} / \operatorname{Tr}(e^{-\beta H})$. Here, $\operatorname{Tr}$ represents the trace over the entire Hilbert space $\mathcal{H}$. We naturally use the canonical basis of $\mathcal{D}_{\mathrm{fin}}$ to compute the trace of a trace-class operator $O$, namely, \begin{equation}\label{trace_Dfin}
	\begin{aligned}[b]
		\operatorname{Tr}(O)=\sum_{\bm{k}\in \mathbb{N}_{0}^{V}}\mel{\bm{k}}{O}{\bm{k}}.
	\end{aligned}
\end{equation}

For the partial trace over $\mathcal{H}_{\Lambda}$ with $\Lambda \subseteq V$, we explicitly write $\operatorname{Tr}_{\Lambda}$, meaning the summation runs over $\bm{k} \in \mathbb{N}_0^{\Lambda}$. We denote the partition function by $\mathcal{Z}(\beta) \coloneq \operatorname{Tr}(e^{-\beta H})$. An operator $O$ is said to be supported on a subset $X \subseteq V$ if it acts trivially on the complement $X^c \coloneq V \setminus X$. Formally, this means $O$ admits a tensor product decomposition $O = O_X \otimes \id_{X^c}$, where $\id_{X^c}$ is the identity operator on $\mathcal{H}_{X^c}$. When no confusion arises, we slightly abuse notation and use $O_X$ to represent the extended operator $O_X \otimes \id_{X^c}$ on the entire space $\mathcal{H}$. The support of $O$, denoted by $\operatorname{supp}(O)$, is the smallest subset $X$ for which the above decomposition holds. For any operator $O$ such that $O e^{-\beta H}$ is trace-class, its thermal expectation value at inverse temperature $\beta$ is defined as $\langle O \rangle_{\beta H} \coloneq \operatorname{Tr}(O \rho_{\beta})$. The following proposition guarantees that this average is well-defined for all operators growing at most polynomially with the particle numbers.

\begin{proposition}\label{pro_thermal_average}
If an operator $O$ whose domain contains the dense subspace $\mathcal{D}_{\text{fin}}$ grows at most polynomially with the total particle number $N$, i.e., there exist constants $C_1, C_2 > 0$ such that $\|O\psi\| \le \|(N+C_1)^{C_2}\psi\|$ for all $\psi \in \mathcal{D}_{\mathrm{fin}}$, then the operator $Oe^{-\beta H}$ is trace-class.
\end{proposition}
\begin{proof}[Sketch of Proof]
	The intuition is that the Hamiltonian grows quadratically with the local particle numbers, and the exponential factor $e^{-\beta n^2}$ dominates any polynomial of $n$ for large $n \in \mathbb{N}_{0}$. Detailed derivations are provided in Appendix \ref{app_well_define}.
\end{proof}
\begin{remark}
	If $O_1$ and $O_2$ are operators growing at most polynomially with $N$, one can show that their product $O_1 O_2$ also exhibits polynomial growth. This holds provided that $O_2$ is a finite linear combination of terms that shift the particle number by a finite amount. That is, $O_2 = \sum_j O_2^{(j)}$, where for each monomial $O_2^{(j)}$ there exists a constant $c_j \in \mathbb{R}$ such that $N O_2^{(j)} = O_2^{(j)} (N+c_j)$. 
\end{remark}

This proposition motivates us to introduce the following \label{reply_poly_operator} notation.
\begin{definition}[Polynomial-Type Operators]
	For a subset $X \subseteq V$, let $\mathcal{P}_X(\bm{a}, \bm{a}^*)$ denote the algebra of polynomials in the creation and annihilation operators $ (a_x)_{x \in X}$ and $ (a^*_x)_{x \in X}$ over $\mathbb{C}$. For a single site $x$, we adopt the shorthand $\mathcal{P}(a_x, a_x^*)$.
\end{definition}
\begin{remark}
	This class covers most physically relevant observables. While one could theoretically extend our results to observables with sub-exponential growth with respect to $\{n_x\}_{x \in V}$, we focus on the polynomial case to maintain clarity and avoid unnecessary technical complications regarding domains of unbounded operators.
\end{remark}

Proposition \ref{pro_thermal_average} enables us to define correlation functions for a rather large class of operators, which is a central focus of this article. Let two disjoint subsets $X, Y \subset V$ be the supports of $O_X$ and $O_Y$, respectively. If $O_X e^{-\beta H}$, $O_Y e^{-\beta H}$, and $O_X O_Y e^{-\beta H}$ are all trace-class, we define their correlation function at inverse temperature $\beta$ as 
\begin{equation}\label{correlation}
	\begin{aligned}[b]
		C_{\beta}(O_X,O_Y)\coloneq \operatorname{Tr}(O_{X} O_{Y}\rho_{\beta})-\operatorname{Tr}(O_X\rho_{\beta})\cdot \operatorname{Tr}(O_Y\rho_{\beta}).
	\end{aligned}
\end{equation} 

On the separable Hilbert space $\mathcal{H}$, for any $p \in [1, \infty)$, let $\mathcal{S}_p(\mathcal{H})$ denote the space of compact operators on $\mathcal{H}$ whose singular values $\{s_{\ell}(O)\}_{\ell=1}^{\infty}$ are $p$-summable, i.e., $\sum_{\ell=1}^{\infty} s_{\ell}(O)^p < \infty$. For $O \in \mathcal{S}_p(\mathcal{H})$, its Schatten $p$-norm is defined as 
\begin{equation}\label{pnorm}
	\begin{aligned}[b]
		\norm{O}_{p}\coloneq \qty[\sum_{\ell=1}^{\infty}s_{\ell}(O)^{p}]^{1/p},\quad p\in [1,\infty).
	\end{aligned}
\end{equation}
Note that for $p = \infty$, the Schatten norm coincides with the standard operator norm, which we will refer to simply as the norm for brevity. In this work, we will frequently use Hölder's inequality 
\begin{equation}\label{holder_inequality}
	\begin{aligned}[b]
\|O O'\|_p \le \|O\|_{p_1} \|O'\|_{p_2}		
	\end{aligned}
\end{equation}
with $p^{-1} = p_1^{-1} + p_2^{-1}$, for $O \in \mathcal{S}_{p_1}(\mathcal{H})$ and $O' \in \mathcal{S}_{p_2}(\mathcal{H})$ \cite{Simon2005}. \label{reply_schatten}In particular, we rely on the Cauchy-Schwarz case $\|O O'\|_1 \le \|O\|_2 \|O'\|_2$ to bound the trace norm of products, as well as the standard estimates $\|O O'\|_1 \le \|O\|_1 \|O'\|$ and $|\operatorname{Tr}(O)| \le \|O\|_1 = \operatorname{Tr}(|O|)$ for expectation values. For convenience, we adopt the following convention: if $O_X$ is an operator with support $X \subseteq V$, its singular values $\{s_{\ell}(O_X)\}_{\ell=1}^{\infty}$ are computed within the  tensor factor $\mathcal{H}_X$ rather than the full space $\mathcal{H}$. Consequently, for disjoint subsets $X, Y \subset V$, we have $\|O_X O_Y\|_p = \|O_X\|_p \|O_Y\|_p$ given $O_X \in \mathcal{S}_p(\mathcal{H}_X)$ and $O_Y \in \mathcal{S}_p(\mathcal{H}_Y)$.

For an unbounded operator $O$, the standard operator norm $\|O\|$ is ill-defined. However, the physically relevant operators in this work typically take the form $O e^{-\beta H}$ for some $O \in \mathcal{P}_X(\bm{a}, \bm{a}^*)$ with $X \subseteq V$. Proposition \ref{pro_thermal_average} guarantees that $O e^{-\beta H}$ is trace-class. Since $\mathcal{S}_1 \subset \mathcal{S}_p$ for all $p > 1$, it follows that $O e^{-\beta H}$ has well-defined Schatten $p$-norms for all $p \ge 1$.

\subsection{Main Results: Precise Statements}\label{section_mainresults_precise}
In this subsection, we state our two main theorems and their immediate applications. \label{reply_main_theorem_page}

\begin{theorem}[Low-Boson-Density Inequality]\label{theorem_low_density}
	Let $\rho_{\beta}$ be the Gibbs state of the Bose-Hubbard model \eqref{bh} defined on a finite graph $(V,E)$ at inverse temperature $\beta>0$. There exist constants $\beta_c > 0$ and $K_{\mathrm{low}} > 0$, depending only on the model parameters $U_{\min}, U_{\max}, J, \mu$, and $\mathfrak{d}$, but independent of the system size $|V|$ and the site $x \in V$, such that
	\begin{equation}\label{low_density_main_present}
		\begin{aligned}[b]
			\operatorname{Tr}(n_{x}^{s}\rho_{\beta}) \leq K_{\mathrm{low}}^{s} \cdot s! \cdot \beta^{-s/2}
		\end{aligned}
	\end{equation}
	holds for all $\beta \in (0, \beta_c]$ and $s \in \mathbb{N}_0$.
\end{theorem}
\begin{remark}
	The temperature scaling of inequality \eqref{low_density_main_present} is optimal, as it coincides with the exact scaling of the on-site case (i.e., $J_{xy} = \widetilde{J}_{xy} = 0$). See Sec.\,\ref{sec_discussion} for detailed discussions. Furthermore, for any $X \subseteq V$, the inequality \eqref{low_density_main_present} allows one to upper bound the thermal expectation value of any operator in $\mathcal{P}_{X}(\bm{a}^{\ast},\bm{a})$ at high temperatures. 
\end{remark}

\begin{remark}
	By invoking Stirling's approximation, inequality \eqref{low_density_main_present} can be recast into the form $\langle n_{x}^{s} \rangle_{\beta H} \leq e^{-1}(e^{-1}\kappa_{1}s^{\kappa_{2}})^{s}$, where $\kappa_{1}$ scales as $\beta^{-1/2}$ and $\kappa_{2}=1$. This form matches exactly the low-density assumptions commonly used in the literature; see, e.g., inequality (3) in \cite{kuwahara2024effective}.
\end{remark}

\begin{theorem}[Boson Clustering Theorem at High Temperatures]\label{theorem_clustering}
	Let $\rho_{\beta}$ be the Gibbs state of the Bose-Hubbard model \eqref{bh} defined on a finite graph $(V,E)$ at inverse temperature $\beta>0$. Let $X, Y \subset V$ be two disjoint regions. Let $C_{\beta}(O_{X},O_{Y})$ be the correlation function of observables $O_{X}\in \mathcal{P}_{X}(\bm{a}^{\ast},\bm{a})$ and $O_{Y}\in \mathcal{P}_{Y}(\bm{a}^{\ast},\bm{a})$, as defined in \eqref{correlation}. 
	Then, there exist constants $\beta_c, K_{\mathrm{cl}}, C_{1}, C_{2} > 0$, depending only on the model parameters $U_{\min}, U_{\max}, J, \mu$, and $\mathfrak{d}$, but independent of the system size $|V|$ and the subsets $X, Y$, such that
	\begin{equation}\label{}
		\begin{aligned}[b]
			|C_{\beta}(O_{X},O_{Y})|\leq K_{\mathrm{cl}}^{|X|+|Y|}\|O_{X}e^{-\sqrt{\beta}N_{X}}\|\|O_{Y}e^{-\sqrt{\beta}N_{Y}}\|e^{-\operatorname{dist}(X,Y)/\xi(\beta)}
		\end{aligned}
	\end{equation}
	holds for all $\beta \in (0, \beta_c]$, where the correlation length is given by $\xi(\beta)=-\qty{\ln [\sigma C_{1}\beta^{1/2} /(1-C_{2}\beta^{1/2})]}^{-1}$.
\end{theorem}
\begin{remark}
We remark that the temperature scaling of the prefactor $\|O_{X}e^{-\sqrt{\beta}N_{X}}\|$ precisely matches the divergent behavior of the physical on-site fluctuations in the on-site case, demonstrating the tightness of this regularization scheme. See Sec.\,\ref{sec_discussion} for details.
\end{remark}

The main theorems established above directly lead to the following physical consequences regarding the specific heat density and the mutual information of the system. Here, the specific heat density is defined directly via energy fluctuations, i.e., 
\begin{equation}\label{def_c_v}
	\begin{aligned}[b]
		\mathcal{C}_{V}(\beta)\coloneq |V|^{-1}\beta^{2}\qty(\langle H^{2} \rangle_{\beta H}- \langle H \rangle^{2}_{\beta H}).
	\end{aligned}
\end{equation}
To introduce the mutual information, we first consider a bipartition of the system into vertex subsets $V=A\sqcup B$. The Bose-Hubbard Hamiltonian \eqref{bh} can be decomposed as 
\begin{equation}\label{eq:ham_decomp}
	H=H_{A}+H_{B}+H_{\partial},
\end{equation}
where $\partial=\{\lambda\in E \colon \lambda \cap A \neq \emptyset, \lambda \cap B \neq \emptyset\}$ represents the set of edges overlapping both subsystems, and $H_{\partial}\coloneq- \sum_{\lambda\in \partial}h_{\lambda}$ describes the coupling between subsystems $A$ and $B$. Let $(\rho_{A},\rho_{B})\coloneq (\operatorname{Tr}_{B}(\rho_{\beta}), \operatorname{Tr}_{A}(\rho_{\beta}))$ denote the reduced density operators of the Gibbs state. The mutual information between the subsystems is defined as
\begin{equation}\label{def_mi}
	\mathcal{I}(A:B)\coloneq S(\rho_{A})+S(\rho_{B})-S(\rho_{\beta}),
\end{equation}
where $S(\cdot)$ is the von Neumann entropy. Then we list the corollaries of the main theorems as follows,

\begin{corollary}[Quasi Dulong-Petit Law]\label{corollary_dulong}
	Let $\mathcal{C}_V(\beta)$ be the specific heat density of the Bose-Hubbard model \eqref{bh} on a finite graph $(V,E)$. There exist a temperature threshold $\beta_c > 0$ and a constant $C > 0$, both independent of the system size $|V|$, such that $\mathcal{C}_V(\beta) \leq C$ holds for all $\beta \in (0, \beta_c]$. Here, $\beta_c$ and $C$ depend only on the model parameters $U_{\min}, U_{\max}, J, \mu$, and $\mathfrak{d}$.
\end{corollary}

\begin{corollary}[Boson Thermal Area Law]\label{corollary_area_law}
	Let $\mathcal{I}(A:B)$ be the mutual information of the Bose-Hubbard model \eqref{bh} on a finite graph $(V,E)$. There exist a temperature threshold $\beta_c > 0$ and a constant $C > 0$, both independent of the system size $|V|$, such that $\mathcal{I}(A:B) \leq C\cdot \beta \cdot |\partial A| $ holds for all $\beta \in (0, \beta_c]$. Here, $\beta_c$ and $C$ depend only on the model parameters $U_{\min}, U_{\max}, J, \mu$, and $ \mathfrak{d}$, with $|\partial A|$ being the boundary size of $A$.
\end{corollary}

\section{Main Technique: Interaction-Picture Cluster Expansion}\label{sec_main_technique}
The expansion of physical quantities with respect to a series of countable objects called clusters, first proposed in \cite{mayer1941molecular}, has proven to be a powerful tool for exploring various problems in statistical mechanics \cite{ruelle1969statistical,alhambra2023quantum,kliesch2014locality}. However, due to the unbounded nature of bosonic operators, the conventional cluster expansion used to prove clustering properties \cite{kliesch2014locality} is not directly applicable. This section is devoted to presenting a detailed description of our interaction-picture cluster expansion technique, which serves as the starting point for the proofs of our main theorems. 

By using functional calculus, one can verify that the Boltzmann weight defined in \eqref{boltzmann_factor_spectrum} satisfies 
\begin{equation}\label{cauchy_problem_H}
	\begin{aligned}[b]
		\frac{\mathrm{d}}{\mathrm{d}t} e^{-t H} \psi = -H e^{-t H} \psi
	\end{aligned}
\end{equation}
for any $\psi\in \mathcal{D}(H)\supset \mathcal{D}_{\fin}$ and $t>0$. Here, the symbol $\mathrm{d}/\mathrm{d}t$ denotes the strong derivative, i.e., taken in the norm topology of the Hilbert space $\mathcal{H}$. Let us define the map $\mathcal{G}(\tau) = e^{-(\beta-\tau)H} e^{-\tau W}\psi$ for $\tau \in [0, \beta]$ and $\psi \in \mathcal{D}_{\mathrm{fin}}$. Note that both $e^{-(\beta-\tau)H}$ and $e^{-\tau W}$ are bounded operators by spectral calculus. The derivative of $\mathcal{G}(\tau)$ with respect to $\tau$ is given by $-e^{-(\beta-\tau)H}Ie^{-\tau W}\psi$ by using \eqref{cauchy_problem_H} with $t=\beta-\tau$ and a similar equation for $W$. Then, by integrating with respect to $\tau$ over $[0,\beta]$, we obtain
\begin{equation}\label{duhamel_psi}
	\begin{aligned}[b]
		e^{-\beta H}\psi = e^{-\beta W}\psi+ \ii{0}{\beta}{\tau}e^{-(\beta-\tau)H}Ie^{-\tau W}\psi.
	\end{aligned}
\end{equation} 
We remark that this integration is well-defined due to the following result. 

\begin{proposition}\label{pro_bochner}
	For the Bose-Hubbard model \eqref{bh} defined on a finite graph $(V, E)$, and for any $\psi \in \mathcal{D}_{\mathrm{fin}}$, the map $[0, \beta] \ni \tau \mapsto e^{-(\beta-\tau)H} I e^{-\tau W} \psi \in \mathcal{H}$ is Bochner integrable.
\end{proposition}
\begin{proof}[Sketch of Proof]
It suffices to show that the norm of the function $\tau \mapsto e^{-(\beta-\tau)H} I e^{-\tau W} \psi$ is integrable over the interval $[0,\beta]$. See details in Appendix \ref{app_well_define}.
\end{proof}
\label{reply2_prop5}

Note that \eqref{duhamel_psi} holds for all $\psi\in \mathcal{D}_{\fin}$. By the bounded linear transformation (BLT) theorem, we can extend it to the entire Hilbert space and obtain 
\begin{equation}\label{duhamel}
	\begin{aligned}[b]
		e^{-\beta H}=e^{-\beta W}+ \ii{0}{\beta}{\tau}e^{-(\beta-\tau)H}Ie^{-\tau W},
	\end{aligned}
\end{equation}
with the integral operator defined as $(\int\mathrm{d}s\, \mathcal{G}(s))\psi\coloneq \int\mathrm{d}s\, \mathcal{G}(s)\psi$. This formula is commonly known as Duhamel's formula. We iterate \eqref{duhamel} to obtain the following Dyson series:
\begin{equation}\label{dyson}
	\begin{aligned}[b]
		e^{-\beta H}=\sum_{m=0}^{\infty}\ii{0}{\beta}{\tau_{1}}\ii{0}{\tau_{1}}{\tau_{2}}\dots\ii{0}{\tau_{m-1}}{\tau_{m}} e^{-(\beta-\tau_{1})W}Ie^{-(\tau_{1}-\tau_{2})W}Ie^{-(\tau_{2}-\tau_{3})W}\dots e^{-(\tau_{m-1}-\tau_{m})W}Ie^{-\tau_{m}W},
	\end{aligned}
\end{equation}
with $m=0$ representing the zero-order term $e^{-\beta W}$. The series \eqref{dyson} corresponds to the imaginary-time interaction picture in physics, and its absolute convergence is guaranteed by the following result.

\begin{proposition}\label{pro_dyson_converge}
	For the Bose-Hubbard model \eqref{bh} defined on a finite graph $(V, E)$, the corresponding Dyson series \eqref{dyson} absolutely converges in norm topology for any $\beta >0$.   
\end{proposition}
\begin{proof}[Sketch of Proof]
	Since $I$ scales as $W^{1/2}$, the singular part of $\|Ie^{-\tau W}\|$ scales as $\tau^{-1/2}$. One can then show that for each fixed $m$, the absolute upper bound of the integral scales as $C_{V}^{m}/\Gamma(m/2+1)$ with $C_{V,\beta}$ depending only on $|V|,\beta$ and model parameters ($U_{\min}, U_{\max}, J, \mu$, and $\mathfrak{d}$), which ensures absolute convergence. See details in Appendix \ref{app_well_define}.
\end{proof}

The absolute convergence of the Dyson series serves as the starting point for the following cluster expansion. For a more compact notation, we denote $O_{X}(\tau)\coloneq e^{\tau W}O_{X}e^{-\tau W}$ as the imaginary-time evolution for $O_{X}\in \mathcal{P}_{X}(\bm{a},\bm{a}^{\ast})$. Since $W = \sum_{x\in V} W_x$ consists of on-site terms, we have $\operatorname{Supp}(O_{X}(\tau))=\operatorname{Supp}(O_{X})=X$, which implies $O_{X}(\tau)= e^{\tau W_{X}}O_{X}e^{-\tau W_{X}}$. Here, for a subset $\Lambda\subseteq V$, we denote $W_{\Lambda}\coloneq \sum_{x\in \Lambda}W_{x}$. 
Recalling that $I=\sum_{\lambda\in E}h_{\lambda}$ [cf.\,\eqref{bh}], we substitute this into \eqref{dyson} to obtain
\begin{equation}\label{dyson_expand}
	\begin{aligned}[b]
		e^{-\beta H}&=\sum_{m=0}^{\infty}\ii{0}{\beta}{\tau_{1}}\ii{0}{\tau_{1}}{\tau_{2}}\dots\ii{0}{\tau_{m-1}}{\tau_{m}} e^{-\beta W} \prod_{i=1}^{m}I(\tau_{i})
		=\sum_{m=0}^{\infty}\int_{\vec{\tau}\in \mathcal{T}_{m}}\mathrm{D}\vec{\tau}\,e^{-\beta W}\prod _{i=1}^{m}\qty[\sum_{\lambda\in E}h_{\lambda}(\tau_{i})]
		\\&=\sum_{m=0}^{\infty}\sum_{\qty{\lambda_{k}\in E}_{k=1}^{m}}\int_{\vec{\tau}\in \mathcal{T}_{m}}\mathrm{D}\vec{\tau}\,e^{-\beta W}\prod_{i=1}^{m}h_{\lambda_{i}}(\tau_{i}),
	\end{aligned}
\end{equation}
where $\prod_{i=1}^{m}$ denotes the ordered product. 
To obtain the second equality in \eqref{dyson_expand}, we introduced the shorthand $\mathrm{D}\vec{\tau}\coloneq \mathrm{d}\tau_{1}\mathrm{d}\tau_{2}\dots\mathrm{d}\tau_{m}$ with $\vec{\tau}\coloneq (\tau_{1},\tau_{2},\dots,\tau_{m})$ and used $\mathcal{T}_{m}\coloneq \{\vec{\tau}\in \mathbb{R}^{m}\colon  0\leq \tau_{m}\leq \dots \leq \tau_{2}\leq \tau_{1}\leq \beta\}$ to denote the hyper-triangular region.
The summation on the right-hand side (RHS) of \eqref{dyson_expand} motivates us to introduce a more compact notation customized for cluster expansion. 

To address this technique in a general form, we first give several new definitions. Following the terminology of free Lie algebras \cite{frost2023lie,reutenauer2003free}, we identify the edges as letters and the edge set $E$ as the alphabet.\label{reply_def_word} Then the key concept is the word. 

\begin{definition}[Word]\label{def_word}
	A word $w$ is an element of $E^{\times m}$ for some $m \in \mathbb{N}_{>0}$. The integer $m$ is called the length of the word and will be denoted by $|w|$. We can represent it as $w=(w_1, w_2, ..., w_{|w|})$ with $w_i \in E$. Let $G_w \subseteq E$ be the set obtained by deleting repeated elements in $w$.
\end{definition}
\begin{remark}
 By construction, $G_{w}$ is also an edge subset for any word $w$. Naturally, we define $V_{w}\coloneq \{x\in V\colon \exists \lambda\in w \text{ s.t. } \lambda\ni x\}$ as the support of the word $w$. We also denote $w\circ u \coloneq (w_{1},w_{2},\dots,w_{|w|},u_{1},u_{2},\dots,u_{|u|})$ as the concatenation of two words $w$ and $u$.
\end{remark}

Recall that letters may repeat multiple times within a word. To rigorously describe this property, we introduce the concept of multiplicity. 
\begin{definition}[Multiplicity]
	For any word $w$ and edge $\lambda$ contained in $w$ (denoted by $\lambda \in w$), we define the edge-wise multiplicity $\mu_{\lambda}(w)$ as the number of occurrences of the edge $\lambda$ in the word $w$.
\end{definition}
\begin{remark}
	Consequently, the length of a word $w$ can be expressed as the sum of all its edge-wise multiplicities:
	\begin{equation}\label{word_length}
		\begin{aligned}[b]
			|w|=\sum_{\lambda\in G_w}\mu_{\lambda}(w),
		\end{aligned}
	\end{equation}
	where the summation over $\lambda\in G_w$ runs over all distinct edges contained in $w$ without repetition. 
	Whenever convenient, we also use $\mu_{k}(w)$ to denote the multiplicities corresponding to the edge labelled by $k=1,2,\dots,|G_{w}|$ in the word $w$. In this notation, \eqref{word_length} can be alternatively expressed as $|w|=\sum_{k=1}^{|G_{w}|}\mu_{k}(w)$. 
\end{remark}

With an auxiliary vector $\vec{\tau}\in \mathcal{T}_{|w|}$ and a given word $w=(w_{1},w_{2},\dots,w_{|w|})$, we denote $h(w,\vec{\tau})\coloneq h_{w_{1}}(\tau_{1})h_{w_{2}}(\tau_{2})\dots h_{w_{|w|}}(\tau_{|w|})$. We use $E^{\star}$ to represent the set of all words of arbitrary length. Likewise, by $G^{\star}$ we denote the set of words with letters strictly in the edge subset $G\subseteq E$, which means any edge subset $G$ can be identified as a subalphabet. These notations enable us to recast \eqref{dyson_expand} as
\begin{equation}\label{interaction_picture_CE}
	\begin{aligned}[b]
		e^{-\beta H}=\sum_{w\in E^{\star}}\int_{\vec{\tau}\in \mathcal{T}_{|w|}}\mathrm{D}\vec{\tau}\,e^{-\beta W}h(w,\vec{\tau})\eqcolon \sum_{w\in E^{\star}}f(w),
	\end{aligned}
\end{equation}
where every term $f(w)$ is a bounded operator, and the summation runs over all possible words. 
This is the desired cluster expansion for our bosonic lattice model, which is in sharp distinction from that for finite-dimensional spin systems \cite{kliesch2014locality}. In the spin case, thanks to the uniform boundedness of each local term in the Hamiltonian, one can typically adopt a direct power series expansion of $e^{-\beta H}$. One can then identify each term in the series with a word in $E^*$. Under this identification, the uniform bound on local terms, combined with a sufficiently small $\beta$, guarantees that the contributions of individual terms are small enough to ensure the absolute convergence of the subsequent cluster summation. \label{reply2_spin_expansion}

In the rest of the paper, we will also consider the function $f_{\Lambda}(w)$ for the subsystem $\Lambda\subset V$, which is defined by the same expression but with the factor $e^{-\beta W}$ replaced by $e^{-\beta W_{\Lambda}}$. Namely, we write
\begin{equation}\label{f_lambda_w}
	\begin{aligned}[b]
		f_{\Lambda}(w)\coloneq \int_{\vec{\tau}\in \mathcal{T}_{|w|}}\mathrm{D}\vec{\tau}\,e^{-\beta W_{\Lambda}}h(w,\vec{\tau}).
	\end{aligned}
\end{equation}
This is well-defined as long as $V_{w}\subseteq \Lambda$, which ensures (with slightly abused notation) $h_{w_{i}}(\tau_{i})=e^{\tau_{i} W}h_{w_{i}}e^{-\tau_{i} W}=e^{\tau_{i} W_{\Lambda}}h_{w_{i}}e^{-\tau_{i} W_{\Lambda}}$. 
We also naturally identify $f_{V}(w)$ with $f(w)$. For the subsystem Hamiltonian $H_{\Lambda}$, we obtain the corresponding expansion
\begin{equation}\label{subsystem_expansion}
	\begin{aligned}[b]
		e^{-\beta H_{\Lambda}}=\sum_{w\in E^{\star}\colon V_{w}\subseteq \Lambda}f_{\Lambda}(w). 
	\end{aligned}
\end{equation}

In applications, one typically decomposes and rearranges the summation with respect to words and employs graph-theoretical tools for analysis. To this end, the following definition of a cluster will be repeatedly used.
\begin{definition}[Cluster]
	A word $w$ is called a cluster if its corresponding edge set $G_{w}$ is connected. 
\end{definition}
We will use the letter $c$ to denote clusters, to distinguish them from general words.

\section{Proofs of the Main Theorems}\label{sec_proof_main_theorem}
To maintain the flow of the main argument, we postpone the proofs of the technical lemmas presented in this section to Section \ref{sec_proof_techinical_lemma}. 
Throughout the proof for both of the two theorems, we assume $\beta$ is sufficiently small such that all the conditions of the invoked lemmas are satisfied.
\subsection{Low-Boson-Density Inequality}\label{sec_proof_low_density}
The goal of this section is to prove Theorem \ref{theorem_low_density}. This result deals with the thermal expectation value of the local particle number moment, $ \langle n_{z}^{s} \rangle_{\beta H}=\mathcal{Z}(\beta)^{-1}\operatorname{Tr}(n_{z}^{s}e^{-\beta H}) $, for a fixed site $z\in V$ and $s\in \mathbb{N}_{>0}$. We first prepare the following lemmas justifying the absolute convergence in the trace norm topology as well as the series rearrangement \label{reply_lemma_trace} hereafter.

\begin{lemma}\label{lemma_trace_norm_n_f}
	Let $f_{\Lambda}(w)$ be defined as in \eqref{f_lambda_w} with $V_{w}\subseteq \Lambda$ being the support of the word $w$. Let $z\in V_{w}, s\in \mathbb{N}_{>0}$ and $\beta_{c}>0$ be fixed. Then, there exists a positive constant $C$ depending only on the model parameters $U_{\min}, U_{\max}, J, \mu, \mathfrak{d}$, and $\beta_{c}$, such that the following bound holds
	\begin{equation}\label{nGexpWfw}
		\begin{aligned}[b]
			\norm{n_{z}^{s}f_{\Lambda}(w)}_{1}
			\leq \int_{\vec{\tau}\in\mathcal{T}_{|w|}} \mathrm{D}\vec{\tau} \, \| n_z^s \, e^{-\beta W} h(w, \vec{\tau}) \|_1\leq  \sqrt{(2s)!} 
			\qty(\frac{C}{\sqrt{\beta}})^{|\Lambda|+s}  \frac{(C\sqrt{\beta})^{|w|}}{|w|!}
			\prod_{\lambda\in G_w}\mu_{\lambda}(w)!
		\end{aligned}
	\end{equation}
	for any $\beta \in (0,\beta_{c}]$. Here $\{\mu_{\lambda}(w)\}_{\lambda\in G_w}$ denotes the edge-wise multiplicities of the word $w$. This inequality still holds for $s=0$ once we identify $n_{z}^{0}=\id_{z}$.
\end{lemma}

The proof relies on factorizing $n_z^s e^{-\beta W} h(w, \vec{\tau})$ over the sites $x \in V_{w}$ and estimating each trace norm using the Cauchy–Schwarz inequality [\eqref{holder_inequality} with $(p,p_{1},p_{2})=(1,2,2)$]. Connecting these estimates to word combinatorics concludes the proof. The highly technical details are deferred to Section \ref{subsection_norm_convergence}. Intuitively, the factorial scaling term $\sqrt{(2s)!}$ and the temperature scaling involving $\beta$ originate from bounding the $s$-th order moment of the local particle number under the Gaussian-like thermal suppression provided by the quadratic on-site potential $W$, which relies centrally on the analytical estimates established in Lemma \ref{lemma_analysis}. The product of factorials $\prod \mu_\lambda(w)!$ accounts for the combinatorics of repeated hopping or squeezing on the same edges. \label{reply2_emphasize}

\begin{lemma}\label{lemma_mu_w_G_sum}
	Let $\{\mu_{\lambda}(w)\}_{\lambda\in G_w}$ be the edge-wise multiplicities of the word $w$. Then, for any $C\in (0,1)$ and any $G\subseteq E$, we have
	\begin{equation}\label{}
		\begin{aligned}[b]
			&\sum_{w\in G^{\star}:G_{w}= G}\qty[\frac{C ^{|w|}}{|w|!}\prod_{\lambda\in G_{w}}\mu_{\lambda}(w)!]=\qty(\frac{C}{1-C})^{|G|},
			\\&
			\sum_{w\in G^{\star}}\qty[\frac{C ^{|w|}}{|w|!}\prod_{\lambda\in G_{w}}\mu_{\lambda}(w)!]
			=\qty(\frac{1}{1-C})^{|G|}.
		\end{aligned}
	\end{equation}
\end{lemma}
The proof relies on standard combinatorial manipulations and is deferred to Section \ref{subsection_combinatorics}.

\begin{lemma}\label{lemma_convergence_trace_norm}
	Let $f(w)$ be defined as in \eqref{interaction_picture_CE}. Then for any $z\in V$ and $s\in \mathbb{N}_{>0}$, there exists a threshold $\beta_{c}>0$ such that the series $\sum_{w\in E^{\star}}n_{z}^{s}f(w)$ converges absolutely in the trace norm, i.e.,
	\begin{equation}\label{}
		\begin{aligned}[b]
			\sum_{w\in E^{\star}}\|n_{z}^{s}f(w)\|_{1}\leq \sum_{w \in E^\star} \int_{\vec{\tau}\in\mathcal{T}_{|w|}} \mathrm{D}\vec{\tau} \, \| n_z^s \, e^{-\beta W} h(w, \vec{\tau}) \|_1<\infty
		\end{aligned}
	\end{equation} 
	for any $\beta \in (0,\beta_{c}]$. This conclusion also extends to the case $s=0$ by identifying $n_{z}^{0}$ with the identity operator $\id_{z}$. 
\end{lemma}
\begin{proof}[Sketch of Proof]
	It follows directly from Lemmas \ref{lemma_trace_norm_n_f} and \ref{lemma_mu_w_G_sum} as detailed in Section \ref{subsection_norm_convergence}.
\end{proof}

Equipped with Lemmas \ref{lemma_trace_norm_n_f}-\ref{lemma_convergence_trace_norm}, we formally begin our proof. Intuitively, once we use the interaction-picture cluster expansion \eqref{interaction_picture_CE} to express the Boltzmann weight $e^{-\beta H}$ as a sum over words, these words can be classified according to whether they contain the site $z$. Here, we say a site $z$ belongs to a word $w$, denoted by $z \in w$, if and only if there exists an edge $\lambda\in G_w$ such that $z\in \lambda$. We then decompose the expectation value as
\begin{equation}\label{n_z^s}
	\begin{aligned}[b]
		\langle n_{z}^{s} \rangle_{\beta H}
		=\frac{1}{\mathcal{Z}(\beta)}\sum_{w\in E^{\star}: w \ni z}\operatorname{Tr}\qty(n_{z}^{s}f(w))+\frac{1}{\mathcal{Z}(\beta)}\operatorname{Tr}\qty(n_{z}^{s}\sum_{w\in E^{\star}: w \not\owns z}f(w)).
	\end{aligned}
\end{equation}
The justification for the rearrangement in the trace norm topology is provided by Lemma \ref{lemma_convergence_trace_norm}, which also enables us to frequently rearrange series in the subsequent derivations. To deal with the second term on the RHS of \eqref{n_z^s}, we rewrite the function $f(w)$ by factoring out the bounded operator $e^{-\beta W_z}$: 
\begin{equation}\label{fw_factor}
	\begin{aligned}[b]
		f(w)=\int_{\vec{\tau}\in \mathcal{T}_{|w|}}\mathrm{D}\vec{\tau}\,e^{-\beta W_{z}}e^{-\beta W_{z^{c}}}h(w,\vec{\tau})=e^{-\beta W_{z}}\int_{\vec{\tau}\in \mathcal{T}_{|w|}}\mathrm{D}\vec{\tau}\,e^{-\beta W_{z^{c}}}h(w,\vec{\tau}).
	\end{aligned}
\end{equation}
Here, $W_{z^{c}}=\sum_{x\in V: x\neq z}W_{x}$ excludes the on-site potential at $z$, with $z^{c}$ being a shorthand for $V\setminus \{z\}$. Since the word $w$ in \eqref{fw_factor} does not contain $z$, the integral on the RHS remains well-defined. By defining the defected Hamiltonian $H_{z^{c}}=-\sum_{\lambda\in E:  \lambda \not\owns z}h_{\lambda}+\sum_{x\in V: x\neq z}W_{x}$ and using the expansion \eqref{interaction_picture_CE} in reverse, one deduces
\begin{equation}\label{}
	\begin{aligned}[b]
		\sum_{w\in E^{\star}: w \not\owns z}f(w)=e^{-\beta W_{z}}e^{-\beta H_{z^{c}}}.
	\end{aligned}
\end{equation}
Evidently, the operator $n_{z}^{s}e^{-\beta W_{z}}e^{-\beta H_{z^{c}}}$ is trace-class, which justifies the decomposition in \eqref{n_z^s}. Then, by invoking the estimate for the subsystem partition function (see Lemma \ref{lemma_subsystem_partition} below), we obtain 
\begin{equation}\label{def_n_z_s}
	\begin{aligned}[b]
		\frac{1}{\mathcal{Z}(\beta)}\operatorname{Tr}\qty(n_{z}^{s}\sum_{w\in E^{\star}: w\not\owns z}f(w))=\frac{\operatorname{Tr}(n_{z}^{s}e^{-\beta W_{z}}e^{-\beta H_{z^{c}}})}{\operatorname{Tr}(e^{-\beta H})}\leq \frac{\operatorname{Tr}_{z}(n_{z}^{s}e^{-\beta W_{z}})}{\operatorname{Tr}_{z}(e^{-\beta W_{z}})} \eqcolon \langle n_{z}^{s} \rangle_{\beta W}.
	\end{aligned}
\end{equation}
The estimate for the moment $\langle n_{z}^{s} \rangle_{\beta W}$ is straightforward, and we summarize it in Lemma \ref{lemma_n_z_s_moment} below.  
\begin{lemma}\label{lemma_subsystem_partition}
	In the Bose-Hubbard model 
	defined in \eqref{bh} over a finite graph $(V,E)$, the following inequality holds for any non-empty subset $\Lambda\subset V$:
	\begin{equation}\label{subsystem_cor}
		\begin{aligned}[b]
			\frac{\operatorname{Tr}_{\Lambda^{c}}(e^{-\beta H_{\Lambda^{c}}})}{\operatorname{Tr}(e^{-\beta H})}\leq \frac{1}{\operatorname{Tr}_{\Lambda}(e^{-\beta W_{\Lambda}})}.
		\end{aligned}
	\end{equation}
	Here, $H_{\Lambda^{c}}$ is the subsystem Hamiltonian on the region $\Lambda^{c} = V \setminus \Lambda$.
\end{lemma}
The proof is based on applying the Gibbs variational principle regarding the free energy to the decoupled reference state $e^{-\beta H_{\Lambda^{c}}}/\operatorname{Tr}_{\Lambda^{c}}(e^{-\beta H_{\Lambda^{c}}}) \otimes e^{-\beta W_{\Lambda}}/\operatorname{Tr}_{\Lambda}(e^{-\beta W_{\Lambda}})$. Because $e^{-\beta W_{\Lambda}}$ only depends on the local particle numbers, the expectations of the off-diagonal hopping and squeezing terms exactly vanish. This elegant cancellation directly leads to the factorization of the total partition function into subsystem ones. The detailed derivation is presented in Section \ref{subsection_ratio}.

\begin{lemma}\label{lemma_n_z_s_moment}
	Let $\beta_{c}>0$ be fixed and $\langle n_{z}^{s} \rangle_{\beta W}$ be defined in \eqref{def_n_z_s}. Then there exists a positive constant $C$ depending only on $\beta_{c}, U_{\min}, U_{\max}$ and $\mu$, such that $\langle n_{z}^{s} \rangle_{\beta W}\leq C^{s}\sqrt{s!}\beta^{-s/2}$ for any $\beta \in (0,\beta_{c}]$ and positive integer $s$.
\end{lemma}
A rough estimate can be obtained by comparing it with the integral $$\int_{0}^{\infty}\mathrm{d}x\, x^{s}e^{-\beta (ax^{2}+bx)}\bigg/\int_{0}^{\infty}\mathrm{d}x\, e^{-\beta (ax^{2}+bx)}$$ for a positive $a$. See Section \ref{subsection_analytical} for a detailed proof. 

Next, we deal with the first term in \eqref{n_z^s}. Note that the summation runs over all words containing the site $z$, regardless of their connectivity or length. To reorganize the summand, we introduce several notations regarding cluster combinatorics for later convenience. We say a word $w'$ is contained in another word $w$ (denoted by $w'\subset w$) if $w'$ can be obtained by deleting a certain set of letters from $w$; i.e., $G_{w'}\subseteq G_{w}$ and $\mu_{\lambda}(w')\leq \mu_{\lambda}(w)$ for all $\lambda\in G_{w'}$. For any edge subset $G\subset E$, we denote by $\mathcal{C}(G)$ the set of words that contain at least one cluster $c$ such that $G_c\cap G\neq \emptyset$. Let $G_{z}\coloneq \{\lambda\in E \colon \lambda \ni  z\}$ denote the set of edges adjacent to the site $z$. We can then formally rewrite the first term on the RHS of \eqref{n_z^s} (which we denote by $C_{\beta}(n_{z}^{s})$ for convenience) as 
\begin{equation}\label{c_beta_n_z_s}
	\begin{aligned}[b]
		C_{\beta}(n_{z}^{s})=\frac{1}{\mathcal{Z}(\beta)}\sum_{w\in \mathcal{C}( G_z)}\operatorname{Tr}\qty(n_{z}^{s}f(w)).
	\end{aligned}
\end{equation}
The summation here still accounts for the edge multiplicities of the words, which complicates further estimation. To counter this difficulty, we reorganize the summation using graph-theoretical concepts. To that end, we denote by 
\begin{equation}\label{}
	\begin{aligned}[b]
		\mathcal{A}(F)\coloneq \{G\subseteq E \text{ connected } \colon G\cap F \neq \emptyset\}
	\end{aligned}
\end{equation}
the set of connected edge subsets that contain at least one edge of $F\subset E$. For each fixed word $w\in \mathcal{C}(G_{z})$, let $G^{(z)}_{w}$ be the maximal connected subset of $G_{w}$ that belongs to $\mathcal{A}(G_{z})$. Crucially, this mapping is unique: 
\begin{figure}
	\centering
	\includegraphics[width=0.8\linewidth]{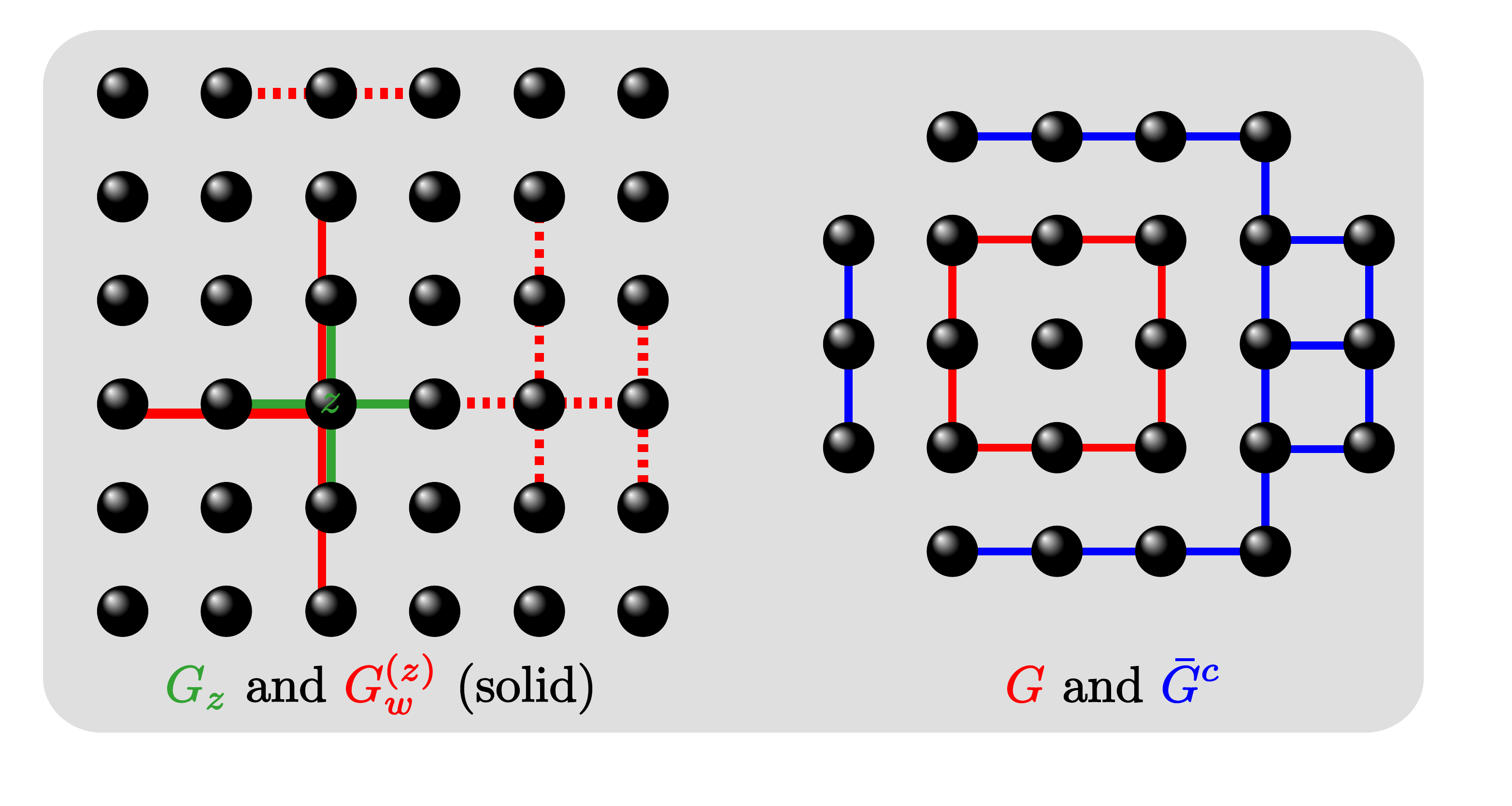}
	\caption{Illustration of $G_{w}$ (red) and $G_{w}^{(z)}$ (red and solid) on a $2$D square lattice. The red dotted lines symbolize the remaining parts of the word $w$ outside $G_w^{(z)}$. This also demonstrates why the inclusion $V_{\overline{G}^{c}}\subseteq (V_{G})^{c}$ is generally possible due to the ``holes'' enclosed by $G$.}
	\label{fig_example}
\end{figure} \label{replt2_figure2}
we can always generate $w$ by using the edges in $G^{(z)}_{w}$ and dress it with edges that do not overlap with $G^{(z)}_{w}$. See \Fig{fig_example} for an illustration. With this intuition, one can verify that the following identity holds: 
\begin{equation}\label{def_rhoG}
	\begin{aligned}[b]
		\sum_{w\in \mathcal{C}(G_{z})}f(w)=\sum_{G\in \mathcal{A}(G_{z})}\sum_{w\in [(\partial G)^{c}]^{\star}:G_{w}\supseteq G}f(w)\eqcolon \sum_{G\in \mathcal{A}(G_{z})}\rho(G).
	\end{aligned}
\end{equation}
Note that the graph weight $\rho(G)$ covers all words that contain every edge in $G$ at least once, since $(\partial G)^{c}= G\sqcup \overline{G}^{c}$. The following lemma provides a more convenient form for the graph weight, and its proof is deferred to Section \ref{subsection_combinatorics}:
\begin{lemma}\label{lemma_rho_G}
	Let $G\in \mathcal{A}(G_{z})$ be fixed. The operator $\rho(G)$ defined in \eqref{def_rhoG} equals
	\begin{equation}\label{}
		\begin{aligned}[b]
			e^{-\beta H_{V_{\overline{G}^{c}}}}e^{-\beta \Delta _{G}}\sum_{w\in G^{\star}: G_{w}=G}f_{V_{G}}(w).
		\end{aligned}
	\end{equation}  
	Here, we define $V_{\mathrm{res}}(G) \coloneq V \setminus (V_G \cup V_{\overline{G}^{c}})$, and $\Delta_{G}=\sum_{x\in V_{\mathrm{res}}(G) }W_{x}$ collects the corresponding on-site potentials. 
\end{lemma}
\begin{remark}
We note that the inclusion $V_{\overline{G}^{c}}\subseteq(V_{G})^{c}$ is always satisfied, thereby ensuring the well-definedness of $\Delta_{G}$. Nevertheless, the strict equality $V_{\overline{G}^{c}}=(V_{G})^{c}$ does not necessarily hold (see \Fig{fig_example} for an illustration).
\end{remark}

Substituting \eqref{def_rhoG} into \eqref{c_beta_n_z_s} and applying Lemma \ref{lemma_rho_G}, we obtain
\begin{equation}\label{}
	\begin{aligned}[b]
		C_{\beta}(n_{z}^{s})=\frac{1}{\mathcal{Z}(\beta)}\sum_{G\in \mathcal{A}(G_{z})}\sum_{w\in G^{\star}: G_{w}=G}\operatorname{Tr}\qty(n_{z}^{s}e^{-\beta H_{V_{\overline{G}^{c}}}}e^{-\beta \Delta _{G}}f_{V_{G}}(w)),
	\end{aligned}
\end{equation}
and therefore
\begin{equation}\label{C_beta_n_z_s}
	\begin{aligned}[b]
		|C_{\beta}(n_{z}^{s})|&\leq \frac{1}{\mathcal{Z}(\beta)}\sum_{G\in \mathcal{A}(G_{z})}\sum_{w\in G^{\star}: G_{w}=G}\norm{n_{z}^{s}e^{-\beta H_{V_{\overline{G}^{c}}}}e^{-\beta \Delta _{G}}f_{V_{G}}(w)}_{1}
		\\&\leq   \sum_{G\in \mathcal{A}(G_{z})}\frac{1}{\operatorname{Tr}_{V_{G}}(e^{-\beta W_{V_{G}}})}\sum_{w\in G^{\star}: G_{w}=G}\norm{n_{z}^{s}f_{V_{G}}(w)}_{1}.
	\end{aligned}
\end{equation}
Here, in the second line, we used Lemma \ref{lemma_subsystem_partition} by setting $\Lambda=(V_{\overline{G}^{c}})^{c}=V_{G}\cup V_{\mathrm{res}}(G)$. To proceed further, we invoke the previously established trace norm bounds alongside the following lemma for the prefactor $1/\operatorname{Tr}_{V_{G}}(e^{-\beta W_{V_{G}}})$.

\begin{lemma}\label{lemma_W_x}
	Let $W_{x}$ be the on-site potential defined in \eqref{bh} and $\beta_{c}>0$ be fixed. Then there exists a positive constant $C$, depending only on $\beta_{c}, U_{\min}, U_{\max}$ and $\mu$, such that the following bound holds 
	\begin{equation}\label{}
		\begin{aligned}[b]
			\frac{1}{\operatorname{Tr}_{x}(e^{-\beta W_{x}})}\leq C \beta^{1/2}
		\end{aligned}
	\end{equation}
	for any $\beta\in (0, \beta_{c}]$.
\end{lemma}
The proof follows from a comparison with the Gaussian integral. See Section \ref{subsection_analytical} for details. 

Equipped with the lemmas above, we can now return to \eqref{C_beta_n_z_s} and obtain
\begin{equation}\label{abs_C_beta_n_z_s}
	\begin{aligned}[b]
		|C_{\beta}(n_{z}^{s})|&\leq \sum_{G\in \mathcal{A}(G_{z})}(K_{1}\beta^{1/2})^{|V_{G}|}\sum_{w\in G^{\star}: G_{w}=G}\sqrt{(2s)!}\qty(\frac{K_{2}}{\sqrt{\beta}})^{|V_{G}|+s}
		\frac{(K_{2}\sqrt{\beta})^{|w|}}{|w|!}
		\prod_{\lambda\in G_w}\mu_{\lambda}(w)!
		\\&=\sqrt{(2s)!}\qty(\frac{K_{2}}{\sqrt{\beta}})^{s}\sum_{G\in \mathcal{A}(G_{z})}\qty(K_{1}K_{2})^{|V_{G}|}\qty(\frac{K_{2}\sqrt{\beta}}{1-K_{2}\sqrt{\beta}})^{|G|}.
	\end{aligned}
\end{equation}
Here, the constants $K_1$ and $K_{2}$ correspond to the constants from Lemmas \ref{lemma_trace_norm_n_f} and \ref{lemma_W_x}, respectively. The final equality follows from the summation formula in Lemma \ref{lemma_mu_w_G_sum} with the parameter $C = K_2\sqrt{\beta}$, provided $\beta$ is small enough such that $K_2\sqrt{\beta} < 1$. We conclude the treatment of $C_{\beta}(n_{z}^{s})$ by showing that the summation on the RHS of \eqref{abs_C_beta_n_z_s} can be bounded by a constant independent of the system size. Note that $|V_{G}|\leq 2|G|$ since each edge contains two sites. By noting that $|\{G\in \mathcal{A}(G_{z})\colon |G|=m\}|\leq |G_{z}|\sigma^{m}\leq \mathfrak{d}\sigma^{m}$, where $\sigma$ is the growth constant [cf.\,\eqref{def_growth_constant}], we obtain
\begin{equation}\label{series_K_1_K_2}
	\begin{aligned}[b]
		\sum_{G\in \mathcal{A}(G_{z})}\qty(K_{1}K_{2})^{|V_{G}|}\qty(\frac{K_{2}\sqrt{\beta}}{1-K_{2}\sqrt{\beta}})^{|G|}\leq 	\sum_{G\in \mathcal{A}(G_{z})} \qty(\frac{K'_{2}\sqrt{\beta}}{1-K_{2}\sqrt{\beta}})^{|G|}\leq \sum_{m=0}^{\infty}\mathfrak{d}\sigma^{m}\qty(\frac{K'_{2}\sqrt{\beta}}{1-K_{2}\sqrt{\beta}})^{m}.
	\end{aligned}
\end{equation}
Here, we denote $K'_{2}\coloneq K_{2}\max\{K_{1}K_{2},1\}^{2}$ and choose $\beta$ to be sufficiently small to ensure $\sigma K_{2}'\sqrt{\beta}/(1-K_{2}\sqrt{\beta})<1$, guaranteeing that the series converges. By substituting \eqref{abs_C_beta_n_z_s} and \eqref{def_n_z_s} back into \eqref{n_z^s}, and incorporating the estimates from Lemma \ref{lemma_n_z_s_moment} and \eqref{series_K_1_K_2}, we finally obtain for $\beta\leq \beta_{c}$:
\begin{equation}\label{low_density_form_1}
	\begin{aligned}[b]
		\langle n_{z}^{s} \rangle_{\beta H}\leq K_{\mathrm{low}}^{s} \cdot s! \cdot \beta^{-s/2},
	\end{aligned}
\end{equation} 
where $K_{\mathrm{low}}$ and $\beta_{c}$ are positive constants depending only on the model parameters $U_{\min}, U_{\max}, J, \mu$, and $\mathfrak{d}$. We have also utilized the elementary bound $\sqrt{(2s)!}\le 2^ss!$ and absorbed the factors into the redefined constant $K_{\mathrm{low}}$. The final threshold $\beta_{c}$ is taken to be the smallest among all the threshold inverse temperatures incurred during the proof. Although an explicit expression for a candidate threshold temperature with respect to the model parameters is complicated, the expression itself is fully computable. It can be given in the form $\beta_{c}=(1-\epsilon)(\sigma K_{2}'+K_{2})^{-2}$ for any $\epsilon\in (0,1)$, which is clearly independent of the system size.

\subsection{Clustering Theorem}\label{sec_proof_clustering}
This subsection is devoted to the proof of Theorem \ref{theorem_clustering}, which establishes an upper bound for the magnitude of the thermal correlation function defined in \eqref{correlation}. Since the original formula for the correlation function is inconvenient to manipulate, we first introduce several new notations. 

We introduce the doubled Hilbert space, defined as $\widetilde{\mathcal{H}} \coloneq \mathcal{H} \otimes \mathcal{H}$. 
For any operator $O$ on $\mathcal{H}$, we introduce the shorthand notations 
$O^{(0)} \coloneq O \otimes \id$, 
$O^{(1)} \coloneq O \otimes \id - \id \otimes O$ and 
$O^{(+)} \coloneq O \otimes \id + \id \otimes O$ 
to represent the corresponding operators on $\widetilde{\mathcal{H}}$. Let $\widetilde{\mathcal{D}}_{\fin} \coloneq \mathcal{D}_{\fin} \otimes \mathcal{D}_{\fin}$ denote the algebraic tensor product of the finite-particle domains, which will be referred to as the common core for all $O_{X}^{(+)}$ with $O_{X}\in \mathcal{P}_{X}(\bm{a},\bm{a}^{\ast})$ for any $X\subseteq V$. 
According to Theorem VIII.33 of \cite{Reed1981}, since the Bose-Hubbard Hamiltonian $H$ is essentially self-adjoint on $\mathcal{D}_{\fin}$, the operators $H^{(0)}, \id\otimes H$, and $H^{(+)}$ are essentially self-adjoint on $\widetilde{\mathcal{D}}_{\fin}$. Therefore, we can use the spectral theorem to define $e^{-\beta H^{(+)}}$ for any $\beta >0$. Furthermore, since $H$ is bounded from below (cf. Proposition \ref{pro_stability}), the operators $H \otimes \id$ and $\id \otimes H$ are strongly commuting self-adjoint operators bounded from below. 
Applying the Trotter product formula for unbounded operators [cf. Theorem VIII.31 of \cite{Reed1981}], we obtain the strict identity:
\begin{equation}
	e^{-\beta H^{(+)}} = e^{-\beta H} \otimes e^{-\beta H}.
\end{equation}
This, together with the notations introduced above, enables one to recast the correlation function defined in \eqref{correlation} as
\begin{equation}\label{correlation_double}
	\begin{aligned}[b]
		C_{\beta}(O_X,O_Y)=	\frac{1}{\mathcal{Z}(\beta)^{2}}\operatorname{Tr}	\qty(O_{X}^{(0)}O_{Y}^{(1)}e^{-\beta H^{(+)}}).
	\end{aligned}
\end{equation}
Here, by a slight abuse of notation, we use $\operatorname{Tr}$ to denote the trace over $\widetilde{\mathcal{H}}$; this will not cause any confusion, as its meaning is clear from the context. Similarly, we use $\operatorname{Tr}_{\Lambda}$ to denote the partial trace over $\widetilde{\mathcal{H}}_{\Lambda}\coloneq \mathcal{H}_{\Lambda}\otimes \mathcal{H}_{\Lambda}$ for $\Lambda\subseteq V$. 

Analogous to the analysis in Section \ref{sec_main_technique}, we establish the interaction-picture cluster expansion for $e^{-\beta H^{(+)}}$, i.e., 
\begin{equation}\label{dyson_til}
	\begin{aligned}[b]
		e^{-\beta H^{(+)}}=&\sum_{m=0}^{\infty}\ii{0}{\beta}{\tau_{1}}\ii{0}{\tau_{1}}{\tau_{2}}\dots\ii{0}{\tau_{m-1}}{\tau_{m}}
 e^{-(\beta-\tau_{1})W^{(+)}}I^{(+)}e^{-(\tau_{1}-\tau_{2})W^{(+)}}I^{(+)}\dots I^{(+)}e^{-\tau_{m}W^{(+)}},
	\end{aligned}
\end{equation}
which can also be recast as 
\begin{equation}\label{interaction_picture_til}
	\begin{aligned}[b]
		e^{-\beta H^{(+)}}=\sum_{w\in E^{\star}}\int_{\vec{\tau}\in \mathcal{T}_{|w|}}\mathrm{D}\vec{\tau}\,e^{-\beta W^{(+)}}\widetilde{h}(w,\vec{\tau})\eqcolon\sum_{w\in E^{\star}}\widetilde{f}(w),
	\end{aligned}
\end{equation}
with the shorthand $\widetilde{h}(w,\vec{\tau})\coloneq h_{w_{1}}(\tau_{1})^{(+)}h_{w_{2}}(\tau_{2})^{(+)}\dots h_{w_{|w|}}(\tau_{|w|})^{(+)}$. The operator $\widetilde{f}(w)$ is bounded, and the expansion is well-defined, as summarized in the following statement.

\begin{proposition}\label{pro_dyson_converge_til}
	For the Bose-Hubbard model \eqref{bh} defined on a finite graph $(V, E)$, the operator $\widetilde{f}(w)$ is bounded, and the corresponding Dyson series \eqref{dyson_til} converges absolutely in the norm topology for any $\beta >0$.   
\end{proposition}
\begin{proof}
	Observe that there exists $c>0$ such that $\|I^{(+)}e^{-\tau W^{(+)}}\| \leq c\|Ie^{-\tau W}\|$ for any $\tau\in (0,\beta)$, since $W$ is bounded from below. The proof follows directly from that of Proposition \ref{pro_dyson_converge}. Since the norm of the $m$-th term in \eqref{dyson} scales as $C^{m}/\Gamma(m/2+1)$ for some $C>0$, the extra $c^{m}$ factor arising from $\|I^{(+)}e^{-\tau W^{(+)}}\|$ does not violate absolute convergence. Here, both $c$ and $C$ depend on $|V|$. 
\end{proof} 

Following a strategy analogous to the previous subsection, we first prepare several lemmas justifying the absolute convergence in the trace norm topology as well as series rearrangements. As the motivation will become clear shortly, we introduce the following regularizing factors to suppress the unboundedness of $O_{X}$ and $O_{Y}$.

\begin{definition}\label{definition_regularization}
	Let $W$ be the on-site potential of the Bose-Hubbard model \eqref{bh} defined on a finite graph. For some $K_{0}>0$, we define
	\begin{equation}\label{check_O_X}
		\begin{aligned}[b]
			\widecheck{W}\coloneq  W- K_{0}\beta^{-1/2}(N_{X}+N_{Y}),\quad \widecheck{O}_{X}^{[0]}\coloneq O_{X}^{(0)}e^{-K_{0}\beta^{1/2} N_{X}^{(+)}},\quad \widecheck{O}_{Y}^{[1]}\coloneq O_{Y}^{(1)}e^{-K_{0}\beta^{1/2} N_{Y}^{(+)}}
		\end{aligned}
	\end{equation} 
	such that both $\widecheck{O}_{X}^{[0]}$ and $\widecheck{O}_{Y}^{[1]}$ are bounded operators. Here, we write the subscripts $[0], [1]$ to distinguish them from the unregularized symbols $(0), (1)$. 
\end{definition}
Throughout this article, we set $K_{0}=1$ for simplicity.\label{reply_remark_k0}
\begin{remark}\label{remark_beta_sqaure_root}
	We note that the presence of the $\beta^{-1/2}$ factor in $\widecheck{W}$ does not affect the previous analysis for the Gibbs state of the regularized Hamiltonian $\widecheck{H}=-I+\widecheck{W}$, provided that the estimates rely on Lemma \ref{lemma_analysis} (such as Lemma \ref{lemma_trace_norm_n_f} and Theorem \ref{theorem_low_density}). Indeed, the expression $\beta\widecheck{H}=\beta H +\beta^{1/2}(N_{X}+N_{Y})$ in the Boltzmann factor merely modifies the linear coefficient of the local particle number operator. At high temperatures $\beta \in (0,\beta_{c}]$, this modified coefficient and the ratio $(\beta\mu +\beta^{1/2})(\beta U)^{-1/2}$ are uniformly bounded from above. Therefore, the assumptions of Lemma \ref{lemma_analysis} remain valid.
\end{remark}

Accordingly, we define 
\begin{equation}\label{def_f_che}
	\begin{aligned}[b]
		\widecheck{f}(w)\coloneq \int_{\vec{\tau}\in \mathcal{T}_{|w|}}\mathrm{D}\vec{\tau}\,e^{-\beta \widecheck{W}^{(+)}}\widetilde{h}(w,\vec{\tau}).
	\end{aligned}
\end{equation}
By similar arguments regarding the boundedness of $\widetilde{f}(w)$ and the fact that $\widecheck{W}$ still grows quadratically with the local particle numbers, the operator $\widecheck{f}(w)$ is also bounded. For a subset $\Lambda\subseteq V$ and a word $w$ such that $V_{w}\subseteq \Lambda$, we define 
\begin{equation}\label{def_f_che_lambda}
	\begin{aligned}[b]	\widecheck{f}_{\Lambda}(w)\coloneq \int_{\vec{\tau}\in \mathcal{T}_{|w|}}\mathrm{D}\vec{\tau}\,e^{-\beta \widecheck{W}_{\Lambda}^{(+)}}\widetilde{h}(w,\vec{\tau}),
	\end{aligned}
\end{equation}
where 
\begin{equation}\label{def_W_x_che}
	\begin{aligned}[b]
		\widecheck{W}_{\Lambda}\coloneq \sum_{x\in \Lambda}\qty(W_{x}-\bm{1}_{X\cup Y}(x)\beta^{-1/2}n_{x}) = W_{\Lambda}-\beta^{-1/2}(N_{X\cap \Lambda}+N_{Y\cap \Lambda}),
	\end{aligned}
\end{equation}
and the indicator function $\bm{1}_{X\cup Y}(x)$ equals $1$ for $x\in X\cup Y$ and vanishes otherwise. In parallel, we define $\widecheck{H}_{\Lambda}\coloneq \widecheck{W}_{\Lambda}-I_{\Lambda}$ and obtain the corresponding expansion:
\begin{equation}\label{subsystem_expansion_til}
	\begin{aligned}[b]
		e^{-\beta \widecheck{H}_{\Lambda}^{(+)}}=\sum_{w\in E^{\star}\colon V_{w}\subseteq \Lambda}\widecheck{f}_{\Lambda}(w). 
	\end{aligned}
\end{equation}
The following two lemmas justify the absolute convergence in the trace norm topology of the doubled Hilbert space, enabling subsequent series rearrangements.

\begin{lemma}\label{lemma_trace_norm_f_che}
	Let $\widecheck{f}_{\Lambda}(w)$ be defined as in \eqref{def_f_che_lambda} with $V_{w}\subseteq \Lambda$ being the support of the word $w$. Let $z\in V_{w}, s\in \mathbb{N}_{>0}$ and $\beta_{c}>0$ be fixed. Then, there exists a positive constant $C$ depending only on the model parameters $U_{\min}, U_{\max}, J, \mu, \mathfrak{d}$ and $\beta_{c}$, such that the following bound holds:
	\begin{equation}\label{}
		\begin{aligned}[b]
			\|\widecheck{f}_{\Lambda}(w)\|_{1}
			\leq \int_{\vec{\tau}\in\mathcal{T}_{|w|}} \mathrm{D}\vec{\tau} \,\norm{e^{-\beta \widecheck{W}_{\Lambda}^{(+)}}\widetilde{h}(w,\vec{\tau})}_{1} \leq
			\qty(\frac{C}{\beta})^{|\Lambda|}  \frac{(C\sqrt{\beta})^{|w|}}{|w|!}
			\prod_{\lambda\in G_w}\mu_{\lambda}(w)!
		\end{aligned}
	\end{equation}
	for any $\beta \in (0, \beta_{c}]$. Here $\{\mu_{\lambda}(w)\}_{\lambda\in G_w}$ denotes the edge-wise multiplicities of the word $w$.
\end{lemma}
The proof relies on splitting $e^{-\beta \widecheck{W}_{\Lambda}^{(+)}}\widetilde{h}(w,\vec{\tau})$ between two factor Hilbert spaces and invoking Lemma \ref{lemma_trace_norm_n_f}. The highly technical details are deferred to Section \ref{subsection_norm_convergence}.

\begin{lemma}\label{lemma_convergence_trace_norm_til}
	Let $\widetilde{f}(w)$  and $\widecheck{f}(w)$ be defined as in  \eqref{interaction_picture_til} and \eqref{def_f_che}. Then there exists a threshold $\beta_{c}>0$ such that for any $\beta \in (0, \beta_{c}]$, the series $\sum_{w\in E^{\star}}O_{X}^{(0)}O_{Y}^{(1)}\widetilde{f}(w)$ converges absolutely in the trace norm, i.e., 
	\begin{equation}\label{}
		\begin{aligned}[b]
			\sum_{w\in E^{\star}}\norm{O_{X}^{(0)}O_{Y}^{(1)}\widetilde{f}(w)}_{1}\leq 	\sum_{w\in E^{\star}}\int_{\vec{\tau}\in \mathcal{T}_{|w|}}\mathrm{D}\vec{\tau}\,\norm{\widecheck{O}_{X}^{[0]}\widecheck{O}_{Y}^{[1]}e^{-\beta \widecheck{W}_{\Lambda}^{(+)}}\widetilde{h}(w,\vec{\tau})}_{1}<\infty,
		\end{aligned}
	\end{equation} 
	and moreover,
	\begin{equation}\label{}
		\begin{aligned}[b]
			\sum_{w\in E^{\star}}\|\widecheck{f}(w)\|_{1}\leq \sum_{w\in E^{\star}}\int_{\vec{\tau}\in\mathcal{T}_{|w|}} \mathrm{D}\vec{\tau} \,\norm{e^{-\beta \widecheck{W}^{(+)}}\widetilde{h}(w,\vec{\tau})}_{1} <\infty.
		\end{aligned}
	\end{equation} 
\end{lemma}
\begin{proof}[Sketch of Proof]
	This follows directly from Lemmas \ref{lemma_mu_w_G_sum} and \ref{lemma_trace_norm_f_che}; see details in Section \ref{subsection_norm_convergence}.
\end{proof}

Lemma \ref{lemma_convergence_trace_norm_til} and Fubini's theorem enable us to recast the correlation function as 
\begin{equation}\label{cor_w_Estar}
	\begin{aligned}[b]
		C_{\beta}(O_{X},O_{Y})=\frac{1}{\mathcal{Z}(\beta)^{2}}\sum_{w\in E^{\star}}\operatorname{Tr}	\qty(O_{X}^{(0)}O_{Y}^{(1)}\widetilde{f}(w)).
	\end{aligned}
\end{equation}
By the following lemma (whose proof is in Section \ref{subsection_combinatorics}), we note that not all terms in \eqref{interaction_picture_til} contribute to the correlation function \eqref{cor_w_Estar}. Specifically, the terms corresponding to words that do not connect $X$ and $Y$ cancel out exactly.

\begin{lemma}\label{lemma_not_connecting_word}
	Let $G\subset E$ be an edge subset that does not connect $X$ and $Y$. Then for any $w\in G^{\star}$, we have
	\begin{equation}\label{eq_lemma_8}
		\begin{aligned}[b]
			\operatorname{Tr}\qty(O_{X}^{(0)}O_{Y}^{(1)}\widetilde{f}(w))=0.
		\end{aligned}
	\end{equation}
\end{lemma}
Mathematically, this exact cancellation arises from the swap symmetry within the doubled Hilbert space. Since $G$ does not connect $X$ and $Y$, the operators spatially factorize. In the component containing $Y$, the Boltzmann weights and hopping/squeezing terms are symmetric under the swapping of Hilbert spaces, whereas the operator $O_{Y}^{(1)}$ is clearly antisymmetric, which strictly leads to a vanishing trace.

Applying Lemma \ref{lemma_not_connecting_word}, we recognize that we only need to consider words that connect $X$ and $Y$. The length of such a word must be at least $\operatorname{dist}(X,Y)\eqcolon L$. 
To rewrite the correlation function, we introduce several notations regarding cluster combinatorics for later convenience. For any edge subset $G\subset E$ and a fixed $L\in \mathbb{N}_{>0}$, we denote by $\mathcal{C}_{\geq L}(G)$ the set of words that contain at least one cluster $c$ such that $c\cap G\neq \emptyset$ (meaning there exists $\lambda\in E$ such that $\lambda\in c$ and $\lambda\in G$) and $|c|\geq L$. Substituting \eqref{interaction_picture_til} into \eqref{correlation_double}, we obtain
\begin{equation}\label{cor_word_c_L}
	\begin{aligned}[b]
		C_{\beta}(O_X,O_Y)=	\frac{1}{\mathcal{Z}(\beta)^{2}}\sum_{w\in \mathcal{C}_{\geq L}(\partial X)}\operatorname{Tr}	\qty(O_{X}^{(0)}O_{Y}^{(1)}\widetilde{f}(w)),
	\end{aligned}
\end{equation}
with the straightforward observation that any word in the complement $\qty[\mathcal{C}_{\geq L}(\partial X)]^{c}$ does not connect $X$ and $Y$. 
To reorganize the summation, let $\mathcal{C}_{\geq L}^{k}(G)$ denote the set of words containing exactly $k$ disjoint clusters of the type defined in $\mathcal{C}_{\geq L}(G)$. Naturally, we have $\bigcup_{k=1}^{\infty}\mathcal{C}_{\geq L}^{k}(G)=\mathcal{C}_{\geq L}(G)$.

\begin{figure}
	\centering
	\includegraphics[width=0.8\linewidth]{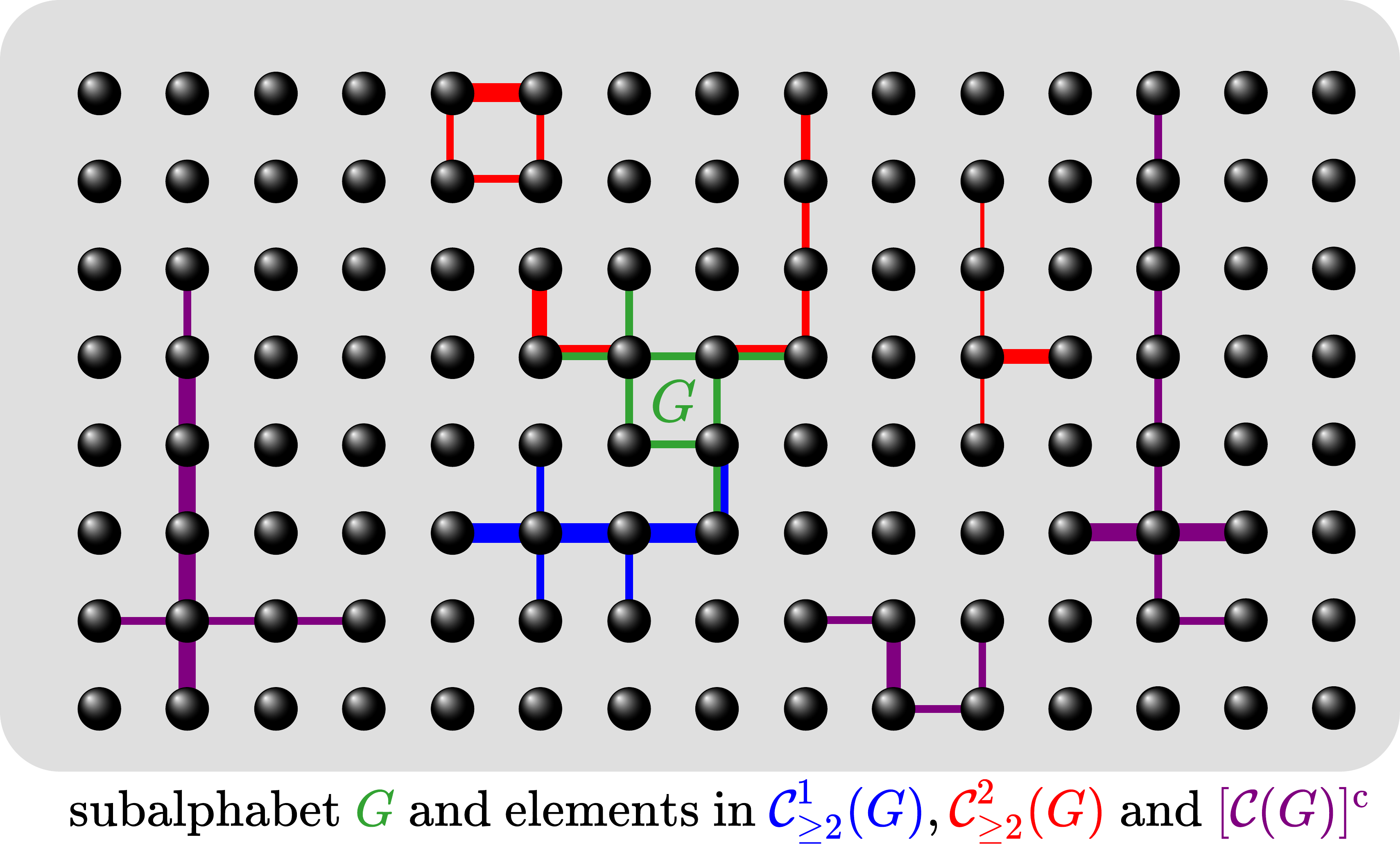}
	\caption{A $2$D square lattice as an illustration of $\mathcal{C}_{\geq L}^{k}(G)$ and $\mathcal{C}(G)$ with the different thickness representing the multiplicities.}
	\label{fig_concept_c_L^k}
\end{figure}

See \Fig{fig_concept_c_L^k} for an intuitive demonstration of these concepts. We then invoke the following combinatorial rearrangement:

\begin{lemma}\label{lemma_rearrange_C_L}
	Let $(\mathcal{X},\|\cdot \|)$ be a Banach space. For any sequence $(b_k)_{k=1}^\infty \subset \mathcal{X}$ satisfying $\sum_{k=1}^{\infty}2^{k}\|b_{k}\|<\infty$, the identity
	\begin{equation}\label{eq_lemma_rearrage}
		\begin{aligned}[b]
			\sum_{k=1}^\infty b_k = - \sum_{m=1}^\infty (-1)^m \sum_{k=m}^\infty \binom{k}{m} b_k
		\end{aligned}
	\end{equation} 
	holds. Moreover, by choosing $\mathcal{X}$ as the space of trace-class operators over $\widetilde{\mathcal{H}}$, the sequence $b_{k} \coloneq \sum_{w\in \mathcal{C}_{\geq L}^{k}( \partial X)}O_{X}^{(0)}O_{Y}^{(1)}\widetilde{f}(w)$ satisfies the precondition:
	\begin{equation}\label{}
		\begin{aligned}[b]
			\sum_{k=1}^{\infty}\sum_{w\in \mathcal{C}_{\geq L}^{k}( \partial X)}2^{k}\|O_{X}^{(0)}O_{Y}^{(1)}\widetilde{f}(w)\|_{1}<\infty.
		\end{aligned}
	\end{equation}
\end{lemma}
The proof relies on exchanging the order of summation and applying Lemma \ref{lemma_trace_norm_f_che} as detailed in Section \ref{subsection_combinatorics}. 

Before implementing the graph-theoretic estimates, we first introduce a regularization factor (see Definition \ref{definition_regularization}) to suppress the unboundedness of the operators $O_X$ and $O_Y$. This allows us to rewrite the correlation function as
\begin{equation}\label{C_beta_check}
	\begin{aligned}[b]
		C_{\beta}(O_X,O_Y)&=\frac{1}{\mathcal{Z}(\beta)^{2}}\sum_{k=1}^{\infty}\sum_{w\in \mathcal{C}^{k}_{\geq L}(\partial X)}\operatorname{Tr}	\qty(O_{X}^{(0)}O_{Y}^{(1)}\widetilde{f}(w))
		\\&=-\frac{1}{\mathcal{Z}(\beta)^{2}}\sum_{m=1}^{\infty}(-1)^{m}\sum_{k=m}^{\infty}\binom{k}{m}\sum_{w\in \mathcal{C}_{\geq L}^{k}( \partial X)}\operatorname{Tr}\qty(O_{X}^{(0)}O_{Y}^{(1)}\widetilde{f}(w))
		\\&=-\frac{1}{\mathcal{Z}(\beta)^{2}}\sum_{m=1}^{\infty}(-1)^{m}\sum_{k=m}^{\infty}\binom{k}{m}\sum_{w\in \mathcal{C}_{\geq L}^{k}( \partial X)}\operatorname{Tr}\qty(\widecheck{O}_{X}^{[0]}\widecheck{O}_{Y}^{[1]}\widecheck{f}(w)).
	\end{aligned}
\end{equation}
The second equality follows from applying Lemma \ref{lemma_rearrange_C_L} to \eqref{cor_word_c_L}. To obtain the last equality, we use the fact that $N_X^{(+)}$ and $N_Y^{(+)}$ commute with $\widecheck{W}^{(+)}$, which allows us to factorize the exponential weight inside the integral defining $\widetilde{f}(w)$ as $e^{-\beta W^{(+)}}=e^{-\beta^{1/2}N_X^{(+)}}e^{-\beta^{1/2}N_Y^{(+)}}e^{-\beta \widecheck{W}^{(+)}}$. Since the operators $e^{-\beta^{1/2}N_X^{(+)}}$ and $e^{-\beta^{1/2}N_Y^{(+)}}$ are bounded and independent of the integration variables, they can be factored out of the integral. According to \eqref{check_O_X} and \eqref{def_f_che}, they are then paired with $O_X^{(0)}$ and $O_Y^{(1)}$ to form $\widecheck{O}_X^{[0]}$ and $\widecheck{O}_Y^{[1]}$, respectively, which reduces $\widetilde{f}(w)$ to $\widecheck{f}(w)$. \label{reply2_regular} 

With this preparation, we note that the summation in \eqref{C_beta_check} over words still concerns edge multiplicities, complicating further estimates. To circumvent this difficulty, we reorganize the summation using graph-theoretical concepts. To that end, for an edge subset $F$, we denote by 
\begin{equation}\label{mcA_L_F}
	\begin{aligned}[b]
		\mathcal{A}_{\geq L}(F)\coloneq \{G\subseteq E \text{ connected } \colon F\cap G \neq \emptyset, |G|\geq L\}
	\end{aligned}
\end{equation}
the set of connected edge subsets that contain at least one edge of $F$ and have \label{reply_define_mcA} a size equal to or greater than $L$. Moreover, we denote by $\mathcal{A}_{\geq L}^{k}(F)$ the corresponding set of $k$-fold nonoverlapping connected edge subsets, i.e.,
\begin{equation}\label{mcA_L_F_k}
	\begin{aligned}[b]
		\mathcal{A}_{\geq L}^{k}(F)\coloneq \qty{\bigcup_{j=1}^{k}G_{j}:(G_{j})_{j=1}^{k}\subseteq \mathcal{A}_{\geq L}(F), V_{G_{j}}\cap V_{G_{j'}}=\emptyset,\, \forall j\neq j'\in \{1,2,\dots,k\} }.
	\end{aligned}
\end{equation} 
For any $G\subseteq E$, we define
\begin{equation}\label{def_rho_G_til}
	\begin{aligned}[b]
		\widecheck{\rho}(G)\coloneq \sum_{w\in [(\partial G)^{c}]^{\star}:G_{w}\supseteq G}\widecheck{f}(w),
	\end{aligned}
\end{equation}
where the summation collects words that contain all the edges in $G$ at least once but do not contain any boundary edges of $G$. We then invoke the following two technical combinatorial lemmas (proved in Section \ref{subsection_combinatorics}) to transform the summation in \eqref{C_beta_check} over words into a summation over series of nonoverlapping connected edge subsets adjacent to $\partial X$, yielding a more convenient form for $\widecheck{\rho}(G)$.

\begin{lemma}\label{lemma_word_to_graph}
	Let $\widecheck{f}(w)$ be the operator defined in \eqref{def_f_che} and $F\subset E$ be an edge subset. The following identity holds:
	\begin{equation}\label{}
		\begin{aligned}[b]
			\sum_{k=m}^{\infty}\binom{k}{m}\sum_{w\in \mathcal{C}^{k}_{\geq L}(F)}\widecheck{f}(w)=\sum_{G\in \mathcal{A}_{\geq L}^{m}(F)}\widecheck{\rho}(G).
		\end{aligned}
	\end{equation}
\end{lemma}
\begin{lemma}\label{lemma_rho_G_til}
	Let $G\in \mathcal{A}_{\geq L}^{m}(F)$ be fixed with disjoint decomposition $G = \bigcup_{j=1}^{m}G_{j}$. The operator $\widecheck{\rho}(G)$ defined in \eqref{def_rho_G_til} equals
	\begin{equation}\label{}
		\begin{aligned}[b]
			e^{-\beta \widecheck{H}^{(+)}_{V_{\overline{G}^{c}}}}e^{-\beta \widecheck{\Delta}^{(+)} _{G}}\sum_{w^{(1)}\in G_{1}^{\star}: G_{w^{(1)}}=G_{1}}\widecheck{f}_{V_{G_{1}}}(w^{(1)})\dots\sum_{w^{(m)}\in G_{m}^{\star}: G_{w^{(m)}}=G_{m}}\widecheck{f}_{V_{G_{m}}}(w^{(m)}).
		\end{aligned}
	\end{equation}  
	Here, we denote $\widecheck{H}_{V_{\overline{G}^{c}}}=\widecheck{W}_{V_{\overline{G}^{c}}}-I_{V_{\overline{G}^{c}}}$ and define $V_{\mathrm{res}}(G) \coloneq V \setminus (V_G \cup V_{\overline{G}^{c}})$,  $\widecheck{\Delta}_{G}=\widecheck{W}_{V_{\mathrm{res}}(G)}$. 
\end{lemma}

Applying Lemmas \ref{lemma_word_to_graph} and \ref{lemma_rho_G_til} to \eqref{C_beta_check}, we obtain  \label{reply_lemma_rho_G_til}
\begin{equation}\label{}
	\begin{aligned}[b]
		C_{\beta}(O_X,O_Y)=-\frac{1}{\mathcal{Z}(\beta)^{2}}\sum_{m=1}^{\infty}(-1)^{m}\sum_{G\in \mathcal{A}^{m}_{\geq L}(\partial X)}\sum_{\qty{\substack{w^{(j)}\in G^{\star}_{j}:\\ G_{w^{(j)}}=G_{j}}}_{j=1}^{m}}\operatorname{Tr}\qty(	\widecheck{O}_{X}^{[0]}\widecheck{O}_{Y}^{[1]}e^{-\beta \widecheck{H}^{(+)}_{V_{\overline{G}^{c}}}}e^{-\beta \widecheck{\Delta}^{(+)} _{G}}\prod_{j=1}^{m}\widecheck{f}_{V_{G_{j}}}(w^{(j)})),
	\end{aligned}
\end{equation}
with $\bigcup_{j=1}^{m}G_{j}$ being the disjoint decomposition of $G$, leading to the estimate
\begin{equation}\label{abs_cor_G_sum}
	\begin{aligned}[b]
		\qty|	C_{\beta}(O_X,O_Y)|&\leq \frac{1}{\mathcal{Z}(\beta)^{2}}\sum_{m=1}^{\infty}\sum_{G\in \mathcal{A}^{m}_{\geq L}(\partial X)}\sum_{\qty{\substack{w^{(j)}\in G^{\star}_{j}:\\ G_{w^{(j)}}=G_{j}}}_{j=1}^{m}}\norm{	\widecheck{O}_{X}^{[0]}\widecheck{O}_{Y}^{[1]}e^{-\beta \widecheck{H}^{(+)}_{V_{\overline{G}^{c}}}}e^{-\beta \widecheck{\Delta}^{(+)} _{G}}\prod_{j=1}^{m}\widecheck{f}_{V_{G_{j}}}(w^{(j)})}_{1}
		\\	&\leq \|\widecheck{O}_{X}^{[0]}\widecheck{O}_{Y}^{[1]}\| \sum_{m=1}^{\infty}\sum_{G\in \mathcal{A}^{m}_{\geq L}(\partial X)}\qty[\frac{\operatorname{Tr}_{(V_{G})^{c}}\qty(e^{-\beta (\widecheck{H}_{V_{\overline{G}^{c}}}+\widecheck{\Delta}_{G})})}{\mathcal{Z}(\beta)}]^{2}\sum_{\qty{\substack{w^{(j)}\in G^{\star}_{j}:\\ G_{w^{(j)}}=G_{j}}}_{j=1}^{m}}\prod_{j=1}^{m}\norm{\widecheck{f}_{V_{G_{j}}}(w^{(j)})}_{1}.
	\end{aligned}
\end{equation}
Note that taking the partial trace of $e^{-\beta \widecheck{H}^{(+)}_{V_{\overline{G}^{c}}}}e^{-\beta \widecheck{\Delta}^{(+)} _{G}}$ over the doubled Hilbert space strictly factorizes into the square of the trace over the single Hilbert space. This accounts for the squared bracket on the RHS of the second line in \eqref{abs_cor_G_sum}.
We then need the following lemmas to proceed.
\begin{lemma}\label{lemma_subsystem_partition_che}
	Let $\beta_c > 0$ be fixed and sufficiently small to ensure the validity of Theorem \ref{theorem_low_density}. Let a non-empty subset $\Lambda\subset V$ be fixed. In the Bose-Hubbard model 
	defined in \eqref{bh} over a finite graph $(V,E)$, there exists a positive constant $C$ depending only on the model parameters $U_{\min}, U_{\max}, J, \mu, \mathfrak{d}$, and $\beta_{c}$, such that the following bound holds:
	\begin{equation}\label{subsystem_cor}
		\begin{aligned}[b]
			\frac{\operatorname{Tr}_{\Lambda^{c}}(e^{-\beta \widecheck{H}_{\Lambda^{c}}})}{\operatorname{Tr}(e^{-\beta H})}\leq \frac{C^{|X|+|Y|}}{\operatorname{Tr}_{\Lambda}(e^{-\beta W_{\Lambda}})}
		\end{aligned}
	\end{equation}
	for any $\beta \in (0, \beta_{c}]$.
	Here, $\widecheck{H}_{\Lambda^{c}}=\widecheck{W}_{\Lambda^{c}}-I_{\Lambda^{c}}$ is the regularized subsystem Hamiltonian on the region $\Lambda^{c}\subset V$.
\end{lemma}
The proof is similar to that of Lemma \ref{lemma_subsystem_partition} and is summarized in Section \ref{subsection_ratio}.

\begin{lemma}[Lemma 8 in \cite{kliesch2014locality}]\label{lemma_y_kfold}
	Let $y\in (0,1)$ and $L\in \mathbb{N}_{>0}$ be fixed, and let $\mathcal{A}_{\geq L}(F)$ and $\mathcal{A}^{k}_{\geq L}(F)$ be defined as in \eqref{mcA_L_F} and \eqref{mcA_L_F_k}. Then for any $F\subset E$, the following inequality holds:
	\begin{equation}\label{y_f_fold_eq1}
		\begin{aligned}[b]
			\sum_{G\in \mathcal{A}_{\geq L}^{k}(F)}y^{|G|}\leq \frac{1}{k!}\qty(\sum_{G\in \mathcal{A}_{\geq L}(F)}y^{|G|})^{k}.
		\end{aligned}
	\end{equation}
\end{lemma}
\begin{proof}
	For completeness, we sketch the proof here. For each $k$-fold nonoverlapping connected edge subset in $\mathcal{A}^{k}_{\geq L}(F)$, we have $k$ choices to decompose it into two disjoint components $G_{1}$ and $G_{2}$, belonging to $\mathcal{A}^{k-1}_{\geq L}(F)$ and $\mathcal{A}_{\geq L}(F)$, respectively. By covering all cases regardless of disjointness, one can bound the left-hand side (LHS) of \eqref{y_f_fold_eq1} by $ \sum_{G_1 \in \mathcal{A}_{\ge L}^{k-1}(F)} \sum_{G_2 \in \mathcal{A}_{\ge L}(F)} y^{|G_1| + |G_2|}$. We factorize this summation and iterate to complete the proof.
\end{proof}

Applying Lemmas \ref{lemma_subsystem_partition_che} (with $\Lambda=(V_{\overline{G}^{c}})^{c}$) and \ref{lemma_trace_norm_f_che} (choosing $\Lambda=V_{G_{j}}$ for $j=1,2,\dots,m$) to \eqref{abs_cor_G_sum}, we obtain
\allowdisplaybreaks[4]
\begin{align*}\label{abs_cor_m_sum}
	&\qty|C_{\beta}(O_X,O_Y)| 
	\leq \|\widecheck{O}_{X}^{[0]}\widecheck{O}_{Y}^{[1]}\|
	\\&\quad\times \sum_{m=1}^{\infty}\sum_{G\in \mathcal{A}^{m}_{\geq L}(\partial X)}\qty[\frac{K_{3}^{|X|+|Y|}}{\operatorname{Tr}_{V_{G}}\qty(e^{-\beta W_{V_{G}}})}]^{2}\sum_{\qty{\substack{w^{(j)}\in G^{\star}_{j}:\\ G_{w^{(j)}}=G_{j}}}_{j=1}^{m}}\prod_{j=1}^{m}	\qty[\qty(\frac{K_{4}}{\beta})^{|V_{G_{j}}|}  \frac{(K_{4}\sqrt{\beta})^{|w^{(j)}|}}{|w^{(j)}|!}
	\prod_{\lambda\in G_w^{(j)}}\mu_{\lambda}(w^{(j)})!]
	\\\leq& \|\widecheck{O}_{X}^{[0]}\widecheck{O}_{Y}^{[1]}\| \sum_{m=1}^{\infty}\sum_{G\in \mathcal{A}^{m}_{\geq L}(\partial X)}\qty[K_{3}^{|X|+|Y|}(K_{5}\beta^{1/2})^{|V_{G}|}]^{2}\qty(\frac{K_{4}}{\beta})^{|V_{G}|} \qty(\frac{K_{4}\sqrt{\beta}}{1-K_{4}\sqrt{\beta}})^{|G|}
	\\\leq& K_{3}^{|X|+|Y|}\|\widecheck{O}_{X}^{[0]}\widecheck{O}_{Y}^{[1]}\| \sum_{m=1}^{\infty}\sum_{G\in \mathcal{A}^{m}_{\geq L}(\partial X)} \qty(\frac{K'_{4}\sqrt{\beta}}{1-K_{4}\sqrt{\beta}})^{|G|}.
	\refstepcounter{equation}\tag{\theequation}
\end{align*}
Here, the constants $K_3, K_{4}$, and $K_{5}$ correspond to the constants in Lemmas \ref{lemma_subsystem_partition_che}, \ref{lemma_trace_norm_f_che}, and \ref{lemma_W_x}, respectively. We have also used the summation formula from Lemma \ref{lemma_mu_w_G_sum} with the parameter $C = K_4\sqrt{\beta}$, provided $\beta$ is small enough such that $K_4\sqrt{\beta} < 1$, and noted from disjointness that $\sum_{j=1}^{m}|G_{j}|=|G|$. In the last line of \eqref{abs_cor_m_sum}, we used $|V_{G}|\leq 2|G|$ and denoted $K_{4}'\coloneq K_{4}\max\{K^{2}_{5}K_{4},1\}^{2}$. 

Let $\beta_{c}$ be sufficiently small such that $y(\beta)\coloneq K_{4}'\beta^{1/2}/(1-K_{4}\beta^{1/2})<\sigma^{-1}$ holds for all $\beta\leq \beta_{c}$. We then proceed with Lemma \ref{lemma_y_kfold} to obtain
\begin{equation}\label{estimate_cor_factor_m}
	\begin{aligned}[b]
		&\sum_{m=1}^{\infty}\sum_{G\in \mathcal{A}^{m}_{\geq L}(\partial X)} \qty(\frac{K'_{4}\sqrt{\beta}}{1-K_{4}\sqrt{\beta}})^{|G|}\leq \sum_{m=1}^{\infty}\frac{1}{m!}\qty[\sum_{G\in \mathcal{A}_{\geq L}(\partial X)}y(\beta)^{|G|}]^{m}\leq \sum_{m=1}^{\infty}\frac{1}{m!}\qty{|\partial X|\sum_{m'=L}^{\infty}[\sigma y(\beta)]^{m'}}^{m}
		\\&= \exp\qty{|\partial X|\frac{[\sigma y(\beta)]^{L}}{1-\sigma y(\beta)}}-1\leq K_{6}^{|\partial X|}e^{-L/\xi(\beta)}
	\end{aligned}
\end{equation}
for some constant $K_{6}>0$ depending only on $U_{\min}, U_{\max}, J, \mu, \mathfrak{d},\sigma$, and $\beta_{c}$, with the correlation length defined as
\begin{equation}\label{def_correlation_length}
	\begin{aligned}[b]
		\xi(\beta)\coloneq -\qty{\ln [\sigma y(\beta)]}^{-1}=-\qty{\ln \qty[\frac{\sigma K_{4}'\beta^{1/2}}{1-K_{4}\beta^{1/2}}]}^{-1}.
	\end{aligned}
\end{equation}
Here, for the second inequality, we invoked $\mathcal{A}_{\geq L}(\partial X)=\bigcup_{m'=L}^{\infty}\{G\in\mathcal{A}(\partial X)\colon |G|=m'\}$ and $|\{G\in \mathcal{A}(\partial X)\colon |G|=m'\}|\leq |\partial X|\sigma^{m'}$. To obtain the final estimate, we notice that $v\colon \mathbb{R}_{>0}\ni x\mapsto (e^{x}-1)/x$ is monotonically increasing, which implies $e^{x}-1\leq x v(x_{0})$ for $x\in (0,x_{0}]$. Using this observation and choosing $x_{0}=|\partial X|/[1-\sigma y(\beta)]$, we find 
\begin{equation}\label{}
	\begin{aligned}[b]
		\exp\qty{|\partial X|\frac{[\sigma y(\beta)]^{L}}{1-\sigma y(\beta)}}-1\leq |\partial X|\frac{[\sigma y(\beta)]^{L}}{1-\sigma y(\beta)} v(x_{0})\leq |\partial X|\frac{v(|\partial X|/[1-\sigma y(\beta_{c})])}{1-\sigma y(\beta_{c})}e^{-L/\xi(\beta)},
	\end{aligned}
\end{equation}
where the prefactor is bounded from above by $K_{6}^{|\partial X|}$ with $K_{6}\coloneq \exp(1/\qty[1-\sigma y(\beta_{c})])$.

Substituting \eqref{estimate_cor_factor_m} into \eqref{abs_cor_m_sum} and noting that $\|\widecheck{O}_{X}^{[0]}\|\leq \|O_{X}e^{-\sqrt{\beta}N_{X}}\|$, $\|\widecheck{O}_{Y}^{[1]}\|\leq 2\|O_{Y}e^{-\sqrt{\beta}N_{Y}}\|$, we finally establish 

\begin{equation}\label{}
	\begin{aligned}[b]
		\qty|C_{\beta}(O_X,O_Y)| &\leq 2 K_{3}^{|X|+|Y|}K_{6}^{|\partial X|}\|O_{X}e^{-\sqrt{\beta}N_{X}}\|\|O_{Y}e^{-\sqrt{\beta}N_{Y}}\| e^{-\operatorname{dist}(X,Y)/\xi(\beta)}
		\\&\leq K_{\mathrm{cl}}^{|X|+|Y|}\|O_{X}e^{-\sqrt{\beta}N_{X}}\|\|O_{Y}e^{-\sqrt{\beta}N_{Y}}\|e^{-\operatorname{dist}(X,Y)/\xi(\beta)},
	\end{aligned}
\end{equation}
recalling that $L=\operatorname{dist}(X,Y)$ by definition. We have used the bound $|\partial X|\leq \mathfrak{d}|X|$ and noted that the constant $K_{\mathrm{cl}}\coloneq 2\max\{K_{6}^{\mathfrak{d}},1\}K_{3}$ depends only on $U_{\min}, U_{\max}, J, \mu, \mathfrak{d},\sigma$, and $\beta_{c}$. 

\section{Technical Lemmas}\label{sec_proof_techinical_lemma}
In this section, we present detailed proofs for the technical lemmas invoked in Section \ref{sec_proof_main_theorem}. \label{reply2_emph18}

\subsection{Auxiliary Results}\label{sec_auxiliary_results}
In this subsection, we summarize several auxiliary results for later convenience. Among these, Lemma \ref{lemma_analysis} stands out as a central technical lemma, which enables us to control the expectation values of unbounded particle number operators and provides the desired analytical scalings.
\begin{lemma}\label{lemma_gibbs}
Let $H$ be a self-adjoint operator (not necessarily bounded) such that $e^{-\beta H}$ is trace class. For any $\beta>0$ and any quantum state $\rho$ (i.e., positive and $\operatorname{Tr}(\rho)=1$), let $F(\rho)\coloneq \operatorname{Tr}(H\rho)-\beta^{-1}S(\rho)$ with $S(\rho)$ being the von-Neumann entropy. Then the inequality $F(\rho)\geq F(\rho_{\beta})$ holds, where $\rho_{\beta}\coloneq e^{-\beta H}/\operatorname{Tr}(e^{-\beta H})$. 
\end{lemma}
\begin{proof}
It follows immediately from $\Tr(O_{1}\ln O_{1} -O_{1}\ln O_{2})\geq\Tr(O_{1}-O_{2})$
for all positive trace class operators $O_{1}, O_{2}$ on complex Hilbert space (See Proposition\,2.5.3 in \cite{ruelle1969statistical}).
\end{proof}
Before stating the next lemma, we introduce several helpful concepts. Let $(G_{j})_{j=0}^{m}$ be a collection of nonoverlapping edge subsets (i.e., $(V_{G_{j}})_{j=0}^{m}$ are disjoint). First, for any word $w \in E^*$ and any edge subset $G$, we define the restriction $w\upharpoonright G$ as the word obtained from $w$ by omitting all letters not contained in $G$. Let $\bm{u} \coloneq (u^{(j)})_{j=0}^{m}$ be a fixed family of words with $u^{(j)}\in G_{j}^{\ast}$. We denote by $[\bm{u}]$ the set of words whose restrictions coincide with the specified family $\bm{u}$, i.e.,
\begin{equation}\label{def_set_u}
	\begin{aligned}[b]
[\bm{u}]\coloneq \{w\in E^{\star}\colon w\upharpoonright G_{j} =u^{(j)},\,\forall j\in \{0,1,2,...,m \} \text{ and } G_{w}=\bigcup_{j=0}^{m}G_{j}\}	\end{aligned}
\end{equation}
and represent it by $[(u^{(0)},u^{(1)},...,u^{(m)})]$. 
Then the following result demonstrates how to simplify the summation of cluster expansion terms over the set $[\bm{u}]$,
\begin{lemma}\label{lemma_shuffle}
Let $f(w)$ be defined as in \eqref{interaction_picture_CE}. For a fixed set $[\bm{u}]$ defined in \eqref{def_set_u} with respect to nonoverlapping edge subsets $(G_{j})_{j=0}^{m}$, the following identities hold:
\begin{equation}\label{}
	\begin{aligned}[b]
		\sum_{w\in [\bm{u}]}f(w)=e^{-\beta \Delta_{\bm{u}}}f_{V_{G_{0}}}(u_{0})f_{V_{G_{1}}}(u_{1})f_{V_{G_{2}}}(u_{2})\dots f_{V_{G_{m}}}(u_{m}).
	\end{aligned}
\end{equation}
Here, we denote $V_{\mathrm{res}}(\bm{u})\coloneq V\setminus\bigcup_{j=0}^{m}V_{G_{j}}$ and define $\Delta_{\bm{u}}=\sum_{x\in V_{\mathrm{res}}(\bm{u})}W_{x}$ to collect the corresponding on-site potentials. Moreover, for $\widecheck{f}(w)$ defined in \eqref{def_f_che},
\begin{equation}\label{}
	\begin{aligned}[b]
		\sum_{w\in [\bm{u}]}\widecheck{f}(w)=e^{-\beta \widecheck{\Delta}^{(+)}_{\bm{u}}}\widecheck{f}_{V_{G_{0}}}(u_{0})\widecheck{f}_{V_{G_{1}}}(u_{1})\widecheck{f}_{V_{G_{2}}}(u_{2})\dots \widecheck{f}_{V_{G_{m}}}(u_{m}),
	\end{aligned}
\end{equation}
where $\widecheck{\Delta}_{\bm{u}}=\sum_{x\in V_{\mathrm{res}}(\bm{u})}\widecheck{W}_{x}$ [cf.\, \eqref{def_W_x_che}].
\end{lemma}
\begin{proof}
We present the proof for $f(w)$ defined in \eqref{interaction_picture_CE}; the derivations for $\widecheck{f}(w)$ are essentially identical.

First, observe that the elements in $[\bm{u}]$ can be generated by ordered permutations of the concatenated word $u^{(0)}\circ u^{(1)}\circ \dots \circ u^{(m)}$, such that the relative order of the letters within each subword $u^{(j)}$ is strictly preserved. To address this structure conveniently, we introduce the concept of a shuffle permutation.  For each $j\in \{0,1,\dots,m\}$, we label the letters in $u^{(j)}$ by the integer subset $I_{j}$. Specifically, we set $I_{0}=\{1,\dots,|u^{(0)}|\}$, $I_{1}=\{|u^{(0)}|+1,\dots,|u^{(0)}|+|u^{(1)}|\}$, and so on. Recall that $|w|=\sum_{j=0}^{m}|u^{(j)}|$. We denote by $\operatorname{Sh}(|u^{(0)}|,|u^{(1)}|,\dots,|u^{(m)}|)\subset S_{|w|}$ the set of all permutations $\pi\in S_{|w|}$ that are strictly increasing when restricted to each subset $I_{j}$. Namely, for all $j\in \{0,1,\dots,m\}$ and all $k,k'\in I_{j}$ with $k<k'$, the inequality $\pi(k)<\pi(k')$ holds. 

For a sequence (either a vector or a word) $b = (b_1, b_{2},\dots, b_{|w|})$ and a permutation $\pi \in S_{|w|}$, we define the permuted sequence as $\pi\cdot b \coloneq (b_{\pi(1)},b_{\pi(2)},\dots, b_{\pi(|w|)})$. 
We can then rewrite the definition of $f(w)$ from \eqref{interaction_picture_CE} as 
\begin{equation}\label{}
	\begin{aligned}[b]
		f(w)=\int_{[0,\beta]^{|w|}} \mathrm{D}\vec{\tau}\,e^{-\beta W}\bm{1}_{|w|}(\vec{\tau})h(w,\vec{\tau}),
	\end{aligned}
\end{equation}
where $\bm{1}_{n}(\vec{\tau})$ is the characteristic function of the region $\mathcal{T}_{n}$, defined for any $n$-dimensional real vector $\vec{\tau}$ as
\begin{equation}\label{}
	\begin{aligned}[b]
		\bm{1}_{n}(\vec{\tau})= \begin{cases}
			\begin{aligned}
				&1	\, && \text{if } \beta > \tau_1 > \tau_2 > \dots > \tau_{n} > 0
				\\     
				&0	\, && \text{otherwise.}
			\end{aligned}
		\end{cases}
	\end{aligned}
\end{equation}

With these preparations, and using the shorthands $\operatorname{Sh}=\operatorname{Sh}(|u^{(0)}|,\dots,|u^{(m)}|)$ and $u=u^{(0)}\circ u^{(1)}\circ \dots \circ u^{(m)}$, we obtain
\begin{equation}\label{sum_w_bmu}
	\begin{aligned}[b]
		\sum_{w\in [\bm{u}]}f(w)&=\sum_{\pi\in \operatorname{Sh}}\int_{[0,\beta]^{|u|}} \mathrm{D}\vec{\tau}\,e^{-\beta W}\bm{1}_{|u|}(\vec{\tau})h(\pi \cdot u,\vec{\tau})
		\\&=\sum_{\pi\in \operatorname{Sh}}\int_{[0,\beta]^{|u|}} \mathrm{D}(\pi\cdot \vec{\tau})\,e^{-\beta W}\bm{1}_{|u|}(\pi\cdot \vec{\tau})h(\pi \cdot u,\pi\cdot \vec{\tau})
		\\&=\sum_{\pi\in \operatorname{Sh}}\int_{[0,\beta]^{|u|}} \mathrm{D}\vec{\tau}\,e^{-\beta W}\bm{1}_{|u|}(\pi\cdot \vec{\tau}) \prod_{j=0}^{m}h(u^{(j)},\vec{\tau}^{(j)}).
	\end{aligned}
\end{equation}
In the second equality, we performed the change of integration variables $\vec{\tau} \mapsto \pi \cdot \vec{\tau}$, noting that the Jacobian determinant is unity, and therefore $\mathrm{D}(\pi\cdot \vec{\tau})= \mathrm{D}\vec{\tau}$. In the final equality, we used the fact that since the edge subsets are disjoint, operators corresponding to different subwords mutually commute. Furthermore, because a shuffle permutation $\pi \in \operatorname{Sh}$ preserves the relative ordering of indices within each block, the permuted operator string completely factorizes into a product of ordered strings for each subword: $h(\pi \cdot u, \pi \cdot \vec{\tau}) = h(u^{(0)}, \vec{\tau}^{(0)}) h(u^{(1)}, \vec{\tau}^{(1)}) \dots h(u^{(m)}, \vec{\tau}^{(m)})$. Here, $\vec{\tau}^{(j)}$ denotes the subset of time variables assigned to the $j$-th word $u^{(j)}$, such that $\vec{\tau}=\vec{\tau}^{(0)}\oplus \vec{\tau}^{(1)}\oplus \dots \oplus \vec{\tau}^{(m)}$. Notice the identity
\begin{equation}\label{shuffle_id}
	\begin{aligned}[b]
		\sum_{\pi\in \operatorname{Sh}}	\bm{1}_{|u|}(\pi\cdot \vec{\tau})=\prod_{j=0}^{m}\bm{1}_{|u^{(j)}|}(\vec{\tau}^{(j)}),
	\end{aligned}
\end{equation}
which follows immediately from the geometric decomposition of the integration domain. Substituting \eqref{shuffle_id} into \eqref{sum_w_bmu} and decomposing the potential as $W=\Delta_{\bm{u}}+\sum_{j=0}^{m}W_{V_{G_{j}}}$, we can completely factorize the integral and thereby finish the proof.

The proof for the second claim regarding $\widecheck{f}(w)$ proceeds in a completely analogous manner.
\end{proof}
We \label{ref_lemma_ana} may also need some analytical bounds to deal with the estimation of different traces. \label{reply2_lemma181}
\begin{lemma}\label{lemma_analysis}
Let $q \in \mathbb{N}_{>0}$ be fixed. Consider a polynomial $P(x) = \sum_{k=0}^q a_k x^k$ of degree $q$ with leading coefficient $a_q > 0$. Let $a_{\max}$ and $r_{\max}$ be positive constants such that $a_q \le a_{\max}$ and $|a_k|a_q^{-k/q} \le r_{\max}$ for all $k \in \{0,1,...,q-1\}$. Then, there exist positive constants $C_1, C_2$ depending only on $a_{\max}$, $r_{\max}$, and $q$ such that for any non-negative integer $p$, the following inequalities hold:
\begin{equation}\label{anaylsis_1}
	\begin{aligned}[b]	 \qty(\frac{C_{1}}{\sqrt[q]{a_{q}}})^{p+1}(p!)^{1/q}\leq \sum_{n=0}^{\infty}n^{p}e^{-P(n)}\leq  \qty(\frac{C_{2}}{\sqrt[q]{a_{q}}})^{p+1}(p!)^{1/q}
	\end{aligned}
\end{equation}
If $q$ is even, there exists a constant $C_3$ and $C_{4}$ (depending only on the same parameters) such that
	\begin{subequations}\label{}
		\begin{align}
			&		\sum_{n=0}^{\infty}(n+p)^{p}e^{-\min_{j:|j-n|\leq p}P(j)}\leq \qty(\frac{C_{3}}{\sqrt[q]{a_{q}}})^{p+1}p! \label{anaylsis_2} \\
			& 	\sum_{n=0}^{\infty}e^{-\min_{j:|j-n|\leq p}P(j)}\leq C_{4}a_{q}^{-1/q}p \label{anaylsis_3}
		\end{align}
	\end{subequations}
\end{lemma}\label{reply2_lemma18}
\begin{proof}
	The proof relies mainly on comparison to the Gamma integral and scaling analysis. We first denote $y\coloneq a_{q}^{1/q}x$ and write $P(x)=Q(y)\coloneq y^{q}+\sum_{k=0}^{q-1}a_{k}a_{q}^{-k/q}y^{k}$. Note that for any $\epsilon > 0$ and for all $k \in \{0, 1, \dots, q-1\}$, there exists a constant $c_{\epsilon, k, q} > 0$ such that $y^k \le \epsilon y^q + c_{\epsilon, k, q}$ holds for all $y \ge 0$. Choosing $\epsilon=(2r_{\max}q)^{-1}$, we obtain $\qty|\sum_{k=0}^{q-1}a_{k}a_{q}^{-k/q}y^{k}|\leq y^{q}/2+c_{1}$ with $c_{1}>0$ depending on $r_{\max}$ and $q$. This estimation leads us to
	\begin{equation}\label{Q_y_estimation}
		\begin{aligned}[b]
			\frac{1}{2}a_{q}x^{q}-c_{1}	=\frac{1}{2}y^{q}-c_{1}\leq Q(y)\leq \frac{3}{2}y^{q}+c_{1}=\frac{3}{2}a_{q}x^{q}+c_{1},
		\end{aligned}
	\end{equation}
	which is a basic tool for all the following analyses.
	
	\noindent\emph{Proof for inequality \eqref{anaylsis_1}.} Suppose $p>0$, for the upper bound in \eqref{anaylsis_1},
	\begin{equation}\label{p_ge_1}
		\begin{aligned}[b]
			&\sum_{n=0}^{\infty}n^{p}e^{-P(n)}
			\leq   \sum_{n=0}^{\infty}n^{p}e^{-\qty(a_{q}n^{q}/2-c_{1})}\leq \ii{0}{\infty}{x}x^{p}e^{-(a_{q}x^{q}/2-c_{1})}+\sup_{x\geq 0}x^{p}e^{-(a_{q}x^{q}/2-c_{1})}
			\\&=e^{c_{1}}a_{q}^{-(p+1)/q}\qty(\ii{0}{\infty}{y}y^{p}e^{-y^{q}/2}+a_{q}^{1/q}\sup_{y\geq 0}y^{p}e^{-y^{q}/2}).
		\end{aligned}
	\end{equation}
	Here the integral on RHS can be bounded from above by using the lower bound in \eqref{Q_y_estimation}, which is a standard Gamma integral upto a constant $e^{c_{1}}$. The Gamma integral $\ii{0}{\infty}{y}y^{p}e^{-y^{q}/2}$ can be readily upper bounded by $c_{2}^{p}(p!)^{1/q}$ with $c_{2}$ depending on $q$. This estimation follows from standard logarithmic convexity of the Gamma function and the Legendre duplication formula. The second term on RHS of \eqref{p_ge_1} is bounded from above by $a_{q}^{1/q}\sup_{y\geq 0}y^{p}e^{-Q(y)}\leq a_{\max}^{1/q}e^{c_{1}}\sup_{y\geq 0}y^{p}e^{-y^{q}/2}\leq c_{3}^{p}(p!)^{1/q}$ with $c_{3}$ depending on $q$ and $a_{\max}$. Combining the estimation for these two terms we complete the proof. For $p=0$, the upper bound in \eqref{anaylsis_1} can be estimated via
	\begin{equation}\label{}
		\begin{aligned}[b]
			\sum_{n=0}^{\infty}e^{-P(n)}\leq 1+\ii{0}{\infty}{x}e^{-(a_{q}x^{q}/2-c_{1})}\leq 1 + a_{q}^{-1/q}e^{c_{1}}\ii{0}{\infty}{y}e^{-y^{q}/2}\leq c_{4}a_{q}^{-1/q}
		\end{aligned}
	\end{equation}
	for some positive constant $c_{4}$ depending on $a_{\max},r_{\max}$ and $q$. 
	
	For the lower bound in \eqref{anaylsis_1} and $p>0$, we use $P(x)\leq 3a_{q}x^{q}/2+c_{1}$, i.e., 
	\begin{equation}\label{}
		\begin{aligned}[b]
			\sum_{n=0}^{\infty}n^{p}e^{-P(n)}\geq e^{-c_{1}}\sum_{n=0}^{\infty}n^{p}e^{-3a_{q}n^{q}/2}\geq e^{-c_{1}}\sum_{n\in [x_{\ast}/2,x_{\ast}]\cap \mathbb{Z}}n^{p}e^{-3a_{q}n^{q}/2}.
		\end{aligned}
	\end{equation}
	Here, we denote $x_{\ast}=[2p/(3qa_{q})]^{1/q}$ as the maximizer of $x^{p}e^{-3a_{q}x^{q}/2}$ for $x\geq 0$. For $x_{\ast}\geq 1$, a rough estimation yields $|[x_{\ast}/2,x_{\ast}]\cap \mathbb{Z}|\geq x_{\ast}/8$, leading us to
	\begin{equation}\label{lower_x_ge_1}
		\begin{aligned}[b]
			\sum_{n=0}^{\infty}n^{p}e^{-P(n)}\geq e^{-c_{1}}\frac{x_{\ast}}{8}\qty(\frac{x_{\ast}}{2})^{p}e^{-3a_{q}x_{\ast}^{q}/2}\geq c_{5}^{p}a_{q}^{-(p+1)/q}(p!)^{1/q}.
		\end{aligned}
	\end{equation}
	for some positive $c_{5}$ depending on $r_{\max}$ and $q$. If $x_{\ast}<1$, then we use 
	\begin{equation}\label{lower_x_le_1}
		\begin{aligned}[b]
			\sum_{n=0}^{\infty}n^{p}e^{-P(n)}\geq e^{-c_{1}}e^{-3a_{q}/2}\geq e^{-c_{1}}e^{-3a_{\max}/2}x_{\ast}^{p+1}\geq \tilde{c}_{5}^{p}a_{q}^{-(p+1)/q}(p!)^{1/q}
		\end{aligned}
	\end{equation}
	for some positive $c_{5}$ depending on $a_{\max},r_{\max}$ and $q$. By combining \eqref{lower_x_ge_1} and \eqref{lower_x_le_1} we complete the proof for \eqref{anaylsis_1}. Similar analysis also works for $p=0$.
	
	\noindent\emph{Proof for inequality \eqref{anaylsis_2} and \eqref{anaylsis_3}.} Let $p>0$, we note that $q$ is even and therefore we use $P(x)\geq a_{q}|x|^{q}/2-c_{1}$ to obtain
	\begin{equation}\label{}
		\begin{aligned}[b]
			\sum_{n=0}^{\infty}(n+p)^{p}e^{-\min_{j:|j-n|\leq p}P(j)}\leq& e^{c_{1}} \sum_{n=0}^{\infty}(n+p)^{p}e^{-\min_{j:|j-n|\leq p}a_{q}|j|^{q}/2}
			\\\leq& e^{c_{1}}\qty[\sum_{n=0}^{p}(2p)^{p} + \sum_{n=p+1}^{\infty}(n+p)^{p}e^{-a_{q}(n-p)^{q}/2}]
			\\=& e^{c_{1}}\qty[\sum_{n=0}^{p}(2p)^{p} + \sum_{k=1}^{\infty}(k+2p)^{p}e^{-a_{q}k^{q}/2}]
			\\\leq & e^{c_{1}}\qty[\sum_{n=0}^{p}(2p)^{p} + 2^{p}\sum_{k=1}^{\infty}k^{p}e^{-a_{q}k^{q}/2}+2^{p}(2p)^{p}\sum_{k=1}^{\infty}e^{-a_{q}k^{q}/2}]
			\\\leq & c_{6}^{p}a_{q}^{-(p+1)/q}p!
		\end{aligned}
	\end{equation}
	for some positive $c_{6}$ depending on $r_{\max},a_{\max}$ and $q$. To obtain the final result we repeatedly use the upper bound in \eqref{anaylsis_1} which we proved above. The analysis of \eqref{anaylsis_3} is similar, we simply use 
	\begin{equation}\label{}
		\begin{aligned}[b]
			\sum_{n=0}^{\infty}e^{-\min_{j:|j-n|\leq p}P(j)}\leq& e^{c_{1}} \sum_{n=0}^{\infty}e^{-\min_{j:|j-n|\leq p}a_{q}|j|^{q}/2}
			\\\leq& e^{c_{1}}\qty[\sum_{n=0}^{p}1 + \sum_{n=p+1}^{\infty}e^{-a_{q}(n-p)^{q}/2}]
			= e^{c_{1}}\qty(p + \sum_{k=1}^{\infty}e^{-a_{q}k^{q}/2})
			\leq \tilde{c}_{6}a_{q}^{-1/q}p
		\end{aligned}
	\end{equation}
	for some positive $c_{6}$ depending on $r_{\max},a_{\max}$ and $q$. For $p=0$, both inequalities \eqref{anaylsis_2} and \eqref{anaylsis_3} reduces to the upper bound of \eqref{anaylsis_1} with $p=0$. 
	
\end{proof}

\begin{lemma}\label{lemma_nplusp_p}
Let $p\in \mathbb{N}_{0}, \as_{\min},\as_{\max},\zeta_{\max},\theta_{\max},\beta_{c}>0$ be fixed. Suppose that $\alpha,\beta, \theta,\zeta \in \mathbb{R}$ satisfy $0 < \alpha_{\min} \le \alpha \le \alpha_{\max}$, $0<\beta<\beta_{c}$, $|\zeta| \le \zeta_{\max}$ and $|\theta|\leq \theta_{\max}$. Then, there exists positive constants $C$ depending only on $\alpha_{\min}, \alpha_{\max}, \theta_{\max}, \zeta_{\max}$ and $\beta_{c}$ such that the following inequalities hold:
\begin{subequations}
	\begin{align}
		&\sum_{n=0}^{\infty} (n+p)^p \exp \left[ -\beta\min_{j:|j-n| \le p} (\alpha j^2 + (\zeta+\theta\beta^{-1/2})j) \right] \le \qty(\frac{C}{\sqrt{\beta}})^{p+1} p! \\
	&\sum_{n=0}^{\infty} \exp \left[ -\beta\min_{j:|j-n| \le p} (\alpha j^2 + (\zeta +\theta\beta^{-1/2})j) \right] \le \frac{C}{\sqrt{\beta}} p.
	\end{align}
\end{subequations}
\end{lemma}
\begin{proof}
This is a special case of Lemma \ref{lemma_analysis}.
\end{proof}
\begin{lemma}\label{lemma_mul_combin}
	For non-negative integer $m_{1},m_{2},...,m_{s}$ we have
	\begin{equation}\label{mul_combin_res}
		\begin{aligned}[b]
			\frac{\qty(\sum\limits_{i=1}^{s}m_{i})!}{\prod\limits_{i=1}^{s}m_{i}!}\leq s^{\sum\limits_{i=1}^{s}m_{i}}
		\end{aligned}
	\end{equation}
\end{lemma}
\begin{proof}
	From the multinomial theorem
	\begin{equation}\label{mul_combin_1}
		\begin{aligned}[b]
			&\sum_{\substack{m_{1}+m_{2}+...+m_{s}=M\\ m_{1},m_{2},...,m_{s}\geq 0}}x_{1}^{m_{1}}x_{2}^{m_{2}}...x_{s}^{m_{s}}\frac{M!}{m_{1}!m_{2}!...m_{s}!}
			=(x_{1}+x_{2}+...+x_{s})^{M}.
		\end{aligned}
	\end{equation}
	By setting $x_{1}=x_{2}=...=x_{s}=1$, we find the RHS of \eqref{mul_combin_res} and \eqref{mul_combin_1} coincide while LHS of \eqref{mul_combin_res} is only one of the terms in the LHS of \eq{mul_combin_1}.
\end{proof}

\begin{lemma}\label{lemma_mx_property}
Let $\{\mu_{\lambda}(w)\}_{\lambda\in G_w}$ denote the multiplicities of a fixed word $w$. Define $m_x(w) \coloneq \sum_{\lambda\in G_w: \lambda\ni x} \mu_{\lambda}(w)$ to be the total number of letters in $w$ adjacent to the site $x$. Then the following relations hold:
\begin{subequations}\label{mx_ineq}
	\begin{align}
		&\sum_{x\in V_w} m_x(w) = 2|w|, 
		\\
		&\prod_{x\in V_w} m_x(w)! \leq \mathfrak{d}^{2|w|}\prod_{\lambda\in G_w} [\mu_{\lambda}(w)!]^{2},
	\end{align}
\end{subequations}
where $V_w$ denotes the support of the word $w$, and $\mathfrak{d}$ denotes the maximum degree of the underlying graph.
\end{lemma}
\begin{proof}
For the first claim, we have
\begin{equation}\label{}
	\begin{aligned}[b]
		\sum_{x\in V_{w}}m_{x}(w)=\sum_{x\in V_{w}}\sum_{\lambda\in G_w: \lambda\ni x}\mu_{\lambda}(w)
		=2\sum_{\lambda\in G_w}\mu_{\lambda}(w)=2|w|,
	\end{aligned}
\end{equation}
where we note that each edge multiplicity is counted exactly twice in the summation (once for each of its two endpoints). 

For the second claim, we obtain 
\begin{equation}\label{mxw_bound}
	\begin{aligned}[b]
		m_{x}(w)!\leq \mathfrak{d}^{\sum_{\lambda\in G_w: \lambda\ni x}\mu_{\lambda}(w)}  \prod_{\lambda\in G_w: \lambda\ni x}\mu_{\lambda}(w)!
	\end{aligned}
\end{equation}
by using the degree bound $|\{\lambda\in G_w: \lambda\ni x\}|\leq |\{\lambda\in E: \lambda\ni x\}| \leq \mathfrak{d}$ and invoking Lemma \ref{lemma_mul_combin}. Then,
\begin{equation}\label{}
	\begin{aligned}[b]
		\prod_{x\in V_{w}}m_{x}(w)!\leq \mathfrak{d}^{2|w|}\prod_{x\in V_{w}}\prod_{\lambda\in G_{w}: \lambda\ni x}\mu_{\lambda}(w)!=\mathfrak{d}^{2|w|} \prod_{\lambda\in G_w}\qty[\mu_{\lambda}(w)!]^{2}.
	\end{aligned}
\end{equation}
Here, we applied the upper bound \eqref{mxw_bound} and used the fact that the factorial of each edge multiplicity appears exactly twice in the product (again, once for each of its two endpoints).
\end{proof}

\subsection{Norm Estimates and Absolute Convergence}\label{subsection_norm_convergence}
\begin{proof}[Proof of Lemma \ref{lemma_trace_norm_n_f}]
It is convenient to rewrite the operator $f_{\Lambda}(w)$ in terms of imaginary-time durations, i.e., 
\begin{equation}\label{f_w_duration}
	\begin{aligned}[b]
		f_{\Lambda}(w)
		=\int_{\vec{\tau}\in \mathcal{T}_{|w|}}\mathrm{D}\vec{\tau}\,
		\qty(\prod_{k=1}^{|w|}e^{-\Delta\tau_{k}W_{\Lambda}}h_{w_{k}}) e^{-\Delta\tau_{|w|+1}W_{\Lambda}}.
	\end{aligned}
\end{equation}
Here, we have defined the durations $\{\Delta\tau_k\}_{k=1}^{|w|+1}$ as functions of the integration variables $\{\tau_k\}_{k=1}^{|w|}$, i.e., $\Delta\tau_1 \coloneq \beta-\tau_1$, $\Delta\tau_{|w|+1} \coloneq \tau_{|w|}$, and $\Delta\tau_k \coloneq \tau_{k-1}-\tau_k$ for $k=2,3,\dots,|w|$. For simplicity, we slightly abuse notation by dropping the explicit arguments in the functions $\Delta \tau_k$. By the definition of the integration domain $\mathcal{T}_{|w|}$ (where $\beta\geq \tau_1\geq \tau_2\geq \dots\geq\tau_{|w|}\geq 0$), it is clear that these durations are non-negative and automatically satisfy the constraint $\sum_{k=1}^{|w|+1}\Delta\tau_k=\beta$. \label{reply2_delta}

We then have
\begin{equation}\label{}
	\begin{aligned}[b]
		&\norm{n_{z}^{s} f_{\Lambda}(w)}_{1}\leq \int_{\vec{\tau}\in \mathcal{T}_{|w|}}\mathrm{D}\vec{\tau}\,
		\norm{n_{z}^{s}	\qty(\prod_{k=1}^{|w|}e^{-\Delta\tau_{k}W_{\Lambda}}h_{w_{k}}) e^{-\Delta\tau_{|w|+1}W_{\Lambda}}}_{1}.
	\end{aligned}
\end{equation}
To proceed, we factorize the product $\prod_{k=1}^{|w|}\qty[e^{-\Delta\tau_{k}W_{\Lambda}}h_{w_{k}}]$ over the sites in $\Lambda$. For each edge $w_{k}\in w$, we explicitly denote it by $\{x_{k},y_{k}\}$. We further define the following components:
\begin{equation}\label{def_nu}
	\begin{aligned}[b]
		h^{(1)}_{w_{k}}=-J_{xy}a_{x}^{\ast}a_{y},\quad h^{(2)}_{w_{k}}=-J_{xy}a_{y}^{\ast}a_{x},\quad
		h^{(3)}_{w_{k}}=-\widetilde{J}_{xy}a_{x}^{\ast}a^{\ast}_{y},\quad
		h^{(4)}_{w_{k}}=-\widetilde{J}_{xy}a_{x}a_{y},
	\end{aligned}
\end{equation}
so that we can write $h_{w_{k}}=\sum_{\nu_{k}=1}^{4}h_{w_{k}}^{(\nu_{k})}$. For a vector $\vec{\nu}=(\nu_{1},\nu_{2},\dots,\nu_{|w|})\in \{1,2,3,4\}^{|w|}$, we obtain
\begin{equation}\label{}
	\begin{aligned}[b]
		&	\qty(\prod_{k=1}^{|w|}e^{-\Delta\tau_{k}W_{\Lambda}}h_{w_{k}}) e^{-\Delta\tau_{|w|+1}W_{\Lambda}}
		=\sum_{\vec{\nu}\in \{1,2,3,4\}^{|w|}}	\qty[\prod_{k=1}^{|w|}e^{-\Delta\tau_{k}W_{\Lambda}}h^{(\nu_{k})}_{w_{k}}] e^{-\Delta\tau_{|w|+1}W_{\Lambda}},
	\end{aligned}
\end{equation}
which leads us to 
\begin{equation}\label{nGexp_beta_W_V_G}
	\begin{aligned}[b]
		\norm{n_{z}^{s}f_{\Lambda}(w)}_{1}
		\leq \int_{\vec{\tau}\in \mathcal{T}_{|w|}}\mathrm{D}\vec{\tau}\,\sum_{\vec{\nu}\in \{1,2,3,4\}^{|w|}}
		\norm{n_{z}^{s}\qty[\prod_{k=1}^{|w|}e^{-\Delta\tau_{k}W_{\Lambda}}h_{w_{k}}^{(\nu_{k})}]e^{-\Delta\tau_{|w|+1}W_{\Lambda}}}_{1}.
	\end{aligned}
\end{equation}

Next, we bound the RHS of \eqref{nGexp_beta_W_V_G} for a fixed vector $\vec{\nu}$. We first consider the factorization of $\prod_{k=1}^{|w|}h^{(\nu_{k})}_{w_{k}}= \prod_{x\in V_{w}}O_{x}(\vec{\nu})$. Here, the operator $O_{x}(\vec{\nu})\in \mathcal{P}(a_{x},a^{\ast}_{x})$ receives contributions from every edge in $w$. Notice that each edge contributes exactly one of the operators in $\{a_{x},a_{x}^{\ast},\id_{x}\}$ to the site $x\in V_{w}$. The analysis above enables us to decompose $O_{x}(\vec{\nu})\propto \prod_{k=1}^{|w|}o_{x,k}(\vec{\nu})$, where $o_{x,k}(\vec{\nu})\in\{a_{x},a_{x}^{\ast},\id_{x}\}$ is contributed by the $k$-th letter in the word (i.e., $w_{k}$). We remark that $o_{x,k}(\vec{\nu})=\id_{x}$ means $w_{k}$ is not adjacent to $x$. Here, we use the symbol $\propto$ to drop the scalars (i.e., products of elements drawn from $\{J_{xy},\widetilde{J}_{xy}\}_{y\in V_{w}}$) for brevity. This yields the following useful factorization for $\sum_{i=1}^{|w|+1}\Delta\tau_{i}=\beta$:
\begin{equation}\label{}
	\begin{aligned}[b]
		&\qty[\prod_{k=1}^{|w|}e^{-\Delta\tau_{k}W_{\Lambda}}h^{(\nu_{k})}_{w_{k}}] e^{-\Delta\tau_{|w|+1}W_{\Lambda}}
		\propto \qty{\prod_{x\in V_{w}}\qty[\prod_{k=1}^{|w|}e^{-\Delta\tau_{k}W_{x}}o_{x,k}(\vec{\nu})]e^{-\Delta\tau_{|w|+1}W_{x}}} \prod_{x\in \Lambda\setminus V_{w}}e^{-\beta W_{x}}.
	\end{aligned}
\end{equation}
Based on this, let $q_{x,k}(\vec{\nu}) \in \{+1, 0, -1\}$ denote the net change in boson number corresponding to the operator $o_{x,k}(\vec{\nu})$, such that $q_{x,k}(\vec{\nu})$ equals $+1,0,-1$ if $o_{x,k}(\vec{\nu})$ equals $a_{x}^{\ast},\id_{x},a_{x}$, respectively.

To proceed, we express the on-site Hamiltonian $W_x$ as a polynomial function of the number operator $n_x$, i.e., $W_x(n_x) = U_{x} n_x(n_x - 1)/2 - \mu_x n_x$. 
Using the canonical commutation relations and functional calculus, we obtain $W_x(n_x) a_x = a_x W_x(n_x - 1)$ and $W_x(n_x)a^{\ast}_{x}=a^{\ast}_{x} W_{x}(n_{x}+1)$. Here and in what follows, scalar terms in operator expressions are implicitly understood as multiples of the identity operator $\id_x$. Defining $q_{x,\geq k}(\vec{\nu})\coloneq \sum_{i=k}^{|w|}q_{x,i}(\vec{\nu})$, we obtain 
\begin{equation}\label{def_mcW_nx}
	\begin{aligned}[b]
		\qty[\prod_{k=1}^{|w|}e^{-\Delta\tau_{k}W_{\Lambda}}h^{(\nu_{k})}_{w_{k}}] e^{-\Delta\tau_{|w|+1}W_{\Lambda}}
		&=\qty[\prod_{x\in \Lambda}O_{x}(\vec{\nu}) e^{-\sum_{k=1}^{|w|}\Delta\tau_{k}W_{x}(n_{x}+q_{x,\geq k}(\vec{\nu}))-\Delta\tau_{|w|+1}W_{x}(n_{x})}]\prod_{x\in \Lambda\setminus V_{w}}e^{-\beta W_{x}}
		\\&\eqcolon \qty[\prod_{x\in \Lambda}O_{x}(\vec{\nu})e^{-\mathscr{W}_{x}(n_{x})}]\prod_{x\in \Lambda\setminus V_{w}}e^{-\beta W_{x}}.
	\end{aligned}
\end{equation} 
This immediately enables us to recast the trace norm on the RHS of \eqref{nGexp_beta_W_V_G} into
\begin{equation}\label{n_z_s_exp_delta_tau}
	\begin{aligned}[b]
		\norm{n_{z}^{s}\qty[\prod_{k=1}^{|w|}e^{-\Delta\tau_{k}W_{\Lambda}}h_{w_{k}}^{(\nu_{k})}]e^{-\Delta\tau_{|w|+1}W_{\Lambda}}}_{1}
		&= \norm{n_{z}^{s}O_{z}(\vec{\nu})e^{-\mathscr{W}_{z}(n_{z})}}_{1}\qty[\prod_{x\in V_{w}\setminus \{z\}}\norm{O_{x}(\vec{\nu})e^{-\mathscr{W}_{x}(n_{x})}}_{1}]
		\\&\quad\times\prod_{x\in \Lambda\setminus V_{w}}\norm{e^{-\beta W_{x}}}_{1}.
	\end{aligned}
\end{equation}
We will establish an upper bound for the first factor on the RHS of \eqref{n_z_s_exp_delta_tau}, which will seamlessly provide upper bounds for all remaining factors as well. 

We first utilize the Cauchy–Schwarz inequality, i.e., setting $(p,p_{1},p_{2})=(1,2,2)$ in \eqref{holder_inequality}, to obtain
\begin{equation}\label{nzs_Oznu_1norm_2}
	\begin{aligned}[b]
		&\norm{n_{z}^{s}O_{z}(\vec{\nu})e^{-\mathscr{W}_{z}(n_{z})}}_{1}^{2}
		\\\leq& \norm{n_{z}^{s}O_{z}(\vec{\nu})e^{-\mathscr{W}_{z}(n_{z})/2}}_{2}^{2} \norm{e^{-\mathscr{W}_{z}(n_{z})/2}}_{2}^{2}
		=\operatorname{Tr}_{z}\qty(e^{-\mathscr{W}_{z}(n_{z})/2}O_{z}(\vec{\nu})^{\ast}n_{z}^{2s}O_{z}(\vec{\nu})e^{-\mathscr{W}_{z}(n_{z})/2})\operatorname{Tr}_{z}\qty(e^{-\mathscr{W}_{z}(n_{z})}).
	\end{aligned}
\end{equation}
Here, the trace is evaluated using the basis in $\mathcal{D}_{\fin}$, which is an invariant domain for both $n_{z}^{s}O_{z}(\vec{\nu})\in \mathcal{P}(a_{z},a_{z}^{\ast})$ and $e^{-\mathscr{W}_{z}(n_{z})}$. Therefore, the identity $\qty[n_{z}^{s}O_{z}(\vec{\nu})e^{-\mathscr{W}_{z}(n_{z})/2}]^{\ast}=e^{-\mathscr{W}_{z}(n_{z})/2}O_{z}(\vec{\nu})^{\ast}n_{z}^{s}$ holds strictly on $\mathcal{D}_{\fin}$. Let $\sum_{k=1}^{|w|}|q_{x,k}(\vec{\nu})|$ be the total number of creation and annihilation operators at site $x$ provided by the word $w$, which exactly equals the total number of creation and annihilation operators in $O_{x}(\vec{\nu})$. We observe that the absolute value $|q_{x,k}(\vec{\nu})|$ actually does not depend on $\vec{\nu}$, and the summation $\sum_{k=1}^{|w|}|q_{x,k}(\vec{\nu})|$ coincides with the number of edges in $w$ that are adjacent to $x$. We denote it by $m_{x}(w)$ and naturally have
\begin{equation}\label{def_m_z_w}
	\begin{aligned}[b]
		m_{x}(w)\coloneq \sum_{k=1}^{|w|}|q_{x,k}(\vec{\nu})|=\sum_{\lambda\in G_w: \lambda\ni x}\mu_{\lambda}(w).
	\end{aligned}
\end{equation}
Then, the operator $n_{z}^{s}O_{z}(\vec{\nu})$ is given by the product of $m_{z}(w)+2s$ creation and annihilation operators. Motivated by this, for any $k\in \{1,2,\dots,m_{x}(w)+2s\}$, we use the symbol $\mathsf{q}_{k}(\vec{\nu})$ to denote the operator type of the $k$-th operator in $n_{z}^{s}O_{z}(\vec{\nu})$. Here, $\mathsf{q}_k(\vec{\nu}) = +1$ denotes a creation operator $a^{\ast}_{z}$, and $\mathsf{q}_k(\vec{\nu}) = -1$ denotes an annihilation operator $a_{z}$. By repeatedly applying the canonical commutation relations, we obtain
\begin{equation}\label{O_z_nu_ast}
	\begin{aligned}[b]
		O_{z}(\vec{\nu})^{\ast}n_{z}^{2s}O_{z}(\vec{\nu})=	\qty[n_{z}^{s}O_{z}(\vec{\nu})]^{\ast}n_{z}^{s}O_{z}(\vec{\nu})\propto\prod_{k=1}^{m_{z}(w)+2s}\qty[n_{z}+\frac{1+\mathsf{q}_{k}(\vec{\nu})}{2}+\mathsf{q}_{\geq k+1}(\vec{\nu})]
	\end{aligned}
\end{equation}
on $\mathcal{D}_{\fin}$. Here, we denote $\mathsf{q}_{\geq k}(\vec{\nu})\coloneq \sum_{i=k}^{m_{z}(w)+2s}\mathsf{q}_{i}(\vec{\nu})$ for any $k\in \{1,2,\dots,m_{z}(w)+2s\}$, with the convention that $\mathsf{q}_{\geq m_{z}(w)+2s+1}=0$. 

Treating $n_{z}$ as a non-negative integer variable $n$, the RHS of \eqref{O_z_nu_ast} can be bounded from above as follows:
\begin{equation}\label{n_mz_nu_2s}
	\begin{aligned}[b]
		\prod_{k=1}^{m_{z}(w)+2s}\qty[n+\frac{1+\mathsf{q}_{k}(\vec{\nu})}{2}+\mathsf{q}_{\geq k+1}(\vec{\nu})]\leq \qty[n+m_{z}(w)+2s]^{m_{z}(w)+2s}.
	\end{aligned}
\end{equation}
Similarly, by treating $W_{z}(n)\coloneq U_{z}n(n-1)/2-\mu_{z}n$ and $\mathscr{W}_{z}(n)$ [cf.\,\eqref{def_mcW_nx}] as scalar functions, we bound the operator $e^{-\mathscr{W}_{z}(n_{z})}$ inside the traces on the RHS of \eqref{nzs_Oznu_1norm_2}:
\begin{equation}\label{mscrW}
	\begin{aligned}[b]
		-\mathscr{W}_{z}(n)
		&\leq -\qty(\sum_{k=1}^{|w|+1}\Delta\tau_{k})\cdot \min\qty{\min_{k\in \{1,2,\dots,|w|\}}W_{z}(n+q_{z,\geq k}(\vec{\nu})),W_{z}(n)}
		\\&\leq -\beta \min_{|k|\leq m_{z}(w)}W_{z}(n+k)
		= -\beta \min_{|j-n|\leq m_{z}(w)}W_{z}(j),
	\end{aligned}
\end{equation}
where in the penultimate step we used $|q_{z,\geq k}(\vec{\nu})|\leq m_{z}(w)$ from definition \eqref{def_m_z_w}. By restoring the scalars (i.e., $J_{xy}$ and $\widetilde{J}_{xy}$) and invoking the uniform bound \eqref{uniformbound}, we deduce from \eqref{n_mz_nu_2s} and \eqref{mscrW} that 
\begin{equation}\label{trace_z_W_n_z}
	\begin{aligned}[b]
		&\operatorname{Tr}_{z}\qty(e^{-\mathscr{W}_{z}(n_{z})/2}O_{z}(\vec{\nu})^{\ast}n_{z}^{2s}O_{z}(\vec{\nu})e^{-\mathscr{W}_{z}(n_{z})/2})
		\\\leq& \max\{1,J\}^{2m_{z}(w)}\sum_{n=0}^{\infty}\qty[n+m_{z}(w)+2s]^{m_{z}(w)+2s}\exp[-\beta \min_{|j-n|\leq m_{z}(w)}W_{z}(j)]
		\\\leq& \max\{1,J\}^{2m_{z}(w)}\sum_{n=0}^{\infty}\qty[n+m_{z}(w)+2s]^{m_{z}(w)+2s}\exp[-\beta \min_{|j-n|\leq m_{z}(w)+2s}W_{z}(j)].
	\end{aligned}
\end{equation}
Applying Lemma \ref{lemma_nplusp_p} with $\alpha = U_z/2, \zeta = U_z/2 + \mu_z, \theta=0$ and observing that the conditions $0 < U_{\min}/2 \le \alpha \le U_{\max}/2$ and $|\zeta| \le U_{\max}/2 + \mu$ are satisfied due to the uniform bound \eqref{uniformbound}, we arrive at
\begin{equation}\label{}
	\begin{aligned}[b]
		\operatorname{Tr}_{z}\qty(e^{-\mathscr{W}_{z}(n_{z})/2}O_{z}(\vec{\nu})^{\ast}n_{z}^{2s}O_{z}(\vec{\nu})e^{-\mathscr{W}_{z}(n_{z})/2})&\leq \max\{1,J\}^{2m_{z}(w)}\qty(\frac{c_{1}}{\sqrt{\beta}})^{m_{z}(w)+2s+1}[m_{z}(w)+2s]!
		\\&\leq 	\qty(\frac{c_{2}}{\sqrt{\beta}})^{2s+m_{z}(w)+1} m_{z}(w)!(2s)!,
	\end{aligned}
\end{equation}
where $c_1$ is a positive constant depending only on $U_{\min}, U_{\max},\mu$, and $\beta_{c}$. We have also used the inequality $(i+j)! \le 2^{i+j} i! j!$ and defined $c_{2}\coloneq 2\max\{1,J\}^{2}c_{1}$ to absorb the combinatorial factor. Lemma \ref{lemma_nplusp_p} and \eqref{mscrW} also imply that $\operatorname{Tr}_{z}(e^{-\mathscr{W}_{z}(n_{z})})\leq c_{1}\beta^{-1/2}m_{z}(w)\leq c_{1}\beta^{-1/2}e^{m_{z}(w)}$, which, together with \eqref{trace_z_W_n_z}, establishes the following [cf.\,\eqref{nzs_Oznu_1norm_2}]:
\begin{equation}\label{nz_Oz_1norm}
	\begin{aligned}[b]
		\norm{n_{z}^{s}O_{z}(\vec{\nu})e^{-\mathscr{W}_{z}(n_{z})}}_{1}\leq \sqrt{(2s)!}\qty(\frac{c_{3}}{\sqrt{\beta}})^{s+m_{z}(w)/2+1} \qty[m_{z}(w)!]^{1/2},
	\end{aligned}
\end{equation}
with $c_{3}\coloneq \max\{ec_{2},\sqrt{c_{1}c_{2}}\}$. The same analysis applies to $\norm{O_{x}(\vec{\nu})e^{-\mathscr{W}_{x}(n_{x})}}_{1}$ in \eqref{n_z_s_exp_delta_tau}, which can be obtained by identifying $n_{z}^{0}=\id_{z}$ and setting $s=0$ on the RHS of \eqref{nz_Oz_1norm}. For $x\in \Lambda\setminus V_{w}$, we estimate $\|e^{-\beta W_{x}(n_{x})}\|_{1}\leq c_{3}\beta^{-1/2}$ by choosing $p=0$ in Lemma \ref{lemma_nplusp_p}. Returning to \eqref{n_z_s_exp_delta_tau}, we obtain
\begin{equation}\label{trace_norm_nzs_mxw}
	\begin{aligned}[b]
		&\norm{n_{z}^{s}\qty[\prod_{k=1}^{|w|}e^{-\Delta\tau_{k}W_{\Lambda}}h_{w_{k}}^{(\nu_{k})}]e^{-\Delta\tau_{|w|+1}W_{\Lambda}}}_{1}
		\leq \sqrt{(2s)!}\qty(\frac{c_{3}}{\sqrt{\beta}})^{s+|\Lambda\setminus V_{w}|}\prod_{x\in V_{w}}\qty(\frac{c_{3}}{\sqrt{\beta}})^{m_{x}(w)/2+1} \qty[m_{x}(w)!]^{1/2}.
	\end{aligned}
\end{equation}
Using Lemma \ref{lemma_mx_property}, we further simplify this to
\begin{equation}\label{trace_norm_nzs_exp_hnuk}
	\begin{aligned}[b]
		\norm{n_{z}^{s}\qty[\prod_{k=1}^{|w|}e^{-\Delta\tau_{k}W_{\Lambda}}h_{w_{k}}^{(\nu_{k})}]e^{-\Delta\tau_{|w|+1}W_{\Lambda}}}_{1}\leq \sqrt{(2s)!}\qty(\frac{c_{3}}{\sqrt{\beta}})^{s+|\Lambda|}\qty(\frac{\mathfrak{d}c_{3}}{\sqrt{\beta}})^{|w|} 	\prod_{\lambda\in G_w}\mu_{\lambda}(w)!.
	\end{aligned}
\end{equation}
Substituting this bound back into \eqref{nGexp_beta_W_V_G}, we complete the proof by noting that
\begin{equation}\label{}
	\begin{aligned}[b]
		\norm{n_{z}^{s}f_{\Lambda}(w)}_{1}
		&\leq \int_{\vec{\tau}\in \mathcal{T}_{|w|}}\mathrm{D}\vec{\tau}\,\sum_{\vec{\nu}\in \{1,2,3,4\}^{|w|}}
		\sqrt{(2s)!}\qty(\frac{c_{3}}{\sqrt{\beta}})^{s+|\Lambda|}\qty(\frac{\mathfrak{d}c_{3}}{\sqrt{\beta}})^{|w|} 	\prod_{\lambda\in G_w}\mu_{\lambda}(w)!
		\\&=\frac{\beta^{|w|}}{|w|!}\cdot 4^{|w|} \cdot  \sqrt{(2s)!}\qty(\frac{c_{3}}{\sqrt{\beta}})^{s+|\Lambda|}\qty(\frac{\mathfrak{d}c_{3}}{\sqrt{\beta}})^{|w|} 	\prod_{\lambda\in G_w}\mu_{\lambda}(w)!
	\end{aligned}
\end{equation}
and identifying the constant $C$ in the statement of the lemma with $4\mathfrak{d}c_3$. 
\end{proof}

\begin{proof}[Proof of Lemma \ref{lemma_convergence_trace_norm}]
	The proof mainly relies on Lemma \ref{lemma_trace_norm_n_f} with choosing $\Lambda=V$ and Lemma \ref{lemma_mu_w_G_sum}, i.e., we have for some $c>0$ that
	\begin{equation}\label{}
		\begin{aligned}[b]
			\sum_{w\in E^{\star}}\|n_{z}^{s}f(w)\|&\leq 	\sum_{w\in E^{\star}}\int_{\vec{\tau}\in \mathcal{T}_{|w|}}\mathrm{D}\vec{\tau}\,\norm{n_{z}^{s}e^{-\beta W}h(w,\vec{\tau})}_{1}
			\\	&\leq \sqrt{(2s)!} 
			\qty(\frac{c}{\sqrt{\beta}})^{|V|+s}\sum_{w\in E^{\star}}  \frac{(c\sqrt{\beta})^{|w|}}{|w|!}
			\prod_{\lambda\in G_w}\mu_{\lambda}(w)!
			\\&=\sqrt{(2s)!} 
			\qty(\frac{c}{\sqrt{\beta}})^{|\Lambda|+s}\qty(\frac{1}{1-c\sqrt{\beta}})^{|E|}<\infty
		\end{aligned}
	\end{equation}
The final equality holds provided that $\beta$ is sufficiently small to ensure $c\sqrt{\beta}<1$. Accordingly, the critical $\beta_{c}$ in the statement of this lemma is chosen to satisfy this convergence requirement. Furthermore, since the Bose-Hubbard model under consideration is defined on a finite graph $(V,E)$, the finiteness of $|V|$ and $|E|$ guarantees that the final upper bound is finite. 
\end{proof}\label{reply2_prooflemma3}

\begin{proof}[Proof of Lemma \ref{lemma_trace_norm_f_che}]
	The estimation mainly relies on splitting the word $w$ into two subwords, with the corresponding local terms (such as $h_{w_k}$) acting on two separate copies of the Hilbert space. We then invoke the inequality \eqref{trace_norm_nzs_exp_hnuk} established in the proof of Lemma \ref{lemma_trace_norm_n_f} by setting $s=0$ to establish the upper bound.  
	
	Following a treatment similar to the proof of Lemma \ref{lemma_trace_norm_n_f}, we introduce the vector $\vec{\nu}\in \{1,2,3,4\}^{|w|}$ and define $h_{w_{k}}^{(\nu_{k})}$ as in \eqref{def_nu}. We then obtain 
	\begin{equation}\label{f_w_duration_che}
		\begin{aligned}[b]
			\widecheck{f}_{\Lambda}(w)
			=\int_{\vec{\tau}\in \mathcal{T}_{|w|}}\mathrm{D}\vec{\tau}\,\sum_{\vec{\nu}\in \{1,2,3,4\}^{|w|}}
			\qty(\prod_{k=1}^{|w|}e^{-\Delta\tau_{k}\widecheck{W}^{(+)}_{\Lambda}}h^{(\nu_{k})(+)}_{w_{k}}) e^{-\Delta\tau_{|w|+1}\widecheck{W}^{(+)}_{\Lambda}}.
		\end{aligned}
	\end{equation}
	To expand the integrand, we introduce the configuration vector $\vec{\chi}=(\chi_{1},\chi_{2},\dots,\chi_{|w|})\in \{\ri,\rii\}^{|w|}$ to indicate whether an operator acts on the first ($\ri$) or the second ($\rii$) Hilbert space copy.  We define the natural embeddings $P_{\ri}(O)\coloneq O\otimes \id$ and $P_{\rii}(O)\coloneq \id \otimes O$, enabling us to write $h_{w_{k}}^{(\nu_{k})(+)}=\sum_{\as\in \{\ri,\rii\}}P_{\as}(h_{w_{k}}^{(v_{k})})$. Naturally, expanding the product of $h_{w_{k}}^{(\nu_{k})(+)}$ yields $2^{|w|}$ different assignment configurations:
	\begin{equation}\label{prod_k_h_nu_k_plus}
		\begin{aligned}[b]
			\prod_{k=1}^{|w|}h^{(\nu_{k})(+)}_{w_{k}}=\sum_{\vec{\chi}\in \{\ri,\rii\}^{|w|}}\prod_{k=1}^{|w|}P_{\chi_{k}}(h_{w_{k}}^{(\nu_{k})}).
		\end{aligned}
	\end{equation}
	Observing that operators acting on different Hilbert space copies commute, i.e., $[P_{\ri}(O_{1}),P_{\rii}(O_{2})]=0$, we can factorize the product on the RHS of \eqref{prod_k_h_nu_k_plus} into a bipartite tensor product, preserving the relative order within each respective tensor factor Hilbert space:
	\begin{equation}\label{}
		\begin{aligned}[b]
			\prod_{k=1}^{|w|}P_{\chi_{k}}(h_{w_{k}}^{(\nu_{k})})=\prod_{k=1}^{|w|}\qty[h_{w_{k}}^{(\nu_{k})}]^{\delta_{\chi_{k},\ri}}\otimes \prod_{k=1}^{|w|}\qty[h_{w_{k}}^{(\nu_{k})}]^{\delta_{\chi_{k},\rii}}.
		\end{aligned}
	\end{equation}
	Here, the symbol $\delta$ represents the Kronecker delta, and we adopt the convention that $O^{0}$ is understood as the identity operator in the corresponding Hilbert space. Furthermore, since the operator $e^{-\Delta\tau_{k}\widecheck{W}^{(+)}_{\Lambda}}$ factorizes as $e^{-\Delta\tau_{k}\widecheck{W}_{\Lambda}} \otimes e^{-\Delta\tau_{k}\widecheck{W}_{\Lambda}}$, we obtain 
	\begin{equation}\label{exp_W_che_hwk}
		\begin{aligned}[b]
			\prod_{k=1}^{|w|}e^{-\Delta\tau_{k}\widecheck{W}^{(+)}_{\Lambda}}h^{(\nu_{k})(+)}_{w_{k}}=\sum_{\vec{\chi}\in \{\ri,\rii\}^{|w|}}\qty{\prod_{k=1}^{|w|}e^{-\Delta\tau_{k}\widecheck{W}_{\Lambda}}\qty[h_{w_{k}}^{(\nu_{k})}]^{\delta_{\chi_{k},\ri}}}\otimes \qty{\prod_{k=1}^{|w|}e^{-\Delta\tau_{k}\widecheck{W}_{\Lambda}}\qty[h_{w_{k}}^{(\nu_{k})}]^{\delta_{\chi_{k},\rii}}}.
		\end{aligned}
	\end{equation}
	Based on \eqref{exp_W_che_hwk} and recalling \eqref{f_w_duration_che}, the remaining task is to estimate
	\begin{equation}\label{trace_norm_Wche_hwknuk_plus}
		\begin{aligned}[b]
			\norm{	\qty(\prod_{k=1}^{|w|}e^{-\Delta\tau_{k}\widecheck{W}^{(+)}_{\Lambda}}h^{(\nu_{k})(+)}_{w_{k}}) e^{-\Delta\tau_{|w|+1}\widecheck{W}^{(+)}_{\Lambda}}}_{1}\leq \sum_{\vec{\chi}\in \{\ri,\rii\}^{|w|}}\prod_{\as\in \{\ri,\rii\}} \norm{\qty{\prod_{k=1}^{|w|}e^{-\Delta\tau_{k}\widecheck{W}_{\Lambda}}\qty[h_{w_{k}}^{(\nu_{k})}]^{\delta_{\chi_{k},\as}}}e^{-\Delta\tau_{|w|+1}\widecheck{W}_{\Lambda}}}_{1}
		\end{aligned}
	\end{equation}
	for a fixed $\vec{\nu}$.
	We can readily obtain the following estimate from inequality \eqref{trace_norm_nzs_exp_hnuk} by choosing $s=0$:
	\begin{equation}\label{szero_trace_norm_nzs_exp_hnuk}
		\begin{aligned}[b]
			\norm{\qty[\prod_{k=1}^{|w|}e^{-\Delta\tau_{k}W_{\Lambda}}h_{w_{k}}^{(\nu_{k})}]e^{-\Delta\tau_{|w|+1}W_{\Lambda}}}_{1}\leq \qty(\frac{c_{1}}{\sqrt{\beta}})^{|\Lambda|}\qty(\frac{c_{1}}{\sqrt{\beta}})^{|w|} 	\prod_{\lambda\in G_w}\mu_{\lambda}(w)!
		\end{aligned}
	\end{equation}
	for some constant $c_{1}>0$ depending only on $U_{\min}, U_{\max}, \mu,\mathfrak{d}$, and $\beta_{c}$. According to the analysis in Remark \ref{remark_beta_sqaure_root}, one immediately deduces the corresponding bound for the regularized potential: 
	\begin{equation}\label{szero_trace_norm_che}
		\begin{aligned}[b]
			\norm{\qty[\prod_{k=1}^{|w|}e^{-\Delta\tau_{k}\widecheck{W}_{\Lambda}}h_{w_{k}}^{(\nu_{k})}]e^{-\Delta\tau_{|w|+1}\widecheck{W}_{\Lambda}}}_{1}\leq \qty(\frac{c_{2}}{\sqrt{\beta}})^{|\Lambda|}\qty(\frac{c_{2}}{\sqrt{\beta}})^{|w|} 	\prod_{\lambda\in G_w}\mu_{\lambda}(w)!
		\end{aligned}
	\end{equation}
	for some $c_{2}>0$ depending only on $U_{\min}, U_{\max}, \mu,\mathfrak{d}$, and $\beta_{c}$. To see this, we note that for $x\in V_{w}\setminus (X\cup Y)$, the analysis leading to \eqref{nz_Oz_1norm} applies directly upon setting $s=0$. For $x\in V_{w}\cap (X\cup Y)$, we simply modify $\mathcal{W}_{x}(n_{x})$ by redefining the operator-valued function $W_{x}(n_{x})$ as $U_{x} n_x(n_x - 1)/2 - \mu_x n_x -\beta^{-1/2}n_{x}$. This modification does not alter the structural form of \eqref{nz_Oz_1norm} when $s=0$, although it may change the generic constant. We can always choose a larger generic constant $c_{2}$ to unify both cases. Combining these two cases and following the analysis used in \eqref{trace_norm_nzs_mxw}, we arrive at \eqref{szero_trace_norm_che}.
	
	To apply \eqref{szero_trace_norm_che} to \eqref{trace_norm_Wche_hwknuk_plus}, we need some further observations. First, a fixed configuration vector $\vec{\chi}$ can always be interpreted as a decomposition of the word $w$ into two subwords, denoted by $w^{\ri}$ and $w^{\rii}$, such that the respective order of letters is preserved. Consequently, we have
	\begin{equation}\label{trans_Wche_to_was}
		\begin{aligned}[b]
			\qty{\prod_{k=1}^{|w|}e^{-\Delta\tau_{k}\widecheck{W}_{\Lambda}}\qty[h_{w_{k}}^{(\nu_{k})}]^{\delta_{\chi_{k},\as}}}e^{-\Delta\tau_{|w|+1}\widecheck{W}_{\Lambda}}=\qty[\prod_{k=1}^{|w^{\as}|}e^{-\Delta\tau_{k}\widecheck{W}_{\Lambda}}h_{w^{\as}_{k}}^{(\nu_{k})}]e^{-\Delta\tau_{|w^{\as}|+1}\widecheck{W}_{\Lambda}}
		\end{aligned}
	\end{equation}
	for $\as\in \{\ri,\rii\}$. By construction, we notice the following relations: $	G_{w^{\ri}}\cup G_{w^{\rii}}=G_{w}$, $|w^{\ri}|+|w^{\rii}|=|w|$, and 
	\begin{equation}\label{}
		\begin{aligned}[b]
			\mu_{\lambda}(w^{\ri})+\mu_{\lambda}(w^{\rii})=\mu_{\lambda}(w),\quad \forall \lambda\in w,
		\end{aligned}
	\end{equation}
	which, combined with the elementary inequality $i!j!\leq (i+j)!$, implies
	\begin{equation}\label{mulambda_wriwrii}
		\begin{aligned}[b]
			\bigg[\prod_{\lambda\in G_{w^{\ri}}}\mu_{\lambda}(w^{\ri})!\bigg]		\bigg[\prod_{\lambda\in G_{w^{\rii}}}\mu_{\lambda}(w^{\rii})!\bigg]=\prod_{\lambda\in G_{w}}\mu_{\lambda}(w^{\ri})!\mu_{\lambda}(w^{\rii})!\leq \prod_{\lambda\in G_{w}}\mu_{\lambda}(w)!=\prod_{\lambda\in G_{w}}\mu_{\lambda}(w)!.
		\end{aligned}
	\end{equation}
	Based on \eqref{szero_trace_norm_che}, \eqref{trans_Wche_to_was}, and \eqref{mulambda_wriwrii}, we deduce from \eqref{trace_norm_Wche_hwknuk_plus} that
	\begin{equation}\label{}
		\begin{aligned}[b]
			\norm{	\qty(\prod_{k=1}^{|w|}e^{-\Delta\tau_{k}\widecheck{W}^{(+)}_{\Lambda}}h^{(\nu_{k})(+)}_{w_{k}}) e^{-\Delta\tau_{|w|+1}\widecheck{W}^{(+)}_{\Lambda}}}_{1}&\leq \sum_{\vec{\chi}\in \{\ri,\rii\}^{|w|}}\qty(\frac{c^{2}_{2}}{\beta})^{|\Lambda|}\qty(\frac{c_{2}}{\sqrt{\beta}})^{|w^{\ri}|+|w^{\rii}|}	\prod_{\lambda\in G_w}\mu_{\lambda}(w)!
			\\&=\qty(\frac{c^{2}_{2}}{\beta})^{|\Lambda|}\qty(\frac{2c_{2}}{\sqrt{\beta}})^{|w|} 	\prod_{\lambda\in G_w}\mu_{\lambda}(w)!,
		\end{aligned}
	\end{equation}
	which, together with \eqref{f_w_duration_che}, leads us to
	\begin{equation}\label{}
		\begin{aligned}[b]
			\|\widecheck{f}_{\Lambda}(w)\|_{1}&\leq \int_{\vec{\tau}\in \mathcal{T}_{|w|}}\mathrm{D}\vec{\tau}\,\sum_{\vec{\nu}\in \{1,2,3,4\}^{|w|}}\norm{
				\qty(\prod_{k=1}^{|w|}e^{-\Delta\tau_{k}\widecheck{W}^{(+)}_{\Lambda}}h^{(\nu_{k})(+)}_{w_{k}}) e^{-\Delta\tau_{|w|+1}\widecheck{W}^{(+)}_{\Lambda}}}_{1}
			\\&\leq \int_{\vec{\tau}\in \mathcal{T}_{|w|}}\mathrm{D}\vec{\tau}\,\sum_{\vec{\nu}\in \{1,2,3,4\}^{|w|}}\qty(\frac{c^{2}_{2}}{\beta})^{|\Lambda|}\qty(\frac{2c_{2}}{\sqrt{\beta}})^{|w|} 	\prod_{\lambda\in G_w}\mu_{\lambda}(w)!
			\\&	=\qty(\frac{c^{2}_{2}}{\beta})^{|\Lambda|}\frac{(8c_{2}\sqrt{\beta})^{|w|}}{|w|!}	\prod_{\lambda\in G_w}\mu_{\lambda}(w)!.
		\end{aligned}
	\end{equation}
	By choosing the constant $C$ in the statement of the lemma as $\max\{c_{2}^{2},8c_{2}\}$, we complete the proof.
\end{proof}

\begin{proof}[Proof of Lemma \ref{lemma_convergence_trace_norm_til}]
	It follows directly from Lemma \ref{lemma_trace_norm_f_che} with choosing $\Lambda=V$ and Lemma \ref{lemma_mu_w_G_sum}. We note for some $c>0$ that
	\begin{equation}\label{}
		\begin{aligned}[b]
			\sum_{w\in E^{\star}}\|O_{X}^{(0)}O_{Y}^{(1)}\widetilde{f}(w)\|_{1}&\leq 	\sum_{w\in E^{\star}}\int_{\vec{\tau}\in \mathcal{T}_{|w|}}\mathrm{D}\vec{\tau}\,\norm{\widecheck{O}_{X}^{[0]}\widecheck{O}_{Y}^{[1]}e^{-\beta \widecheck{W}_{\Lambda}^{(+)}}\widetilde{h}(w,\vec{\tau})}_{1}
			\\&\leq \|\widecheck{O}_{X}^{[0]}\| \|\widecheck{O}_{Y}^{[1]}\|\qty(\frac{c}{\beta})^{|V|}\sum_{w\in E^{\star}}\frac{(c\sqrt{\beta})^{|w|}}{|w|!}	\prod_{\lambda\in G_w}\mu_{\lambda}(w)!
			\\&\leq \|\widecheck{O}_{X}^{[0]}\| \|\widecheck{O}_{Y}^{[1]}\|\qty(\frac{c}{\beta})^{|V|}\qty(\frac{1}{1-c\sqrt{\beta}})^{|E|}<\infty,
		\end{aligned}
	\end{equation}
	and the second claim follows similarly.
\end{proof}

\subsection{Ratios of Subsystem Partition Functions}\label{subsection_ratio}
\begin{proof}[Proof of Lemma \ref{lemma_subsystem_partition}]
	For fixed $\Lambda\subset V$, we introduce  $\rho_{1}\coloneq \rho_{H_{\Lambda^{c}}}\otimes\rho_{W_{\Lambda}}\coloneq   e^{-\beta H_{\Lambda^{c}}}/\operatorname{Tr}_{\Lambda^{c}}(e^{-\beta H_{\Lambda^{c}}}) \otimes e^{-\beta W_{\Lambda}}/\operatorname{Tr}_{\Lambda}(e^{-\beta W_{\Lambda}})$ and recall $\rho_{\beta}=e^{-\beta H}/\operatorname{Tr}(e^{-\beta H})$.
	From Lemma \ref{lemma_gibbs} we have
	\begin{equation}\label{free_energy}
		\begin{aligned}[b]
			F(\rho_{1})\geq F(\rho_{\beta}).
		\end{aligned}
	\end{equation} 
	To proceed, we first note that 
	\begin{equation}\label{tr_H_rho_1}
		\begin{aligned}[b]
			\operatorname{Tr}\qty(H\rho_{1})
			=&\operatorname{Tr}\qty(\qty(H_{\Lambda^{c}}+H_{\Lambda}+H_{\partial \Lambda})\rho_{1})
			=\operatorname{Tr}_{\Lambda}\qty(H_{\Lambda}\rho_{W_{\Lambda}})+\operatorname{Tr}_{\Lambda^{c}}\qty(H_{\Lambda^{c}}\rho_{H_{\Lambda^{c}}})+\operatorname{Tr}\qty(H_{\partial \Lambda}\rho_{1})
		\end{aligned}
	\end{equation}
	Here, the operator $H_{\partial \Lambda}\coloneq -\sum_{\lambda\cap\Lambda,\Lambda^{c}\neq \emptyset}h_{\lambda}$ captures the hoppings and squeezing terms on the boundary of $\Lambda$.
	On the other hand, we have
	\begin{equation}\label{S_rho_1}
		\begin{aligned}[b]
			\beta^{-1}S(\rho_{1})
			=&\beta^{-1}S(\rho_{W_{\Lambda}})+\beta^{-1}S(\rho_{H_{\Lambda^{c}}})
			\\=&\operatorname{Tr}_{\Lambda}(W_{\Lambda}\rho_{W_{\Lambda}})+\operatorname{Tr}_{\Lambda^{c}}(H_{\Lambda^{c}}\rho_{H_{\Lambda^{c}}})+\beta^{-1}\ln \operatorname{Tr}_{\Lambda}(e^{-\beta W_{\Lambda}})
			+\beta^{-1}\ln \operatorname{Tr}_{\Lambda^{c}}(e^{-\beta H_{\Lambda^{c}}}).
		\end{aligned}
	\end{equation}
	By substituting  \eqref{tr_H_rho_1}and \eqref{S_rho_1} into \eqref{free_energy} and reorganize the terms, we obtain
	\begin{equation}\label{free_energy_results}
		\begin{aligned}[b]
			\ln\frac{\operatorname{Tr}_{\Lambda}(e^{-\beta W_{\Lambda}})\operatorname{Tr}_{\Lambda^{c}}(e^{-\beta H_{\Lambda^{c}}})}{\operatorname{Tr}(e^{-\beta H})}\leq \beta \qty[-\operatorname{Tr}_{\Lambda}(I_{\Lambda}\rho_{W_{\Lambda}})+\operatorname{Tr}\qty(H_{\partial \Lambda}\rho_{1})].
		\end{aligned}
	\end{equation}
	Then we note that
	\begin{equation}\label{tr_h_prime}
		\begin{aligned}[b]
			\operatorname{Tr}_{\Lambda}(I_{\Lambda}\rho_{W_{\Lambda}})=\operatorname{Tr}\qty(H_{\partial \Lambda}\rho_{1})=0,
		\end{aligned}
	\end{equation}
	since the hopping and squeezing terms do not preserve local particle number at any site, while $W_{\Lambda}$ commutes with any local particle number operators. By putting \eqref{tr_h_prime} into \eqref{free_energy_results} we clearly finish the proof.
\end{proof}

\begin{proof}[Proof of Lemma \ref{lemma_subsystem_partition_che}]
The estimation of this ratio relies primarily on Lemma \ref{lemma_subsystem_partition} and a standard interpolation argument. We begin with the decomposition:
\begin{equation}\label{}
	\begin{aligned}[b]			\frac{\operatorname{Tr}_{\Lambda^{c}}(e^{-\beta \widecheck{H}_{\Lambda^{c}}})}{\operatorname{Tr}(e^{-\beta H})}=\frac{\operatorname{Tr}_{\Lambda^{c}}(e^{-\beta \widecheck{H}_{\Lambda^{c}}})}{\operatorname{Tr}_{\Lambda^{c}}(e^{-\beta H_{\Lambda^{c}}})}\cdot \frac{\operatorname{Tr}_{\Lambda^{c}}(e^{-\beta H_{\Lambda^{c}}})}{\operatorname{Tr}(e^{-\beta H})}\leq \frac{\operatorname{Tr}_{\Lambda^{c}}(e^{-\beta \widecheck{H}_{\Lambda^{c}}})}{\operatorname{Tr}_{\Lambda^{c}}(e^{-\beta H_{\Lambda^{c}}})}\cdot \frac{1}{\operatorname{Tr}_{\Lambda}(e^{-\beta W_{\Lambda}})}.
	\end{aligned}
\end{equation}
To bound the first factor, we temporarily drop the subscript $\Lambda^{c}$ for brevity and let $Z \coloneq (X\cup Y)\cap \Lambda^{c}$. We then introduce a one-parameter family of operators $H(\eta)\coloneq H+\eta O_{Z}= H-\eta \beta^{-1/2}N_{Z}$. Since $N_{Z}$ is infinitesimally $H$-bounded, Theorem 2.19 in Sec.\,IX of \cite{Kato1966} establishes the following identity for any $\phi \in \mathcal{D}_{\fin}$:
\begin{equation}\label{duhamel_strong}
	\begin{aligned}[b]
		\frac{\mathrm{d}}{\mathrm{d}\eta}e^{-\beta H(\eta)}\phi =-\ii{0}{\beta}{s}e^{-(\beta-s)H(\eta)}O_{Z}e^{-s H(\eta)}\phi,
	\end{aligned}
\end{equation}
where the integrand $e^{-(\beta-s)H(\eta)}O_{Z}e^{-s H(\eta)}$ is continuous with respect to $s$ in the norm topology of $\mathcal{B}(\mathcal{H})$. Similar to \eqref{cauchy_problem_H}, the strong derivative $\mathrm{d}/\mathrm{d}\eta$ is taken in the norm topology of $\mathcal{H}$. Equation \eqref{duhamel_strong} naturally motivates evaluating the trace: 
\begin{equation}\label{duhamel_weak}
	\begin{aligned}[b]
		\operatorname{Tr}\qty(\frac{\mathrm{d}}{\mathrm{d}\eta}e^{-\beta H(\eta)})=-\operatorname{Tr}\qty(\ii{0}{\beta}{s}e^{-(\beta-s)H(\eta)}O_{Z}e^{-s H(\eta)}).
	\end{aligned}
\end{equation}
On the RHS, the integrand is continuous for $s\in (0,\beta)$ in the trace norm, owing to the smoothing effect of the exponential suppression factor $e^{-\beta H(\eta)}$. At the boundaries $s=0$ and $s=\beta$, the integrand remains bounded in the trace norm due to Theorem \ref{theorem_low_density} for $\beta\leq \beta_{c}$. These observations guarantee that the integrand is absolutely integrable in the trace norm, thereby justifying the interchange of the trace and the integral. 

On the LHS of \eqref{duhamel_weak}, rigorously interchanging the trace and the derivative requires establishing the analyticity of the operator-valued map $\eta \mapsto e^{-\beta H(\eta)}$ in the trace norm topology. We present the proof in the vicinity of $\eta=0$. The general case for any finite $\eta_{0} \in [0,1]$ follows identically via a standard translation argument, treating $H+\eta_{0}O_{Z}$ as the unperturbed Hamiltonian. We examine the corresponding Dyson series:
\begin{equation}\label{exp_eta_expansion}
	\begin{aligned}[b]
			e^{-\beta H(\eta)} &= \sum_{m=0}^{\infty}(-\eta)^{m}\ii{0}{\beta}{\tau_{1}}\ii{0}{\tau_{1}}{\tau_{2}}\dots \ii{0}{\tau_{m-1}}{\tau_{m}}   e^{-(\beta-\tau_{1})H}O_{Z}e^{-(\tau_{1}-\tau_{2})H}\dots e^{-(\tau_{m-1}-\tau_{m})H}O_{Z}e^{-\tau_{m}H} \\
		&\eqcolon \sum_{m=0}^{\infty}(-\eta)^{m}\ii{0}{\beta}{\tau_{1}}\ii{0}{\tau_{1}}{\tau_{2}}\dots \ii{0}{\tau_{m-1}}{\tau_{m}}   S_{m}  
	\end{aligned}
\end{equation}

Following a similar notation used in the proof of Lemma \ref{lemma_trace_norm_n_f}, we estimate the trace norm of the integrand $S_m$ by introducing $\Delta\tau_1\coloneq \beta-\tau_1$, $\Delta\tau_{m+1}\coloneq \tau_m$, and $\Delta\tau_k\coloneq\tau_{k-1}-\tau_k$ for $k=2,3,\dots,m$. Since these non-negative durations satisfy $\sum_{k=1}^{m+1}\Delta \tau_{k}=\beta$, there exists at least one index $\ell\in \{1,2,\dots,m+1\}$ such that $\Delta \tau_{\ell} > \beta/[2(m+1)]$. We denote this lower threshold by $\epsilon \coloneq \beta/[2(m+1)]$. In what follows, $c_{1},c_{2,\beta},c_{2,\beta}',c_{3},c_{4},c_{5}$ and $c_{6,\beta}$ denote generic positive constants independent of $m$ and the integration variables $\{\tau_{k}\}_{k=0}^{m}$ (but may depend on $\beta$). Applying H\"older's inequality for Schatten norms to $S_{m}$ gives
\begin{equation}\label{tracenorm_S_m}
	\begin{aligned}[b]
		\|S_{m}\|_{1}
		&= \|e^{-\Delta\tau_{1} H}O_{Z}\dots e^{-\Delta\tau_{\ell} H}\dots O_{Z}e^{-\Delta\tau_{m+1}H}\|_{1} \\
		&\leq \beta^{-m/2}\|e^{-\Delta\tau_{1}H}N_{Z}\|\dots \|e^{-\Delta\tau_{\ell-1}H}N_{Z}\| \|e^{-\Delta\tau_{\ell}H}\|_{1} \|N_{Z}e^{-\Delta\tau_{\ell+1}H}\|\dots \|N_{Z}e^{-\Delta\tau_{m+1}H}\|.
	\end{aligned}
\end{equation}
The trace-norm factor can be bounded by splitting the duration:
\begin{equation}\label{tnorm_ell}
	\begin{aligned}[b]
		\|e^{-\Delta\tau_{\ell}H}\|_{1} \leq \|e^{-\epsilon H}\|_{1}\|e^{-(\Delta\tau_{\ell}-\epsilon)H}\| \leq \qty(\frac{c_{1}}{\sqrt{\epsilon}})^{|V|}c_{2,\beta} = c'_{2,\beta}(m+1)^{|V|/2}.
	\end{aligned}
\end{equation}
Here, the first norm is controlled using bounds analogous to \eqref{upper_parition_function} and Lemma \ref{lemma_analysis}. For the second norm, the uniform lower boundedness of $H$ implies via functional calculus that $\|e^{-(\Delta\tau_{\ell}-\epsilon)H}\| \leq \max\{e^{-(\Delta\tau_{\ell}-\epsilon)c_{3}},1\} \leq \max\{e^{-\beta c_{3}},1\} \eqcolon c_{2,\beta}$.
To control the remaining operator norms in \eqref{tracenorm_S_m}, we employ the operator inequality $c^{2}_{4}(H+c_{5})\geq N^{2}$ from \eqref{lower_bound_hamlitonian_2}. In the sense of quadratic forms on $\mathcal{D}_{\mathrm{fin}}$, this implies
\begin{equation}\label{graph_norm_N_H}
	\begin{aligned}[b]
		\|N_{Z}\psi\| \leq c_{4}\|(H+c_{5})^{1/2}\psi\|
	\end{aligned}
\end{equation}
for all $\psi\in \mathcal{D}_{\mathrm{fin}}$. Since $\mathcal{D}_{\mathrm{fin}}$ is a core for $(H+c_{5})^{1/2}$, this relative bound extends to $\mathcal{D}((H+c_{5})^{1/2})$. As a result, the operator $N_{Z}(H+c_{5})^{-1/2}$ is bounded on the entire Hilbert space $\mathcal{H}$ with norm bounded by $c_{4}$. Consequently, for any $\Delta\tau_{k} \in (0,\beta)$, we have
\begin{equation}\label{n_NZexp}
	\begin{aligned}[b]
		\|N_{Z}e^{-\Delta\tau_{k}H}\| \leq \|N_{Z}(H+c_{5})^{-1/2}\| \|(H+c_{5})^{1/2}e^{-\Delta\tau_{k}H}\| \leq \frac{c_{4}e^{c_{5}\Delta\tau_{k}}}{\sqrt{2e\Delta\tau_{k}}} \leq \frac{c_{6,\beta}}{(\Delta\tau_{k})^{1/2}}.
	\end{aligned}
\end{equation}
By passing to the adjoint, we symmetrically obtain $\|e^{-\Delta\tau_{k}H}N_{Z}\| \leq c_{6,\beta}(\Delta\tau_{k})^{-1/2}$.

Inserting \eqref{tnorm_ell} and \eqref{n_NZexp} into \eqref{tracenorm_S_m} yields
\begin{equation}\label{sm_bound_final}
	\begin{aligned}[b]
		\|S_{m}\|_{1} \leq \max_{\ell:\Delta\tau_{\ell}>\epsilon} \beta^{-m/2}c'_{2,\beta}(m+1)^{|V|/2}\prod_{\substack{k=1 \\ k\neq \ell}}^{m+1}\frac{c_{6,\beta}}{(\Delta\tau_{k})^{1/2}}\leq \sum_{\ell=1}^{m+1}\beta^{-m/2}c'_{2,\beta}(m+1)^{|V|/2}\prod_{\substack{k=1 \\ k\neq \ell}}^{m+1}\frac{c_{6,\beta}}{(\Delta\tau_{k})^{1/2}}.
	\end{aligned}
\end{equation}
Evaluating the corresponding Liouville-Dirichlet-type simplex integral, one finds for any $\eta\in (0,1)$ that
\begin{equation}\label{dyson_convergence}
	\begin{aligned}[b]
		\sum_{m=0}^{\infty}\eta^{m}\ii{0}{\beta}{\tau_{1}}\dots \ii{0}{\tau_{m-1}}{\tau_{m}}  \| S_{m}\|_{1} \leq \sum_{m=0}^{\infty} \beta^{-m/2}c'_{2,\beta}(m+1)^{|V|/2+1}\frac{(c_{6,\beta}\pi \beta)^{m/2}}{\Gamma(\frac{m}{2} + 1)}.
	\end{aligned}
\end{equation}
The Gamma function in the denominator ensures that the RHS converges, thereby guaranteeing the absolute and uniform convergence of the power series on the LHS. This establishes the analyticity of $e^{-\beta H(\eta)}$ in the trace norm, thereby rigorously justifying the commutation of the trace and the parameter derivative. \label{reply2_analycity}

Based on the analysis above, and utilizing the cyclicity of the trace, we obtain  
\begin{equation}\label{}
	\begin{aligned}[b]
		\frac{\mathrm{d}}{\mathrm{d}\eta}	\operatorname{Tr}\qty(e^{-\beta H(\eta)})=-\ii{0}{\beta}{s}\operatorname{Tr}\qty(e^{-(\beta-s)H(\eta)}O_{Z}e^{-s H(\eta)})=-\beta\operatorname{Tr}\qty(O_{Z}e^{-\beta H(\eta)})
		\leq c_{7} |Z|\operatorname{Tr}\qty(e^{-\beta H(\eta)})
	\end{aligned}
\end{equation}
for some positive constant $c_{7}$ depending only on the model parameters $U_{\min}, U_{\max}, J, \mu, \mathfrak{d}$, and $\beta_{c}$. Here, we invoked Theorem \ref{theorem_low_density}, noting that the effective parameter satisfies $\eta\beta^{1/2}\leq \beta^{1/2}_{c}$ [cf.\,Remark \ref{remark_beta_sqaure_root}]. Finally, restoring the subscript $\Lambda^{c}$ and integrating over $\eta \in [0, 1]$, we complete the proof:
\begin{equation}\label{}
	\begin{aligned}[b]
		\frac{\operatorname{Tr}(e^{-\beta \widecheck{H}_{\Lambda^{c}}})}{\operatorname{Tr}\qty(e^{-\beta H_{\Lambda^{c}}})}=\exp\qty{\ii{0}{1}{\eta}\frac{\mathrm{d}}{\mathrm{d}\eta}\ln \operatorname{Tr}\qty(e^{-\beta H_{\Lambda^{c}}(\eta)})}\leq e^{c_{7}|(X\cup Y)\cap \Lambda^{c}|}\leq e^{c_{7}(|X|+|Y|)}.
	\end{aligned}
\end{equation}
\end{proof}

\subsection{Combinatorics Results}\label{subsection_combinatorics}
\begin{proof}[Proof of Lemma \ref{lemma_mu_w_G_sum}]
First we denote $m\coloneq |G|$ and adopt a shorthand $\mu_{k}$ for $\mu_{k}(w)$ for brevity. Then,
\allowdisplaybreaks[4]
\begin{align*}\label{proof_lemma_mu_w_G_sum_1}
	\sum_{w\in G^{\star}:G_{w}= G}\qty[\frac{C ^{|w|}}{|w|!}\prod_{k=1}^{|G|}\mu_{k}!]
	=&	\sum_{l=m}^{\infty}\sum_{\substack{\mu_{1},\mu_{2},...,\mu_{m}\geq 1\\\mu_{1}+\mu_{2}+...+\mu_{m}=l }}\binom{l}{\mu_{1},\mu_{2},...,\mu_{m}}\frac{C ^{l}}{l!}\prod_{k=1}^{m}\mu_{k}!
	\\=&	\sum_{l=m}^{\infty}\sum_{\substack{\mu_{1},\mu_{2},...,\mu_{m}\geq 1\\\mu_{1}+\mu_{2}+...+\mu_{m}=l }}\frac{l!}{\prod_{k=1}^{m}\mu_{k}!}\frac{C ^{l}}{l!}\prod_{k=1}^{m}\mu_{k}!
		\\=&	\sum_{l=m}^{\infty}\sum_{\substack{\mu_{1},\mu_{2},...,\mu_{m}\geq 1\\\mu_{1}+\mu_{2}+...+\mu_{m}=l }}C^{\mu_{1}}C^{\mu_{2}}...C^{\mu_{m}}
		\\=&	\sum_{\mu_{1}=1}^{\infty}C^{\mu_{1}}\sum_{\mu_{2}=1}^{\infty}C^{\mu_{2}}...\sum_{\mu_{m}=1}^{\infty}C^{\mu_{m}}
		=\qty(\frac{C}{1-C})^{m}.
		\refstepcounter{equation}\tag{\theequation}
	\end{align*}
	In the last line of \eqref{proof_lemma_mu_w_G_sum_1}, we need to use $C\in (0,1)$ to ensure the convergence of the series.
	
By a similar argument, one can obtain
	\begin{align*}\label{}
		\sum_{w\in G^{\star}}\qty[\frac{C ^{|w|}}{|w|!}\prod_{k=1}^{|G|}\mu_{k}!]
		=&	\sum_{l=0}^{\infty}\sum_{\substack{\mu_{1},\mu_{2},...,\mu_{m}\geq 0\\\mu_{1}+\mu_{2}+...+\mu_{m}=0 }}\binom{l}{\mu_{1},\mu_{2},...,\mu_{m}}\frac{C ^{l}}{l!}\prod_{k=1}^{m}\mu_{k}!
		\\=&	\sum_{\mu_{1}=0}^{\infty}C^{\mu_{1}}\sum_{\mu_{2}=0}^{\infty}C^{\mu_{2}}...\sum_{\mu_{m}=0}^{\infty}C^{\mu_{m}}
		=\qty(\frac{1}{1-C})^{m}.
		\refstepcounter{equation}\tag{\theequation}
	\end{align*}
\end{proof}

\begin{proof}[Proof Lemma \ref{lemma_rho_G}]
	We first recall the definition of $\rho(G)$ below for reader's convenience:
	\begin{equation}\label{rho_G_re}
		\begin{aligned}[b]
			\rho(G)= \sum_{w\in [(\partial G)^{c}]^{\star}:G_{w}\supseteq G}f(w).
		\end{aligned}
	\end{equation}
	We observe that any required word $w$ in the summation here belongs to $[(v,v')]$ with $v,v'$ being some words such that the relations $v\in G^{\star}, G_{v}=G$ and $v'\in (\overline{G}^{c})^{\star}$ are satisified [cf.\,\eqref{def_set_u}]. Therefore, the summation in \eqref{rho_G_re} can be rewritten and calculated as
	\begin{equation}\label{}
		\begin{aligned}[b]
			\rho(G)=\sum_{v\in G^{\star}: G_{v}=G}\sum_{v'\in (\overline{G}^{c})^{\star}}\sum_{w\in [(v,v')]}f(w)=\sum_{v\in G^{\star}: G_{v}=G}\sum_{v'\in (\overline{G}^{c})^{\star}}e^{-\beta \Delta_{G}}f_{V_{G}}(v)f_{V_{\overline{G}^{c}}}(v')
		\end{aligned}
	\end{equation}
	Here, we first used Lemma \ref{lemma_shuffle} with choosing $m=1$, $(G_{0},G_{1})=(\overline{G}^{c},G)$ and identify $V_{\mathrm{res}}([(v,v')])=V\setminus(V_{G}\cup V_{\overline{G}^{c}})=V_{\mathrm{res}}(G)$ as defined in the statement of Lemma \ref{lemma_rho_G}. Then by invoking \eqref{subsystem_expansion}, choosing $\Lambda=V_{\overline{G}^{c}}$, we complete the proof.
\end{proof}

\begin{proof}[Proof of Lemma \ref{lemma_not_connecting_word}]
	For notational brevity, let $m=|w|$ be the length of the word. By assumption, there exists a permutation $\{\pi_{1},\pi_{2},...,\pi_{m}\}=\{1,2,...,m\}$ such that 
	\begin{equation}\label{overlap_X_Y_word}      
		(X \cup w_{\pi_1} \cup w_{\pi_2} \cup \dots \cup w_{\pi_s}) \cap (Y \cup w_{\pi_{s+1}} \cup w_{\pi_{s+2}} \cup \dots \cup w_{\pi_m}) = \emptyset
	\end{equation}
	For further notational brevity and without loss of any generality, we can always relabel the letters such that $\{\pi_1, \pi_2, \dots, \pi_s\} = \{1, 2, \dots, s\}$ and $\{1, 2, \dots, s\} \setminus \{\pi_1, \pi_2, \dots, \pi_m\} = \{s+1, s+2, \dots, m\}$. Using this, we introduce the shorthand
	\begin{equation}\label{}
		\begin{aligned}[b]
			h_{w_{1}}(\tau_{1})^{(+)}h_{w_{2}}(\tau_{2})^{(+)}...h_{w_{s}}(\tau_{s})^{(+)}\eqcolon\widetilde{O}_{1}, \quad h_{w_{s+1}}(\tau_{s+1})^{(+)}h_{w_{s+2}}(\tau_{s+2})^{(+)}...h_{w_{m}}(\tau_{m})^{(+)}\eqcolon\widetilde{O}_{2}
		\end{aligned}
	\end{equation}
	and recognize from \eqref{overlap_X_Y_word} that
	\begin{equation}
		(X \cup \operatorname{Supp}\widetilde{O}_1) \cap (Y \cup \operatorname{Supp}\widetilde{O}_2) = \emptyset. 
	\end{equation}
	With the preparation above, we can factorize the trace in \eqref{eq_lemma_8} to region $\Lambda_{1}\coloneq X \cup \operatorname{Supp}\widetilde{O}_1$ and $\Lambda_{1}^{c}$, i.e., we move forward with 
	\begin{align*}\label{}
		&\operatorname{Tr}\qty(O_{X}^{(0)}O^{(1)}_{Y}\widetilde{f}(w))
		\\=&\operatorname{Tr}\qty(\ii{0}{\beta}{\tau_{1}}\ii{0}{\tau_{1}}{\tau_{2}}...\ii{0}{\tau_{m-1}}{\tau_{m}}\,O_{X}^{(0)}O^{(1)}_{Y}e^{-\beta W^{(+)}}h_{w_{1}}(\tau_{1})^{(+)}h_{w_{2}}(\tau_{2})^{(+)}...h_{w_{m}}(\tau_{m})^{(+)})
		\\=&\ii{0}{\beta}{\tau_{1}}\ii{0}{\tau_{1}}{\tau_{2}}...\ii{0}{\tau_{m-1}}{\tau_{m}}\,\operatorname{Tr}\qty( O_{X}^{(0)}O^{(1)}_{Y}e^{-\beta W^{(+)}} h_{w_{1}}(\tau_{1})^{(+)}h_{w_{2}}(\tau_{2})^{(+)}...h_{w_{m}}(\tau_{m})^{(+)})
		\\=&\ii{0}{\beta}{\tau_{1}}\ii{0}{\tau_{1}}{\tau_{2}}...\ii{0}{\tau_{m-1}}{\tau_{m}}\,\operatorname{Tr}_{\Lambda_{1}}\qty(O_{X}^{(0)}e^{-\beta W_{\Lambda_{1}}^{(+)}}\widetilde{O}_{1}) \cdot \operatorname{Tr}_{\Lambda_{1}^{c}}\qty(O_{Y}^{(1)}e^{-\beta W_{\Lambda_{1}^{c}}^{(+)}}\widetilde{O}_{2}).
		\refstepcounter{equation}\tag{\theequation}
	\end{align*}
	Here, the exchange of iterated integral and trace will be justified by Fubini's theorem, since the integrand is absolutely integrable in trace norm as established in Lemma \ref{lemma_convergence_trace_norm_til}. Let $\mathcal{S}$ be the swap operator on $\mathcal{H}_{\Lambda_{1}^c} \otimes \mathcal{H}_{\Lambda_{1}^c}$, we observe that the operators $\widetilde{O}_{2}$ and $e^{-\beta W_{\Lambda_{1}^{c}}^{(+)}}=e^{-\beta W_{\Lambda_{1}^{c}}}\otimes  e^{-\beta W_{\Lambda_{1}^{c}}}$ are symmetric under $\mathcal{S}$ while the operator $O_Y^{(1)}$ is antisymmetric. Consequently, the trace $\operatorname{Tr}_{\Lambda_{1}^{c}}\qty(O_{Y}^{(1)}e^{-\beta W_{\Lambda_{1}^{c}}^{(+)}}\widetilde{O}_{2})$ vanishes. We thus complete the proof.
\end{proof}

\begin{proof}[Proof of Lemma \ref{lemma_rearrange_C_L}]
The proof is finished by exchanging the order of summation and applying the binomial identity, 
	\begin{equation}\label{}
		\begin{aligned}[b]
			- \sum_{k=1}^K b_k \qty[ \sum_{m=1}^k (-1)^m \binom{k}{m} ]
			= - \sum_{k=1}^K b_k \qty[ \sum_{m=0}^k \binom{k}{m} (-1)^m 1^{k-m} - \binom{k}{0} ] 
			= - \sum_{k=1}^K b_k (-1) = \sum_{k=1}^K b_k.
		\end{aligned}
	\end{equation}
Since $\|\binom{k}{m}b_{k}\|\leq 2^{k}\|b_{k}\|$, the limit of RHS in \eqref{eq_lemma_rearrage} can be written as $- \sum_{m=1}^{\infty} (-1)^m \sum_{k=m}^{\infty} \binom{k}{m} b_k$. For $b_{k}=\sum_{w\in \mathcal{C}_{\geq L}^{k}( \partial X)}O_{X}^{(0)}O_{Y}^{(1)}\widetilde{f}(w)$, the following rather rough estimation invoking Lemmas \ref{lemma_trace_norm_f_che} and \ref{lemma_mu_w_G_sum} is sufficient, we have for some $C>0$ that 
\begin{equation}\label{}
	\begin{aligned}[b]
		\sum_{k=1}^{\infty}2^{k}\|b_{k}\|_{1}&\leq \sum_{k=1}^{\infty}\sum_{w\in \mathcal{C}^{k}_{\geq L}(\partial X)} \|O_{X}^{(0)}O_{Y}^{(1)}\widetilde{f}(w)\|_{1}\leq \|\widecheck{O}_{X}^{[0]}\|\|\widecheck{O}_{Y}^{[1]}\|\sum_{k=1}^{\infty}2^{k}\sum_{w\in \mathcal{C}^{k}_{\geq L}(\partial X)} \|\widecheck{f}(w)\|_{1}
		\\&\leq \|\widecheck{O}_{X}^{[0]}\|\|\widecheck{O}_{Y}^{[1]}\|\sum_{w\in E^{\star}}2^{|w|} \|\widecheck{f}(w)\|_{1} \leq \|\widecheck{O}_{X}^{[0]}\|\|\widecheck{O}_{Y}^{[1]}\|\sum_{w\in E^{\star}}	\qty(\frac{C}{\beta})^{|\Lambda|}  \frac{(2C\sqrt{\beta})^{|w|}}{|w|!}
		\prod_{\lambda\in G_w}\mu_{\lambda}(w)!
		\\&<\infty,
	\end{aligned}
\end{equation}
noting that the number of clusters $k \le |w|$.
\end{proof}
\begin{proof}[Proof of Lemma \ref{lemma_word_to_graph}]
	The proof is motivated by Lemmas 6 and 10 in \cite{kliesch2014locality}, and here we adapt it within our notation and sketch the proof of standard combinatorics trick used in Lemma 10 of \cite{kliesch2014locality} here.  Essentially, we need to prove the following identity
	\begin{equation}\label{}
		\begin{aligned}[b]
			\sum_{k=m}^{\infty}\binom{k}{m}\sum_{w\in \mathcal{C}^{k}_{\geq L}(F)}\widecheck{f}(w)=\sum_{G\in \mathcal{A}_{\geq L}^{m}(F)}\sum_{w\in [(\partial G)^{c}]^{\star}:G_{w}\supseteq G}\widecheck{f}(w),
		\end{aligned}
	\end{equation}
	which follows from a standard double-counting argument. Specifically, let us fix $k \ge m$ and a word $w \in \mathcal{C}_{\geq L}^{k}(F)$. The underlying edge subset $G_w$ contains exactly $k$ maximal clusters overlapping with $F$, which we denote by $\{c^{(i)}\}_{i=1}^{k}$. Due to the constraints $w \in [(\partial G)^{c}]^{\star}$ and $G_{w}\supseteq G$, this fixed word $w$ is counted on the RHS  exactly as many times as there are ways to choose $m$ clusters from $\{c^{(i)}\}_{i=1}^{k}$ to constitute an element in $\mathcal{A}_{\geq L}^{m}(F)$. This count is exactly given by the binomial coefficient $\binom{k}{m}$. This observation completes the proof.
	
	The absolute convergence from Lemma \ref{lemma_convergence_trace_norm_til} justifies the reordering of summation.
\end{proof}

\begin{proof}[Proof of Lemma \ref{lemma_rho_G_til}]
	It follows in a similar way to the proof of Lemma \ref{lemma_rho_G} by using the second claim in Lemma \ref{lemma_shuffle}. We first choose $G_{0}=\overline{G}^{c}$ and $(G_{j})_{j=1}^{m}$ to be exactly the disjoint decomposition of $G$. Then one observes 
	\begin{equation}\label{}
		\begin{aligned}[b]
			&\widecheck{\rho}(G)= \sum_{w\in [(\partial G)^{c}]^{\star}:G_{w}\supseteq G}\widecheck{f}(w)
			\\&=\sum_{u_{0}\in (\overline{G}^{c})^{\star}}\sum_{u_{1}\in G^{\star}_{1}: G_{u_{1}}=G_{1}}\sum_{u_{2}\in G^{\star}_{2}: G_{u_{2}}=G_{2}}...\sum_{u_{m}\in G^{\star}_{m}: G_{u_{m}}=G_{m}}e^{-\beta \widecheck{\Delta}^{(+)}_{\bm{u}}}\widecheck{f}_{V_{G_{0}}}(u_{0})\widecheck{f}_{V_{G_{1}}}(u_{1})\widecheck{f}_{V_{G_{2}}}(u_{2})...\widecheck{f}_{V_{G_{m}}}(u_{m})
		\end{aligned}
	\end{equation}
	By identifying $V_{\mathrm{res}}(\bm{u})=V_{\mathrm{res}}(G)$ and invoking \eqref{subsystem_expansion_til} we complete the proof.
\end{proof}

\subsection{Analytical Bounds}\label{subsection_analytical}
\begin{proof}[Proof of Lemma \ref{lemma_n_z_s_moment}]
It directly follows from Lemma \ref{lemma_analysis} by identifying the polynomial as $P(n) = \beta (U_z n(n-1)/2 - \mu_z n)$ with $q=2$. Since $\beta \in (0, \beta_c]$, we choose the parameters as $a_{\max} = \beta_c U_{\max}/2$ and $r_{\max} = \beta_c^{1/2} (\mu + U_{\max}/2) (U_{\min}/2)^{-1/2}$.
\end{proof}
\begin{proof}[Proof of Lemma \ref{lemma_W_x}]
It directly follows from Lemma \ref{lemma_analysis} by identifying the polynomial as $P(n) = \beta (U_x n(n-1)/2 - \mu_x n)$ with $q=2$. Since $\beta \in (0, \beta_c]$, we choose the parameters as $a_{\max} = \beta_c U_{\max}/2$ and $r_{\max} = \beta_c^{1/2} (\mu + U_{\max}/2) (U_{\min}/2)^{-1/2}$.
\end{proof}
\label{reply2_analyicalbounds}

\section{Implications of the Main Theorems}\label{sec_application}
The low-boson-density inequality and clustering theorem established in the previous section serve as fundamental tools for analyzing bosonic lattice models. In this section, we apply these results to the Bose-Hubbard class of Hamiltonians to derive a quasi Dulong-Petit law \cite{landau1980statistical,fitzgerel1960law} for the specific heat and provide a proof of the bosonic version of the thermal area law \cite{wolf2008area}.

\subsection{Quasi Dulong-Petit Law}\label{sec_dulong}
We show that in the high-temperature regime, the specific heat density $\mathcal{C}_{V}(\beta)$ defined in \eqref{def_c_v} of the Bose-Hubbard model is upper-bounded by a constant independent of the system size, which can be viewed as a weak version of the Dulong–Petit law. 
We restate the precise statement of this result here for the reader's convenience. \label{reply_dulong}

\noindent\textbf{Corollary \ref{corollary_dulong} (Restated).} \textit{Let $\mathcal{C}_V(\beta)$ be the specific heat density of the Bose-Hubbard model \eqref{bh} on a finite graph $(V,E)$. There exist a temperature threshold $\beta_c > 0$ and a constant $C > 0$, both independent of the system size $|V|$, such that $\mathcal{C}_V(\beta) \leq C$ holds for all $\beta \in (0, \beta_c]$. Here, $\beta_c$ and $C$ depend only on the model parameters $U_{\min}, U_{\max}, J, \mu$, and $\mathfrak{d}$.}
\begin{proof}
	We choose a sufficiently small $\beta_{c}$ such that Theorems \ref{theorem_low_density} and \ref{theorem_clustering} hold for all $\beta\in (0, \beta_{c}]$. Let $\mathcal{R}\coloneq V\cup E$. We decompose the Hamiltonian as $H=\sum_{r\in \mathcal{R}}h_{r}$, where $h_{r}=W_{x}$ for $r=x\in V$, and $h_{r}=-h_{\lambda}$ for $r=\lambda\in E$. The specific heat density is then given by $\mathcal{C}_{V}(\beta)=|V|^{-1}\beta^{2}\sum_{r,r'\in \mathcal{R}}C_{\beta}(h_{r},h_{r'})$. We split the summation into overlapping ($r\cap r'\neq \emptyset$) and disjoint ($r\cap r'=\emptyset$) pairs:
	\begin{equation}\label{cv_upperbound}
		\mathcal{C}_{V}(\beta)\leq |V|^{-1}\beta^{2}\sum_{\substack{r,r'\in \mathcal{R}:\\r\cap r'\neq \emptyset}}|C_{\beta}(h_{r},h_{r'})| + |V|^{-1}\beta^{2} \sum_{\substack{r,r'\in \mathcal{R}:\\ r\cap r'=\emptyset}}|C_{\beta}(h_{r},h_{r'})|.
	\end{equation}
	
	For the disjoint term, Theorem \ref{theorem_clustering} implies that $|C_{\beta}(h_{r},h_{r'})|$ is bounded by $$c_{1}^{|r|+|r'|}\|h_{r}e^{-\sqrt{\beta}N_{r}}\|\|h_{r'}e^{-\sqrt{\beta}N_{r'}}\|e^{-\operatorname{dist}(r,r')/\xi(\beta)},$$ 	for some $c_{1}>0$ depending only on $U_{\min}, U_{\max}, J, \mu, \mathfrak{d}$, and $\beta_{c}$. 	
Since $h_{r}$ is at most quadratic in the local particle numbers, the elementary bound $\|n^{s}_{x}e^{-\sqrt{\beta}n_{x}}\|\leq (s/e)^{s}\beta^{-s/2}$ yields $\beta\|h_{r}e^{-\sqrt{\beta}N_{r}}\|\leq c_{2}$ for some $c_{2}>0$ depending only on $U_{\max}, J$, and $\mu$. Given $|\mathcal{R}|\leq (\mathfrak{d}+1)|V|$, the summation over the graph can be bounded by a convergent geometric series by choosing $\beta$ small enough such that $\mathfrak{d}e^{-1/\xi(\beta)}<1$. Thus, the disjoint part is bounded by a constant independent of $|V|$.
	
	For the overlapping term, the Cauchy-Schwarz inequality gives
	\begin{equation}\label{}
		\begin{aligned}[b]
			|C_{\beta}(h_{r},h_{r'})|\leq \sqrt{C_{\beta}(h_{r},h_{r})C_{\beta}(h_{r'},h_{r'})}\leq\frac{1}{2} \qty[C_{\beta}(h_{r},h_{r})+C_{\beta}(h_{r'},h_{r'})].
		\end{aligned}
	\end{equation}
Notice that $C_{\beta}(h_{r},h_{r})$ can be bounded by the thermal moments of $h_{r}$, i.e.,  $C_{\beta}(h_{r},h_{r})\leq \operatorname{Tr}(h_{r}^{2}\rho_{\beta})$. Since $h_{r}$ is at most quadratic in the particle number operators, $h_{r}^{2}$ is at most quartic. Theorem \ref{theorem_low_density} then ensures the existence of some $c_{3}>0$ depending only on $U_{\min}, U_{\max}, J, \mu, \mathfrak{d}$, and $\beta_{c}$, such that $\beta^{2}C_{\beta}(h_{r},h_{r})\leq c_{3}$ for all $r\in \mathcal{R}$. Finally, for any fixed $r\in \mathcal{R}$, there are at most $2(\mathfrak{d}+1)$ elements $r'\in \mathcal{R}$ adjacent to it. Consequently, the summation over overlapping pairs is also bounded by some constant independent of $|V|$. This completes the proof.
\end{proof}

\subsection{Boson Thermal Area Law}
This subsection presents a proof of the thermal area law for the Bose-Hubbard model, which establishes an upper bound for the mutual information $\mathcal{I}(A\colon B)$ defined in \eqref{def_mi} between two bipartite subsystems ($V=A\sqcup B$) at thermal equilibrium.  
While initially established for spin systems \cite{wolf2008area}, the thermal area law has recently been extended to the standard (i.e., without $a^{\ast}a^{\ast}+\mathrm{h.c.}$ terms) homogeneous Bose-Hubbard model \cite{lemm2023thermal}. Here, we provide an improved result that yields a tighter temperature scaling at high temperatures for the inhomogeneous Bose-Hubbard model defined in \eqref{bh}. 

For the standard Bose-Hubbard model \eqref{standard_BHM} defined on a finite periodized $d$-dimensional lattice, recent work \cite{lemm2023thermal} showed that for any $\beta>0$,
\begin{equation}\label{eq:boson_bound}
	\mathcal{I}(A:B)\leq C\max\{1,\beta\} |\partial A|
\end{equation}
for some $C>0$ depending only on $J,U,\mu$, and $d$.  Below, we show that by using Theorem \ref{theorem_clustering}, we can straightforwardly generalize \eqref{eq:boson_bound} to the Bose-Hubbard class \eqref{bh} while obtaining an improved scaling at high temperatures.

\noindent\textbf{Corollary \ref{corollary_area_law} (Restated).} \textit{Let $\mathcal{I}(A:B)$ be the mutual information of the Bose-Hubbard model \eqref{bh} on a finite graph $(V,E)$. There exist a temperature threshold $\beta_c > 0$ and a constant $C > 0$, both independent of the system size $|V|$, such that $\mathcal{I}(A:B) \leq C\cdot \beta \cdot |\partial A| $ holds for all $\beta \in (0, \beta_c]$. Here, $\beta_c$ and $C$ depend only on the model parameters $U_{\min}, U_{\max}, J, \mu, \mathfrak{d}$, and $|\partial A|$ is the boundary size of $A$.}
\begin{proof}
	By Lemma \ref{lemma_gibbs}, we bound the mutual information by the average differences of the boundary term: 
	\begin{equation}\label{IAB_HI}
		\mathcal{I}(A:B) \leq \beta \operatorname{Tr}(H_{\partial}(\rho_{A}\otimes \rho_{B}-\rho_{\beta})).
	\end{equation}
	For the Bose-Hubbard model, by definition we have
	\begin{equation}\label{}
		\begin{aligned}[b]
			H_{\partial}=-\sum_{x\in A,y\in B : x\sim y}(J_{xy}a_{x}^{\ast}a_{y}+\widetilde{J}_{xy}a_{x}^{\ast}a^{\ast}_{y}+\mathrm{h.c.}),
		\end{aligned}
	\end{equation}
	which, together with \eqref{IAB_HI}, leads us to
	\begin{equation}\label{mi_est}
		\begin{aligned}[b]
			\mathcal{I}(A\colon B)\leq \beta \sum_{\substack{x\in A,y\in B:\\ x\sim y}}\qty{J_{xy}\qty[C_{\beta}(a_{x}^{\ast},a_{y})+C_{\beta}(a_{y}^{\ast},a_{x})]+\widetilde{J}_{xy}\qty[C_{\beta}(a_{x}^{\ast},a_{y}^{\ast})+C_{\beta}(a_{y}^{\ast},a_{x}^{\ast})]},
		\end{aligned}
	\end{equation}
	where we used the fact that for $x\in A$, $\operatorname{Tr}_{A}(a_{x}\rho_{A})=\operatorname{Tr}(a_{x}\rho_{\beta})$, and similarly for $y\in B$ or the annihilation operators. We choose $\beta_{c}$ to be sufficiently small such that Theorem \ref{theorem_clustering} holds for all $\beta\in (0, \beta_{c}]$. Notice that for $\operatorname{dist}(x,y)=1$ and $\beta\leq \beta_{c}$, the following estimate holds: 
	\begin{equation}\label{mi_estimation_1}
		\begin{aligned}[b]
			C_{\beta}(a_{x}^{\ast},a_{y})\leq c_{1}^{2}\|a_{x}^{\ast}e^{-\sqrt{\beta}n_{x}}\|\|a_{y}e^{-\sqrt{\beta}n_{y}}\|e^{-1/\xi(\beta)}\leq c_{1}^{2}\cdot(c_{2}\beta^{-1/4})^{2}\cdot c_{3}\beta^{1/2}=c_{1}^{2}c_{2}^{2}c_{3}
		\end{aligned}
	\end{equation}
	for some constants $c_{1},c_{2},c_{3}>0$. We remark that $c_{1}, c_{3}$ depend only on $U_{\min}, U_{\max}, J, \mu$, and $\mathfrak{d}$, while $c_{2}$ depends only on $\beta_{c}$. Here we have used the explicit expression for the thermal correlation length \eqref{def_correlation_length}.  The remaining terms in \eqref{mi_est} can be analyzed accordingly. By noticing that the number of terms in \eqref{mi_est} can be bounded linearly with respect to the boundary size $|\partial A|$, owing to the uniform upper bound for the graph degree, we complete the proof.  
\end{proof}

\section{Discussions}\label{sec_discussion}
\subsection{Optimality and Comparisons}\label{sec_discussion_op}
Several further observations regarding our main theorems are summarized in this subsection.

Firstly, we investigate the sharpness of the temperature scaling for the RHS of the inequalities in Theorem \ref{theorem_low_density} and for the prefactor in Theorem \ref{theorem_clustering}. For Theorem \ref{theorem_low_density}, we consider the thermal expectation value of $n_{x}^{s}$ in the on-site case at high temperatures $\beta \le \beta_{c}$, i.e., $J=0$, where the Gibbs state factorizes. By invoking Lemma \ref{lemma_analysis}, we can estimate the upper and lower bounds of the moment to obtain
\begin{equation}\label{optimal_1}
	\begin{aligned}[b]
		c_{1}^{s}\cdot \beta^{-s/2} \cdot \sqrt{s!}\leq \langle n_{x}^{s} \rangle_{\beta W}=\frac{\sum_{n=0}^{\infty}n^{s}e^{-\beta(U_{x}n(n-1)/2-\mu_{x}n)}}{\sum_{n=0}^{\infty}e^{-\beta(U_{x}n(n-1)/2-\mu_{x}n)}} \leq c_{2}^{s}\cdot \beta^{-s/2} \cdot \sqrt{s!}
	\end{aligned}
\end{equation}
for any $\beta \le \beta_{c}$ and some constants $c_{1}, c_{2}$ depending only on $U_{\min}, U_{\max}, \mu$, and $\mathfrak{d}$. The temperature scaling in this estimate saturates the bound given by the low-boson-density inequality at high temperatures. However, the scaling with respect to $s$ becomes sharper in the purely on-site case. For Theorem \ref{theorem_clustering}, as a heuristic justification, we consider the correlation function $C_{\beta}(n_{x},n_{x})$ in the on-site case, which is given by $\langle n_{x}^{2} \rangle_{\beta W} - \langle n_{x} \rangle^{2}_{\beta W}$ and scales as $\beta^{-1}$ at high temperatures. Although Theorem \ref{theorem_clustering} is not applicable to overlapping regions, we can still see that this physical scaling exactly matches the temperature scaling of the prefactor $\|n_{x}e^{-\sqrt{\beta}n_{x}}\|^{2}$ inferred from Theorem \ref{theorem_clustering}. \label{reply_temp_scaling}

Secondly, setting aside the requirement of particle number conservation, we compare our method with the pioneering work of Ueltschi \cite{Ueltschi1999}. In that work, the clustering of correlations is treated within the abstract cluster expansion framework for polymer models \cite{Koteckỳ1986}. Divergences arising from unbounded operators are tamed by introducing a specific ``boson norm'' scaled by local number operators to guarantee the convergence of the expansion. However, this norm remains finite only for observables that grow at most linearly with the local particle number. In contrast, our approach relies on the exponential regularization scheme $O_X e^{-\sqrt{\beta}N_X}$, which systematically suppresses observables exhibiting arbitrary polynomial growth. Furthermore, while the combinatorial techniques in \cite{Ueltschi1999} are powerful for establishing the analyticity of the free energy density, they are less convenient for extracting precise temperature scalings. Finally, we note that the explicit proofs in \cite{Ueltschi1999} are restricted to the standard Bose-Hubbard model with strictly homogeneous parameters (i.e., $J_{xy}=J, U_{x}=U, \mu_{x}=\mu$ for all $x,y\in V$). Our method, conversely, readily accommodates inhomogeneous hoppings/interactions, requiring only the uniform bounds specified in \eqref{uniformbound}. 

We close this subsection by briefly comparing our bosonic clustering theorem with well-established results for spin or fermionic systems. Translating to our notation, Theorems 2 and 4 of \cite{kliesch2014locality} consider finite-range spin/fermionic models with uniformly bounded interactions defined on a finite graph, characterized by finite-dimensional local Hilbert spaces. For two bounded operators $O_{X}$ and $O_{Y}$ supported on $X, Y \subset V$, it is shown that below a threshold inverse temperature $\beta_{c}$ (independent of the system size), the bound 
\begin{equation}\label{spin_clustering}
	\begin{aligned}[b]
		|C_{\beta}(O_{X},O_{Y})|\leq c_{5}\min\{|X|,|Y|\}\|O_{X}\|\|O_{Y}\|e^{-\operatorname{dist}(X,Y)/\xi'(\beta)}
	\end{aligned}
\end{equation} holds for some $c_{5}>0$ independent of the system size. The thermal correlation length is given by $\xi'(\beta) \coloneq \{- \ln [\sigma e^{c_{6}\beta }(e^{c_{6}\beta }-1)]\}^{-1}$ (which is positive for sufficiently small $\beta$) for some system-size-independent constant $c_{6}>0$. Apart from the inherent boundedness of the operator norms in spin systems, the differences between this result and Theorem \ref{theorem_clustering} lie in two main aspects. First, the scaling with respect to the region sizes ($|X|$ and $|Y|$) is linear in the spin case, whereas Theorem \ref{theorem_clustering} exhibits an exponential scaling due to the regularization method required to suppress the divergence of unbounded bosonic operators. Second, the implementation of this regularization method, along with the trace-norm estimation of the Dyson series, naturally affects the functional form of the correlation length.  

\subsection{Generalizations and Geometric Embeddings}\label{sec_generalization_geometric}
We now briefly discuss the generalization of our main results to a broader class of models and provide more details regarding geometric settings. 

Let $d_{c}>0$ and an even integer $q \in 2\mathbb{N}_{0}$ be fixed. The framework presented in this work can be extended to the following class of models with finite-range hopping and squeezing: 
\begin{equation}\label{BHclass}
	\begin{aligned}[b]
		H=-\sum_{\substack{x,y\in V:
				\operatorname{dist}(x,y)\leq d_{c}}}(J_{xy}a_{x}^{\ast}a_{y}+\widetilde{J}_{xy}a_{x}^{\ast}a^{\ast}_{y}+\mathrm{h.c.}) + \sum_{k=0 }^{q}U^{(k)}_{x}n^{k},
	\end{aligned}
\end{equation}
subject to the uniform bounds \begin{equation}\label{}
	\begin{aligned}[b]
		&0< U_{\min}\leq  U_{x}^{(q)} \leq U_{\max} \eqcolon U^{(q)},\quad
		|U_{x}^{(k)}|\leq U^{(k)}  \quad \text{for } k=0,1,...,q-1.
	\end{aligned}
\end{equation}
This generalized model differs from the canonical Bose-Hubbard model by including beyond-nearest-neighbor hopping terms and higher-order polynomial interactions in the local particle number. Neither modification fundamentally affects our proofs. The additional hopping and squeezing terms merely require us to embed the original simple graph $(V,E)$ into a new simple graph with the same vertex set $V$ but a larger edge set. The new maximum degree can be bounded by $\mathfrak{d}_{c} \coloneq \sum_{k=1}^{\lfloor d_{c}\rfloor}\mathfrak{d}^{k}$, which remains independent of the system size $|V|$. The higher-order polynomial interactions are naturally accommodated by Lemma \ref{lemma_analysis}. Consequently, the low-boson-density inequality remains valid at high temperatures in the modified form $\operatorname{Tr}(n_{x}^{s}\rho_{\beta}) \le K_{\mathrm{low},q}^{s} \cdot s! \cdot \beta^{-s/q}$. For the generalized version of Theorem \ref{theorem_clustering}, we modify the regularization transformation in Definition \ref{definition_regularization} by setting $K_{0}=1$ and defining $\widecheck{W} \coloneq W - \beta^{-1/q}(N_{X}+N_{Y})$, $\widecheck{O}_{X}^{[0]} \coloneq O_{X}^{(0)}e^{-\beta^{1/q} N_{X}^{(+)}}$, and $\widecheck{O}_{Y}^{[1]} \coloneq O_{Y}^{(1)}e^{-\beta^{1/q} N_{Y}^{(+)}}$. This yields the same form for the clustering theorem, but with the correlation length replaced by $\xi_{q}(\beta) \coloneq -\{\ln [\sigma_{c} K_{4,q}'\beta^{1-1/q} / (1-K_{4,q}\beta^{1-1/q})]\}^{-1}$, where $\sigma_{c}$ is the new growth constant. The constants $K_{\mathrm{low},q}$, $K'_{4,q}$, and $K_{4,q}$ remain independent of the system size $|V|$. 

In our proofs, we relied only on the graph distance $\operatorname{dist}(x,y)$, ensuring that our clustering theorems apply to general graph structures regardless of specific geometric embeddings. The physical distance $d(x,y)$ typically enters through a natural finite-range hopping constraint $d(x,y) \le \ell_{\mathrm{hop}}$. By the triangle inequality, this implies $\ell_{\mathrm{hop}} \operatorname{dist}(x,y) \ge d(x,y)$, yielding the corresponding clustering \label{reply_discussion} bound 
\begin{equation}\label{}
	\begin{aligned}[b]
		|C_{\beta}(O_{X},O_{Y})| \le K_{\mathrm{cl}}^{|X|+|Y|} \|O_{X}e^{-\sqrt{\beta}N_{X}}\| \|O_{Y}e^{-\sqrt{\beta}N_{Y}}\| e^{-d(X,Y)/[\ell_{\mathrm{hop}}\xi(\beta)]},
	\end{aligned}
\end{equation}
where $d(X,Y) \coloneq \min_{x \in X, y \in Y} d(x,y)$.

As discussed in a subsequent work \cite{Tong2025} by two of the authors of this article, our framework can indeed be extended to the Bose-Hubbard class of Hamiltonians with power-law decaying long-range couplings. This extension requires a substantial amount of work to handle the much more intricate combinatorics of cluster expansions in the presence of non-local couplings and unbounded bosonic operators. Specifically, in \eqref{BHclass}, the first summation runs over all $x, y \in V$, with the coupling parameters satisfying
\begin{equation}\label{long_range_coupling}
	\begin{aligned}[b]
		\max\{J_{xy},\widetilde{J}_{xy}\}\leq \frac{J}{\qty[1+\operatorname{dist}(x,y)]^{\mathfrak{a}}}
	\end{aligned}
\end{equation}
for some sufficiently large $\mathfrak{a}>0$. For such models at high temperatures, we have rigorously established a power-law clustering theorem and a low-boson-density inequality. It is worth noting that a structurally similar low-density condition has frequently appeared as a fundamental assumption for proving Lieb-Robinson bound and propagation bound for long-range boson models \cite{Lemm2023,Lemm2025}.

\section{Conclusion and Outlook}\label{sec_conclusion}
In this article, by combining cluster expansions with the interaction picture, we established a clustering theorem for the high-temperature Gibbs state of the Bose-Hubbard model. We rigorously bound the moments of the local particle number, thereby providing an analytical justification for the low-boson-density assumption frequently used in previous works. As direct applications, we derived the quasi Dulong-Petit law and the thermal area law for such bosonic systems.

We conclude this paper by outlining several prospective research directions closely related to our work. A primary open question is the rigorous extension of Araki's one-dimensional results \cite{araki1969gibbs} to establish clustering properties and the absence of phase transitions for one-dimensional bosonic chains at arbitrary temperatures. Furthermore, our techniques may be adapted to investigate more complex correlation structures, such as uniform mixing and conditional mutual information \cite{bluhm2022exponential,kuwahara2024clustering}, as well as the high-temperature sudden death of entanglement \cite{bakshi2024high}. Finally, extending these static low-density bounds to dynamical regimes (e.g., Lindblad evolutions) and exploring their implications for the complexity of bosonic thermal states \cite{alhambra2021locally,kliesch2014locality} remain compelling subjects for future investigation.
 
\section*{Acknowledgements}
The authors thank Prof.\,Naomichi Hatano and Prof.\,Masahito Yamazaki for
fruitful discussions. X.-H.T. was supported by the FoPM,
WINGS Program, the University of Tokyo. Z.G. acknowledges support from the University of Tokyo Excellent Young Researcher Program and
from JST ERATO Grant Number JPMJER2302, Japan. T. K. acknowledges the Hakubi projects of RIKEN. T. K. was supported by JST PRESTO (Grant No.
JPMJPR2116), ERATO (Grant No. JPMJER2302), and JSPS Grants-in-Aid for Scientific Research (No. JP23H01099, JP24H00071), Japan.

\section*{Data Availability}
This work did not include the generation or analysis of datasets.

\section*{Conflict of Interest}
The authors have no conflict of interest to declare.
\clearpage
\appendix
\section{Appendix: Operator-Theoretic Foundations}\label{app_well_define}
In this appendix, we provide rigorous functional analysis details that ensure the well-definedness of several concepts in the main text.

We first show that the bosonic Hamiltonian  \eqref{bh} defined over a finite lattice is essentially self-adjoint on $\mathcal{D}_{\fin}$. Recall that the on-site potential grows quadratically with $n_{x}$, dominating the hopping and the squeezing terms. \label{reply_proposition_1}
\begin{proof}[Proof of Proposition \ref{pro_self_adjoint_bh}]
We first note that the graph $(V,E)$ is finite and the on-site repulsive interaction strength is uniformly bounded from below by a strictly positive constant, i.e., $U_{x} \geq U_{\min} > 0$ for all $x\in V$. By the Kato-Rellich theorem (see Theorem X.\,12 in \cite{Reed1975}), it suffices to show that $I$ is $W$-bounded with a relative bound strictly less than one. In fact, the relative bound here can be chosen to be arbitrarily small. To see this, for any $\psi\in \mathcal{D}_{\mathrm{fin}}$, we have 
\begin{equation}\label{Ipsi_relative_bound}
	\begin{aligned}[b]
		\|I\psi\| \leq 2J|E|(\|N\psi\| + \|\psi\|),
	\end{aligned}
\end{equation}
which, although crude, is sufficient. 
Because the on-site potential $W$ grows quadratically with the local particle numbers, $N$ is infinitesimally bounded with respect to $W$. Therefore, for any $\epsilon > 0$, there exists a constant $c_{\epsilon,V}$ (depending only on $\epsilon, |V|, U_{\min}, \mu$, and $J$) such that 
\begin{equation}\label{I_relative_bounded_W}
	\begin{aligned}[b]
		\|I\psi\| \leq \epsilon \|W\psi\| + c_{\epsilon,V} \|\psi\|.
	\end{aligned}
\end{equation}
Since this inequality holds for any $\psi$ in the core $\mathcal{D}_{\mathrm{fin}}$ of $W$, we can extend it to the domain $\mathcal{D}(\overline{W})$. Invoking the Kato-Rellich theorem then completes the proof. 
\end{proof}

With the spectrum of $H$ bounded from below, we now show that the partition function is finite.

\begin{proof}[Proof of Proposition \ref{pro_trace_class}]
Proposition \ref{pro_stability} confirms that $H$ is bounded from below, ensuring that $e^{-\beta H}$ is a well-defined positive bounded operator. It then suffices to show that $\operatorname{Tr}\qty(e^{-\beta H}) < \infty$. Since $\mathcal{D}_{\mathrm{fin}}$ is a core for $H$ (Proposition \ref{pro_self_adjoint_bh}) and trivially for $H_{\mathrm{low}}\coloneq U_{\min} \sum_{x \in V} n_x^2 / 4 - C_{\mathrm{sta}, V}$, the variational min-max principle (see Theorem XIII.\,1 in \cite{Michael1978}) implies the ordering of their eigenvalues: $E_{k}(H) \geq E_{k}(H_{\mathrm{low}})$ for all $k \geq 1$. By Theorem XIII.\,64 of \cite{Reed1978}, this inequality also ensures that $H$ has a purely discrete spectrum, since clearly $E_{k}(H_{\mathrm{low}}) \rightarrow \infty$ as $k \rightarrow \infty$.
Here, $E_{k}(\cdot)$ denotes the $k$-th eigenvalue arranged in non-decreasing order, counting multiplicities. 
Consequently, since the lattice $V$ is finite, we obtain
\begin{equation}\label{upper_parition_function}
	\operatorname{Tr}(e^{-\beta H}) \leq \operatorname{Tr}(e^{-\beta H_{\mathrm{low}}}) 
	= \prod_{x\in V} \qty( \sum_{n=0}^{\infty}e^{-\beta U_{\min}n^{2}/4+\beta C_{\mathrm{sta},V}} ) < \infty,
\end{equation}
which concludes the proof.
\end{proof}
Using Proposition \ref{pro_stability} and the finiteness of $V$, we first establish
\begin{equation}\label{lower_bound_hamlitonian_2}
	\begin{aligned}[b]
		H\geq \frac{U_{\min}}{4|V|}N^{2}-C_{\mathrm{sta},V}.
	\end{aligned}
\end{equation}
We use this lower bound to achieve the following:

\begin{proof}[Proof of Proposition \ref{pro_thermal_average}]
	It suffices to show that $Oe^{-\epsilon H}$ is a bounded operator for any $\epsilon>0$; then $Oe^{-\beta H}=Oe^{-\epsilon H}\cdot e^{-(\beta-\epsilon)H}$ is automatically trace-class due to Proposition \ref{pro_trace_class}. The remainder of the proof is devoted to showing that each factor in the following decomposition is a bounded operator:
	\begin{equation}\label{}
		\begin{aligned}[b]
			O e^{-\epsilon H}&=O(N+c_{1})^{-c_{2}}\cdot (N+c_{1})^{c_{2}} (c_{3}N^{2}+c_{4})^{-k} \cdot  (c_{3}N^{2}+c_{4})^{k} (H+c_{5})^{-k} \cdot (H+c_{5})^{k} e^{-\epsilon H}
			\\&= T_{1}\cdot T_{2}\cdot T_{3} \cdot T_{4}
		\end{aligned}
	\end{equation}
	for some $c_{4},c_{5}>0$ and a positive integer $k$, where we temporarily denote $c_{3}=U_{\min}/(4|V|)$. 
	
	For $T_{1}$, we recognize that $(N+c_{1})^{-c_{2}}$ is well-defined on $\mathcal{D}_{\mathrm{fin}}$. Therefore, for any $\eta\in \mathcal{D}_{\mathrm{fin}}$, there exists $\psi\in \mathcal{D}_{\mathrm{fin}}$ such that $(N+c_{1})^{-c_{2}} \eta= \psi$. By assumption, we have $\|T_{1}\eta\|= \|O(N+c_{1})^{-c_{2}} \eta\|\leq \|(N+c_{1})^{c_{2}}(N+c_{1})^{-c_{2}}\eta\|=\|\eta\|$. This establishes the boundedness of $T_{1}$ on $\mathcal{D}_{\mathrm{fin}}$, and by the bounded linear transformation (BLT) theorem, this conclusion extends to the entire $\mathcal{H}$. For $T_{2}$, we choose sufficiently large $k$ and $c_{4}$ to ensure its boundedness on $\mathcal{D}_{\mathrm{fin}}$ and apply the BLT theorem. The same analysis applies to $T_{4}$ via the spectral mapping theorem. For $T_{3}$, using \eqref{lower_bound_hamlitonian_2}, we first choose a sufficiently large $c_{5}$ such that $H+c_{5}\geq c_{3}N^{2}+c_{5}-C_{\mathrm{sta},V} >0$. Then $(H+c_{5})^{-k}$ is defined over the entire Hilbert space $\mathcal{H}$ with range $\mathcal{D}(H^{k})$. By noting that $\mathcal{D}(H^{k})=\mathcal{D}(N^{2k})$ (the proof of which is presented below), we confirm that $T_{3}$ is defined on the entire $\mathcal{H}$. Note that $(c_{3}N^{2}+c_{4})^{k}$ is self-adjoint and therefore closed, leading to the closedness of $T_{3}$. Then, by the closed graph theorem, $T_{3}$ is bounded. 
	
	A standard analysis of graph norm equivalence establishes $\mathcal{D}(H^{k})=\mathcal{D}(N^{2k})$ for all $k\in \mathbb{N}_{>0}$. The key intuition here is that the on-site potential $W$ (which scales quadratically in $N$) dominates $I$ (which scales linearly in $N$). With this in mind, we can show that for any positive integer $k$, the inequalities $c_{k}(\|N^{2k}\psi\|+\|\psi\|)\leq \|H^{k}\psi\|+\|\psi\| \leq c'_{k}(\|N^{2k}\psi\|+\|\psi\|)$ hold for all $\psi \in \mathcal{D}_{\mathrm{fin}}$ and for some positive constants $c_{k}$ and $c'_{k}$, thereby establishing the graph norm equivalence between $\|\cdot \|_{N^{2k}}$ and $\|\cdot \|_{H^{k}}$. It then suffices to show that $\mathcal{D}_{\mathrm{fin}}$ is a core for both $N^{2k}$ (which is trivially true) and $H^{k}$. This can be easily verified by writing $H^{k}$ as the sum of $W^{k}$ and lower-order remaining terms, which are infinitesimally $W^k$-bounded. Since $\mathcal{D}_{\mathrm{fin}}$ is a core for $W^{k}$, it follows from the Kato-Rellich theorem that it is also a core for $H^{k}$. Finally, by taking the closure, we obtain $\mathcal{D}({N^{2k}})=\overline{\mathcal{D}_{\mathrm{fin}}}^{\|\cdot \|_{N^{2k}}}=\overline{\mathcal{D}_{\mathrm{fin}}}^{\|\cdot \|_{H^{k}}}=\mathcal{D}({H^{k}})$.
\end{proof}

\begin{proof}[Proof of Proposition \ref{pro_bochner}]
	We simply note that $\mathcal{D}_{\mathrm{fin}}$ is invariant under $I$ and $e^{-\tau W}$. Using the boundedness of $e^{-(\beta-\tau)H}$ and $e^{-\tau W}$, and noting that for any fixed $\phi\in \mathcal{D}_{\mathrm{fin}}$ we have $\|I\phi\|\leq c_{V} \|\phi\|$ [cf.\,\eqref{Ipsi_relative_bound}] for some constant $c_{V}$ depending on $J, |V|$ and $\phi$, we establish a uniform upper bound for $\|e^{-(\beta-\tau )H} I e^{-\tau W}\psi\|$ over $\tau \in (0,\beta)$. Therefore,
	\begin{equation}\label{}
		\begin{aligned}[b]
			\ii{0}{\beta}{\tau}\|e^{-(\beta-\tau)H} I e^{-\tau W}\psi\|<\infty,
		\end{aligned}
	\end{equation}
	which completes the proof.
\end{proof}

\begin{proof}[Proof of Proposition \ref{pro_dyson_converge}]
	We first estimate the norm of $Ie^{-\tau W}$ for positive $\tau$. For positive $c_{1}$ and sufficiently large $c_{V}$ (depending on $|V|$ and model parameters), the operator $(c_{1}W+c_{V})^{-1/2}$ is well-defined, and we can write
	\begin{equation}\label{I_exp_a_W}
		\begin{aligned}[b]
			Ie^{-\tau W}= I (c_{1}W+c_{V})^{-1/2} \cdot (c_{1}W+c_{V})^{1/2} e^{-\tau W}.
		\end{aligned}
	\end{equation}  
	It should be emphasized that $\mathcal{D}_{\mathrm{fin}}$ is invariant under $(c_{1}W+c_{V})^{-1/2}$. Therefore, the operation in \eqref{I_exp_a_W} is mathematically rigorous, and the RHS is a densely defined operator with domain containing $\mathcal{D}_{\mathrm{fin}}$. By functional calculus, the norm of $(c_{1}W+c_{V})^{1/2} e^{-\tau W}$ is bounded by $(c_{1}/(2e \tau))^{-1/2}\exp(c_{V}\tau/c_{1})$. By establishing that $\|I\psi\|\leq c'_{V}\|(c_{1}W+c_{V})^{1/2}\psi\|$ for all $\psi\in \mathcal{D}_{\mathrm{fin}}$, we deduce that $I (c_{1}W+c_{V})^{-1/2}$ is bounded on $\mathcal{D}_{\mathrm{fin}}$ and therefore extends to the entire $\mathcal{H}$. Although the constant $c_{V}'$ could be optimized to be independent of $|V|$, this looser bound is sufficient for our purposes. Consequently, we obtain 
	\begin{equation}\label{}
		\begin{aligned}[b]
			\|Ie^{-\tau W}\|\leq \frac{c''_{V}}{\sqrt{\tau}}e^{\frac{c_{V}\tau}{c_{1}}},
		\end{aligned}
	\end{equation}
	for some $c_{V}''>0$ depending only on $|V|$ and model parameters. 
	Then, for a fixed $m$, let $c_{\beta,V}<\infty$ denote the upper bound for $\|e^{-(\beta-\tau_{1})W}\|$, which also relies only on  $|V|$ and model parameters. We deduce
	\allowdisplaybreaks[4]
	\begin{align*}\label{}
		&\ii{0}{\beta}{\tau_{1}}\ii{0}{\tau_{1}}{\tau_{2}}\dots\ii{0}{\tau_{m-1}}{\tau_{m}}\|e^{-(\beta-\tau_{1})W}Ie^{-(\tau_{1}-\tau_{2})W}Ie^{-(\tau_{2}-\tau_{3})W}\dots e^{-(\tau_{m-1}-\tau_{m})W}Ie^{-\tau_{m}W}\|
		\\\leq& c_{\beta,V} (c''_{V})^{m}e^{\frac{c_{V}\beta}{c_{1}}}\ii{0}{\beta}{\tau_{1}}\ii{0}{\tau_{1}}{\tau_{2}}\dots\ii{0}{\tau_{m-1}}{\tau_{m}}(\tau_{1}-\tau_{2})^{-1/2}(\tau_{2}-\tau_{3})^{-1/2}\dots (\tau_{m-1}-\tau_{m})^{-1/2}\tau_{m}^{-1/2}
		\\= &c_{\beta,V}(c''_{V})^{m}e^{\frac{c_{V}\beta}{c_{1}}}\frac{(\pi \beta)^{m/2}}{\Gamma(\frac{m}{2} + 1)}.
		\refstepcounter{equation}\tag{\theequation}
	\end{align*}
	Summing this bound over $m$ yields
	\begin{equation}\label{}
		\begin{aligned}[b]
			\sum_{m=0}^{\infty}c_{\beta,V}(c''_{V})^{m}e^{\frac{c_{V}\beta}{c_{1}}}\frac{(\pi \beta)^{m/2}}{\Gamma(\frac{m}{2} + 1)}<\infty,
		\end{aligned}
	\end{equation}
	which converges for any $\beta>0$ as long as $|V|$ is finite. Note that the integral here is a standard Liouville-Dirichlet-type integral. Therefore, the corresponding Dyson series converges absolutely.
\end{proof}

\nocite{Bibtexkey}
\bibliographystyle{unsrt}
\bibliography{ref}
\end{document}